\documentclass[reprint,prd,nofootinbib,superscriptaddress,11pt]{revtex4}
\usepackage{amssymb,wrapfig,amsmath,graphicx,epsfig,bm,color,verbatim,slashed,mathtools,xparse,adjustbox,dcolumn,float,relsize,tikz,cancel}
\usepackage[hidelinks]{hyperref}
\usepackage{caption,booktabs,array}
\usepackage{subcaption,multirow}
\usepackage[normalem]{ulem}
\usepackage[english]{babel}
\usepackage{soul}
\usepackage{placeins}
\usepackage[ntheorem]{empheq}

\newcolumntype{P}[1]{>{\centering\arraybackslash}p{#1}}
\newcolumntype{C}{>{$}c<{$}}
\AtBeginDocument{
\heavyrulewidth=.08em
\lightrulewidth=.05em
\cmidrulewidth=.03em
\belowrulesep=.65ex
\belowbottomsep=0pt
\aboverulesep=.4ex
\abovetopsep=0pt
\cmidrulesep=\doublerulesep
\cmidrulekern=.5em
\defaultaddspace=.5em
}

\pdfoutput=1
\newcommand{\func}{\operatorname}
\DeclareUnicodeCharacter{2212}{-}

\newcommand{\degree}{\ensuremath{^\circ}}

\begin{document}

\title{Global analysis of a minimally extended scotogenic model}

\author{Huchan Lee}
\email{huchanlee@seoultech.ac.kr}
\affiliation{School of Natural Science, Seoul National University of Science and Technology, Gongreung-ro 232, Seoul 01811, Republic of Korea}

\author{Sin Kyu Kang}
\email{skkang@seoultech.ac.kr}
\affiliation{School of Natural Science, Seoul National University of Science and Technology, Gongreung-ro 232, Seoul 01811, Republic of Korea}

\date{\today }

\begin{abstract}
    We perform a global analysis of a minimally extended scotogenic model motivated by observed non-zero neutrino masses, viable dark matter (DM) candidates, and the instability of the Standard Model (SM) vacuum at high-energies. We examine the bounded-from-below conditions, vacuum stability, and RG-driven perturbativity bounds arising from the extended scalar sector, alongside a comprehensive set of flavor and electroweak (EW) precision observables - including the muon anomalous magnetic moment $\Delta a_{\mu}$, the radiative decays $\ell_{\alpha} \rightarrow \ell_{\beta} \gamma$ and $\ell_{\alpha} \rightarrow 3\ell_{\beta}$, and the $\mu \rightarrow e$ conversion rate, the oblique parameters, and leptonic decays of $Z$ and $H$ bosons. A numerical scan reveals four notable features: the DESI BAO bound would rule out the inverted hierarchy if confirmed by other experiments; the oblique parameters are projected to be within the reach of future precision measurements; the viable fermionic DM candidate mass lies in the range $120-350 \operatorname{GeV}$, while the CP-odd scalar is constrained to $350-600 \operatorname{GeV}$; and our result on $Z \rightarrow \operatorname{Invisible}$ is compatible with the world average at the $3\sigma$ level and is favored by the recent ATLAS measurement at the $3\sigma$ level.
\end{abstract}

\maketitle

\newpage

\tableofcontents

\newpage
\section{Introduction}\label{sec:introduction}

The Standard Model (SM) has successfully described a wide range of phenomena with remarkable precision. Despite its great success, several fundamental observations remain unexplained within the SM framework, most notably the experimentally established neutrino masses and mixing, and the existence of dark matter (DM). The original scotogenic model~\cite{Ma:2006km}, which simultaneously accounts for both of these phenomena, has therefore attracted significant attention. The scotogenic model, however, inherits a theoretical tension already present in the SM: the metastability of the electroweak (EW) vacuum~\cite{Hiller:2024zjp}. For the experimentally measured values of the Higgs boson and top quark masses, the renormalization group (RG) evolution of the SM Higgs quartic coupling drives it to negative values at scales of order $10^{9} \func{GeV}$, signaling an instability of the scalar potential at high energies. In this work, we address this issue by extending the scotogenic scalar sector with an additional scalar field that mixes with the SM Higgs. The presence of this new scalar modifies the RG running of the quartic couplings, allowing them to remain positive across the energy scales where the theory remains perturbative. Consequently, EW vacuum stability is restored through the combined effects of the scalar mixing and RG evolution of the quartic couplings.


This limitation motivates a minimal extension of the original scotogenic model, designed to simultaneously address the EW vacuum stability issue while preserving the model's predictivity and phenomenological viability. We begin by discussing the neutrino mass generation mechanism under both normal and inverted hierarchies. Cosmological observations from the Planck collaboration place a stringent $95\%$ confidence level (C.L.) upper bound on the sum of the three active neutrino masses~\cite{Planck:2018vyg}:
\begingroup
\begin{equation}\label{eqn:cos_numass}
    \sum_{i=1}^{3} m_{\nu_{i}} \leq 10^{-10}\func{GeV}.
\end{equation}
\endgroup
We then identify the viable parameter space for each hierarchy subject to all relevant theoretical constraints, including bounded-from-below (BFB) conditions, vacuum stability, and perturbativity bounds induced by RG running. Within this framework, we compute the precision and flavor observables --- namely, the radiative decays $\ell_{\alpha} \rightarrow \ell_{\beta} \gamma$ and $\ell_{\alpha} \rightarrow 3\ell_{\beta}$, the conversion ratio of $\mu \rightarrow e$, the oblique parameters, and the leptonic and invisible decay modes of the $Z$ and Higgs bosons ($Z/H \rightarrow \overline{\ell}_{\alpha}\ell_{\beta},\, \overline{\nu}\nu$) --- while performing a full renormalization analysis without imposing simplifying assumptions on the scalar sector. In this model, the SM scalar sector is extended by an additional scalar field that acquires a nonzero vacuum expectation value (VEV) and mixes with the SM Higgs. Consequently, tadpole contributions cannot be absorbed solely through counter-terms, as is conventionally done in the SM. We therefore adopt an alternative tadpole scheme~\cite{Fleischer:1980ub} in which one-particle-irreducible (1PI) contributions are consistently combined with explicit tadpole insertions wherever they arise. Within this framework, we compute each precision observable in both the SM and the present model, and take their differences as a measure of the predicted new physics (NP) effects.

Given the flavor observables and constraints arising from both the lepton and scalar sectors, we perform a numerical parameter scan. In view of the complexity and high-dimensionality of the parameter space, we employ a Markov Chain Monte Carlo (MCMC) algorithm~\cite{Markov1906_reprint} constrained by two key observables - the SM Higgs mass at one-loop and the relic density - rather than a random scan. A distinguishing feature of this work is the absence of a large-Yukawa driven analysis. Numerous previous studies have attempted to accommodate the muon anomalous magnetic moment at the $4.2\sigma$ level or above, based on the 2020 White paper~\cite{Aoyama:2020ynm}, which generically requires Yukawa couplings of order unity and consequently produces a significant enhancement of other flavor observables. Such a large-Yukawa-driven analysis is, however, disfavored by the RG analysis detailed in the following section. Furthermore, the 2025 WP paper~\cite{Aliberti:2025beg}, incorporating a revised hadronic vaccum polarization (HVP) contribution evaluated from lattice QCD, reports an updated SM muon $g-2$ prediction yielding:
\begingroup
\begin{equation}
    \Delta a_{\mu} = a_{\mu}^{\func{exp}} - a_{\mu}^{\func{SM}} = 38(63) \times 10^{-11},
\end{equation}
\endgroup
corresponding to a $0.6\sigma$ deviation, which indicates no statistically significant tension between the SM prediction and the experimental world average. It is therefore of considerable importance to identify viable directions for New Physics (NP) searches under these circumstances, and in this work we investigate observables accessible in near-future experiments without invoking artificially enhanced parameters.
\\~\\
This work is organized as follows: Section~\ref{sec:model} introduces the model under study, including the neutrino mass generation mechanism for both mass hierarchies and the DM phenomenology. Section~\ref{sec:scalar_potential} examines the constraints arising from the extended scalar sector, with particular emphasis on the BFB conditions, vacuum stability, and RG-driven perturbativity bounds. Section~\ref{sec:flavor_EWobs} is devoted to flavor and EW precision observables. In Section~\ref{sec:numerical_analysis}, we present the numerical analysis and discuss the resulting phenomenological implications. Finally, Section~\ref{sec:conclusion} summarizes our main findings. Appendix~\ref{subapp:CI} provides details of the Casas-Ibarra parameterization employed in this work. Appendix~\ref{app:SE} presents the self-energy contributions incorporating tadpole insertions. Appendix~\ref{app:bfb} details the derivation of the BFB conditions. Finally, Appendix~\ref{app:RGEs} collects all the RGEs used in this analysis.

\section{Model}\label{sec:model}
The model under study is seen in Table~\ref{tab:BSM_model}.
\begingroup
\begin{table}[t]
\setlength{\tabcolsep}{5pt}
\centering\renewcommand{\arraystretch}{1.5} 
\begin{tabular}{c|ccccc|c|ccc}
\toprule
\toprule
Field & $Q_{i}$ & $u_{i}^{c}$ & $d_{i}^{c}$ & $L_{i}$ & $e_{i}^{c}$ & $N_{i}^{c}$ & $H$ & $\eta$ & $S$ \\ 
\midrule
$SU(3)_C$ & $\mathbf{3}$ & $\mathbf{\bar{3}}$ & $\mathbf{\bar{3}}$ & $\mathbf{1}$ & $\mathbf{1}$ & $\mathbf{1}$ & $\mathbf{1}$ & $\mathbf{1}$ & $\mathbf{1}$ \\ 
$SU(2)_L$ & $\mathbf{2}$ & $\mathbf{1}$ & $\mathbf{1}$ & $\mathbf{2}$ & $\mathbf{1}$ & $\mathbf{1}$ & $\mathbf{2}$ & $\mathbf{2}$ & $\mathbf{1}$ \\ 
$U(1)_Y$ & $\frac{1}{6}$ & $-\frac{2}{3}$ & $\frac{1}{3}$ & $-\frac{1}{2}$ & $1$ & $0$ & $\frac{1}{2}$ & $\frac{1}{2}$ & $0$ \\
$Z_{2}$ & $1$ & $1$ & $1$ & $1$ & $1$ & $-1$ & $1$ & $-1$ & $1$ \\
$U(1)_X$ & $0$ & $0$ & $0$ & $-1$ & $1$ & $1$ & $0$ & $0$ & $-2$ \\ 
\bottomrule
\bottomrule
\end{tabular}
\caption{Particle content of the minimally extended scotogenic model. The new fermions $N_{iR}$ are right-handed (RH) neutrinos, with the index $i=1,2,3$. The $Z_{2}$ charge $1(-1)$ corresponds to even(odd) parity, respectively. Compared to the original scotogenic model, the present model is extended by an additional gauge-singlet scalar $S$, which acquires a nonzero VEV, as well as a global $\func{U}\left( 1 \right)_{X}$ symmetry. All the fermions are two-component left-chiral Weyl spinors.} 
\label{tab:BSM_model}
\end{table}
\endgroup
The renormalizable Lagrangian with the particle content reads in Equation~\ref{eqn:ren_Lag}:
\begingroup
\begin{equation}\label{eqn:ren_Lag}
    \mathcal{L_{\func{ren}}} = y_{u} Q_{i} \widetilde{H} u_{i}^{c} - y_{d} Q_{i} H d_{i}^{c} - y_{e} L_{i} H e_{i}^{c} + g_{X} L_{i} \eta N_{i}^{c} + \frac{g_{R}}{2} S N_{i}^{c} N_{i}^{c} + \func{h.c.}.
\end{equation}
\endgroup
where $y_{u,d,e}$ denote the SM Yukawa couplings while $g_{X,R}$ are new Yukawa couplings associated with the extended sector.  These new couplings are constrained by neutrino oscillation data. Here, we introduce the global $U\left( 1 \right)_{X}$ symmetry in order to generate RH neutrino mass terms through Yukawa-like interactions and to make explicit how masses are assigned to the new scalar fields. This approach has the advantage that the associated Yukawa couplings can be consistently included in the RG evolution of all free parameters, which will be detailed in Section~\ref{subsec:RG_driven_perturbativity}, thereby enabling a more comprehensive analysis. The mass matrix for the RH neutrinos $N_{iR}$ is parameterized in the flavor basis and the diagonalized RH mass matrix is written in the physical basis of $\chi_{i} \left( i = 1, 2, 3 \right)$:
\begingroup
\begin{equation}\label{eqn:diag_mchi}
\begin{split}
    M_{\chi}^{\func{flavor}} = \frac{1}{\sqrt{2}} g_{R}^{ij} v_{S} \quad \text{$i,j = 1, 2, 3$}, \quad U^{\chi,*} M_{\chi}^{\func{flavor}} U^{\chi,\dagger} = M_{\chi}^{\func{diag}}.
\end{split}
\end{equation}
\endgroup
where $U^{\chi}$ is a unitary mixing matrix which diagonalize the Majorana RH neutrino mass matrix. The extended scalar potential is given in Equation~\ref{eqn:ren_sp}:
\begingroup
\begin{equation}\label{eqn:ren_sp}
\begin{split}
    V_{\func{ren}} &= -\mu_{H}^{2} H^{\dagger} H + \mu_{\eta}^{2} \eta^{\dagger} \eta - \mu_{S}^{2} S^{*} S - \frac{1}{2} \mu_{\func{sb}}^{2} (S^{2} + S^{*,2}) \\
    &+ \frac{1}{4} \lambda_{1} ( H^{\dagger} H )^{2} + \frac{1}{4} \lambda_{2} ( \eta^{\dagger} \eta )^{2} + \lambda_{3} ( H^{\dagger} H ) ( \eta^{\dagger} \eta ) + \lambda_{4} ( H^{\dagger} \eta ) ( \eta^{\dagger} H ) \\
    &+ \frac{1}{4} \lambda_{5} \left[ (\eta^{\dagger} H)^2 + h.c. \right] + \frac{1}{4} \lambda_{6} ( S^{*} S )^{2} + \lambda_{7} ( H^{\dagger} H ) ( S^{*} S ) + \lambda_{8} ( \eta^{\dagger} \eta ) ( S^{*} S )
\end{split}
\end{equation}
\endgroup
where $\mu_{H,\eta,S}$ are dimensionful mass parameters, and $\mu_{\func{sb}}$ is a soft-breaking term that gives rise to nonzero masses for the CP-odd scalar. The scalar fields are decomposed as follows after spontaneous symmetry breaking (SSB):
\begingroup
\begin{equation}\label{eqn:higgses}
\begin{split}
H = \begin{pmatrix}
    G^{+} \\
    \frac{1}{\sqrt{2}} \left( v + \phi_{H^{0}} + i \sigma_{H^{0}} \right)    
\end{pmatrix}, \quad
\eta = \begin{pmatrix}
    \eta^{+} \\
    \frac{1}{\sqrt{2}} \left( \eta_{R} + i \eta_{I} \right)    
\end{pmatrix}, \quad
S = \frac{1}{\sqrt{2}} \begin{pmatrix}
    v_{S} + \phi_{S} + i \sigma_{S}
\end{pmatrix},
\end{split}
\end{equation}
\endgroup
here, $G^{+}$ is the would-be Goldstone boson associated with the SM $W$ gauge boson, $\eta^{+}$ is a physical charged scalar field, $v$ is the SM VEV, $v_{S}$ is the VEV of the new singlet scalar $S$. The fields $\phi$ and $\sigma$ correspond to the CP-even and -odd components of the scalars, respectively. The mass matrix for the CP-even scalars in the flavor basis $\left( \phi_{H^{0}}, \phi_{S} \right)$ is:
\begingroup
\begin{equation}\label{eqn:diag_CPeven}
\begin{split}
    M_{\func{CP-even}}^{2} &= \begin{pmatrix}
        \frac{1}{4} (-4 \mu_{H}^{2} + 3 \lambda_{1} v^{2} + 2 \lambda_{7} v_{S}^{2}) & \lambda_{7} v v_{S} \\
        \lambda_{7} v v_{S} & \frac{1}{4} (-4 (\mu_{S}^{2} + \mu_{\func{sb}}^{2}) + 2 \lambda_{7} v^{2} + 3 \lambda_{6} v_{S}^{2})
    \end{pmatrix}
\end{split}
\end{equation}
\endgroup
and the diagonalized CP-even mass matrix by a unitary mixing matrix $U^{H}$ is given in the physical basis of $h_{i} \left( i = 1, 2 \right)$:
\begingroup
\begin{equation}\label{eqn:diag_CPeven}
    U^{H} M_{\func{CP-even}}^{2} U^{H,T} = M_{\func{CP-even}}^{\func{diag},2}, \quad 
    \left( h_{1}, h_{2} \right)^{T} = U_{H}^{T} \left( \phi_{H^{0}}, \phi_{S} \right)^{T}.
\end{equation}
\endgroup
In the CP-odd sector, the mixing matrix is the identity matrix, so the CP-odd scalars in the flavor basis $\left( \sigma_{H^{0}}, \sigma_{S} \right)$ coincide with the mass eigenstates $a_{1}$ and $a_{2}$, where $a_{1}$ is the would-be Goldstone boson of the $Z$ and $a_{2}$ is a physical CP-odd scalar of the extended sector.

\subsection{Neutrino mass generation}\label{subsec:numass}
The active neutrino masses are generated at the one-loop level through interactions with new particles shown in Figure~\ref{fig:numass_generation}.
\begingroup
\begin{figure}[t]
    \centering
    \includegraphics[width=0.3\textwidth]{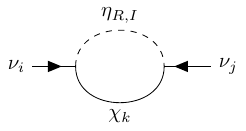}
    \caption{Neutrino mass generation diagram at the one-loop level. Here, $\chi_{k}$ denote the physical RH Majorana neutrinos $\left( k = 1, 2, 3 \right)$ and $\eta_{R,I}$ are the physical CP-even and -odd components of the scalar $\eta$, respectively}\label{fig:numass_generation}
\end{figure}
\endgroup
The one-loop neutrino mass matrix $M_{\nu}$ can be written as:
\begingroup
\begin{equation}\label{eqn:oneloop_Mnu}
    M_{\nu} = g_{X}^{T} M_{L} g_{X}.
\end{equation}
\endgroup
where $g_{X}$ is the $3 \times 3$ Yukawa coupling constants and $M_{L}$ is a symmetric $3 \times 3$ one-loop mass matrix given in Equation~\ref{eqn:oneloop_ML}:
\begingroup
\begin{equation}\label{eqn:oneloop_ML}
\begin{split}
    M_{L}^{11} &= \frac{1}{2} \sum_{k=1}^{3} \left( b_{\func{loop}} (\eta_{R}, k) - b_{\func{loop}} (\eta_{I}, k) \right) \left( U_{(k,1)}^{\chi,*} \right)^{2}, \\
    M_{L}^{22} &= \frac{1}{2} \sum_{k=1}^{3} \left( b_{\func{loop}} (\eta_{R}, k) - b_{\func{loop}} (\eta_{I}, k) \right) \left( U_{(k,2)}^{\chi,*} \right)^{2}, \\
    M_{L}^{33} &= \frac{1}{2} \sum_{k=1}^{3} \left( b_{\func{loop}} (\eta_{R}, k) - b_{\func{loop}} (\eta_{I}, k) \right) \left( U_{(k,3)}^{\chi,*} \right)^{2}, \\
    M_{L}^{12} = M_{L}^{21} &= \frac{1}{2} \sum_{k=1}^{3} \left( b_{\func{loop}} (\eta_{R}, k) - b_{\func{loop}} (\eta_{I}, k) \right) \left( U_{(k,1)}^{\chi,*} \right) \left( U_{(k,2)}^{\chi,*} \right), \\
    M_{L}^{13} = M_{L}^{31} &= \frac{1}{2} \sum_{k=1}^{3} \left( b_{\func{loop}} (\eta_{R}, k) - b_{\func{loop}} (\eta_{I}, k) \right) \left( U_{(k,1)}^{\chi,*} \right) \left( U_{(k,3)}^{\chi,*} \right), \\
    M_{L}^{23} = M_{L}^{32} &= \frac{1}{2} \sum_{k=1}^{3} \left( b_{\func{loop}} (\eta_{R}, k) - b_{\func{loop}} (\eta_{I}, k) \right) \left( U_{(k,2)}^{\chi,*} \right) \left( U_{(k,3)}^{\chi,*} \right), 
\end{split}
\end{equation}
\endgroup
and the loop function $b_{\func{loop}}$ is defined as:
\begingroup
\begin{equation}\label{eqn:bloop}
    b_{\func{loop}} \left( \phi, k \right) = \frac{1}{16 \pi^2} \frac{M_{\chi}^{k}}{M_{\phi}^{2} - M_{\chi}^{k,2}} \left( M_{\phi}^{2} \ln{\frac{M_{\phi}^{2}}{M_{\chi}^{k,2}}} \right)
\end{equation}
\endgroup
where $\phi = \eta_{R,I}$. The coupling matrix $g_{X}$ can be expressed using the Casas-Ibarra parameterization~\cite{Casas:2001sr}, which is detailed in Appendix~\ref{subapp:CI}. The recent neutrino experimental data is given in Table~\ref{tab:exp_neutrino_data}.
\begingroup
\begin{table}[t]
\setlength{\tabcolsep}{5pt}
\centering\renewcommand{\arraystretch}{1.5} 
\begin{tabular}{c|cc|cc}
\toprule
\toprule
 & \multicolumn{2}{c}{Normal hierarchy} & \multicolumn{2}{c}{Inverted hierarchy} \\ \cmidrule(lr){2-3} \cmidrule(lr){4-5}
 & $1\sigma$ range & $3\sigma$ range & $1\sigma$ range & $3\sigma$ range \\
 \midrule
 $\sin^2_{\theta_{12}}$ & $0.3088_{-0.0066}^{+0.0067}$ & $0.2893 \rightarrow 0.3295$ & $0.3088_{-0.0066}^{+0.0067}$ & $0.2893 \rightarrow 0.3295$ \\
  $\theta_{12}/\degree$ & $33.76_{-0.41}^{+0.42}$ & $32.54 \rightarrow 35.03$ & $33.76_{-0.41}^{+0.42}$ & $32.54 \rightarrow 35.03$ \\
 $\sin^2_{\theta_{23}}$ & $0.470_{-0.014}^{+0.017}$ & $0.432 \rightarrow 0.587$ & $0.555_{-0.016}^{+0.013}$ & $0.437 \rightarrow 0.590$ \\
  $\theta_{23}/\degree$ & $43.27_{-0.82}^{+1.00}$ & $41.11 \rightarrow 50.02$ & $48.15_{-0.92}^{+0.75}$ & $41.40 \rightarrow 50.21$ \\ 
 $\sin^2_{\theta_{13}}$ & $0.02249_{-0.00057}^{+0.00057}$ & $0.02070 \rightarrow 0.02420$ & $0.02261_{-0.00056}^{+0.00056}$ & $0.02091 \rightarrow 0.02433$ \\
  $\theta_{13}/\degree$ & $8.62_{-0.11}^{+0.11}$ & $8.27 \rightarrow 8.95$ & $8.65_{-0.11}^{+0.11}$ & $8.31 \rightarrow 8.97$ \\ 
 $\delta_{CP}/\degree$ & $207_{-20}^{+23}$ & $114 \rightarrow 405$ & $283_{-28}^{+24}$ & $202 \rightarrow 347$ \\ 
 $\frac{\Delta m_{21}^{2}}{10^{-5} \func{eV}^{2}}$ & $7.537_{-0.100}^{+0.094}$ & $7.236 \rightarrow 7.823$ & $7.537_{-0.100}^{+0.094}$ & $7.236 \rightarrow 7.822$ \\ 
 $\frac{\Delta m_{3l}^{2}}{10^{-3} \func{eV}^{2}}$ & $2.521_{-0.018}^{+0.026}$ & $2.454 \rightarrow 2.592$ & $-2.500_{-0.023}^{+0.024}$ & $-2.569 \rightarrow -2.430$ \\ 
\bottomrule
\bottomrule
\end{tabular}
\caption{Recent neutrino experimental data given by NuFit 6.1~\cite{Esteban:2024eli}} 
\label{tab:exp_neutrino_data}
\end{table}
\endgroup

\subsection{Normal and inverted hierarchies}\label{subsec:nhih}
In this work, we also investigate the viable parameter space for both neutrino mass hierarchies, normal (NH) and inverted (IH). The hierarchies are implemented using the recent NuFit data given in Table~\ref{tab:exp_neutrino_data}, with the corresponding implementation summarized in Table~\ref{tab:nuhierarchy_implemented}.l
\begingroup
\begin{table}[t]
\setlength{\tabcolsep}{5pt}
\centering\renewcommand{\arraystretch}{1.5} 
\begin{tabular}{c|c|c|c}
\toprule
\toprule
 & Normal hierarchy &  & Inverted hierarchy \\
\midrule
$m_{\nu_{3}}$ & $\sqrt{m_{\nu_{2}}^{2}+\Delta m_{32}^{2}}$ & $m_{\nu_{2}}$ & $\sqrt{m_{\nu_{1}}^{2} + \Delta m_{21}^{2}}$ \\
$m_{\nu_{2}}$ & $\sqrt{m_{\nu_{1}}^2+\Delta m_{21}^{2}}$ & $m_{\nu_{1}}$ & $\sqrt{m_{\nu_{3}}^{2}+|\Delta m_{31}^{2}}|$ \\
$m_{\nu_{1}}$ & $[10^{-19}, 10^{-10}]$ & $m_{\nu_{3}}$ & $[10^{-19}, 10^{-12}]$ \\
\bottomrule
\bottomrule
\end{tabular}
\caption{Neutrino mass hierarchy implemented in this work. All masses are in unit of $\func{GeV}$.} 
\label{tab:nuhierarchy_implemented}
\end{table}
\endgroup
The squared neutrino mass differences will be considered within the $3\sigma$ bounds, as detailed in the numerical section~\ref{sec:numerical_analysis}.

\subsection{Dark Matter candidates and constraints}\label{subsec:dm_candi_const}

The model under study features three potential DM candidates: the lightest fermionic RH neutrino $\chi_{1}$, the CP-even scalar $\eta_{R}$, and the CP-odd scalar $\eta_{I}$, depending on the parameter space. If the DM candidate is fermionic, the New Physics (NP) effects on most flavor observables, discussed in detail in Section~\ref{sec:flavor_EWobs}, arising from the extended fermionic sector are strongly constrained by known experimental inputs. The new coupling constants $g_{X}$ are fixed by neutrino oscillation data, while the new mass scale $m_{\chi_{1}}$ is determined by the relic density, thereby minimizing the number of free parameters. The experimental value of the relic density at $68\%$ confidence level (C.L.) is reported by the Planck Collaboration~\cite{Planck:2018vyg}:
\begingroup
\begin{equation}\label{eqn:relicCDM}
    \Omega_{\func{CDM}} h^{2} = 0.1200 \pm 0.0012.
\end{equation}
\endgroup
The $1\sigma$ error bar of the relic density can be broadened by including EW radiative corrections~\cite{Boudjema:2014gza,Harz:2016dql}. In this work, we adopt the corrected value in order to allow for a larger parameter space.
\begingroup
\begin{equation}\label{eqn:relicrenew}
    \Omega_{\func{CDM}} h^{2} = 0.1200 \pm 0.0120.
\end{equation}
\endgroup
To compute the relic density, we use the public HEP tool \texttt{micrOMEGAs}~\cite{Alguero:2023zol}, which allows for the evaluation of DM candidates, the relic abundance, and spin-dependent and spin-independent direct (and indirect) cross sections. In this work, we primarily constrain the parameter space using the relic density. However, if the direct or indirect cross sections provide additional constraining power, they will be taken into account.

\section{Scalar potential constraints}\label{sec:scalar_potential}

In this section, we discuss the theoretical constraints arising from the extended scalar sector, including the BFB conditions, vacuum stability, and perturbativity bounds induced by RG running of each coupling constant.

\subsection{Bounded-from-below conditions}\label{subsec:bfb}

As the scotogenic model contains an extended SM scalar potential, it is important to ensure that the scalar potential is bounded from below (BFB). This requires analyzing all possible directions in field space along which the scalar fields can take arbitrarily large values, and ensuring that the potential remains positive in each such direction. The resulting BFB conditions can be expressed in terms of the quartic coupling constants, and are given in Equation~\ref{eqn:bfb}:

\begingroup
\begin{equation} \label{eqn:bfb}
    \begin{split}
        \lambda_{1} &> 0 \\
        \lambda_{2} &> 0 \\
        \lambda_{6} &> 0 \\
        \lambda_{8} + \frac{1}{2} \sqrt{\lambda_{2} \lambda_{6}} &> 0 \\
        \lambda_{7} + \frac{1}{2} \sqrt{\lambda_{1} \lambda_{6}} &> 0 \\
        \lambda_{3} + \frac{1}{2} \sqrt{\lambda_{1} \lambda_{2}} &> 0 \\
        \lambda_{3} + \lambda_{4} + \frac{1}{2} \sqrt{\lambda_{1} \lambda_{2}} &> \frac{1}{2} \left| \lambda_{5} \right| \\
        \lambda_{a} \frac{\sqrt{\lambda_{1} \lambda_{2}}}{\lambda_{6}} + \lambda_{4} &> \frac{1}{2} \left| \lambda_{5} \right|
    \end{split}
\end{equation}
\endgroup
where $\lambda_{a}$ is 
\begingroup
\begin{equation}
    \frac{3}{4} \lambda_{6} + \frac{\lambda_{3} \lambda_{6}}{\sqrt{\lambda_{1} \lambda_{2}}} + \lambda_{7} \sqrt{\frac{\lambda_{6}}{\lambda_{1}}} + \lambda_{8} \sqrt{\frac{\lambda_{6}}{\lambda_{2}}}.
\end{equation}
\endgroup
The BFB conditions in Equation~\ref{eqn:bfb} are necessary and sufficient for the potential to be bounded-from-below, as they are derived by requiring $V \rightarrow \infty$ for all possible field directions in the large field limit. Deriving the BFB conditions is detailed in Appendix~\ref{app:bfb}.

\subsection{Vacuum stability}\label{subsec:vacuum_stability}

The SM scalar potential is known to lie near the border between stability and metastability, primarily due to the experimental uncertainties in the top quark mass and the strong coupling constant~\cite{Buttazzo:2013uya,Hiller:2024zjp}. In particular, Ref~\cite{Hiller:2024zjp} quantified this proximity using two top quark mass determinations: the pole mass $M_{t}^{\sigma} = \left( 172.4 \pm 0.7 \right) \func{GeV}$ and the Monte-Carlo mass $M_{t}^{\func{MC}} = \left( 172.57 \pm 0.29 \right) \func{GeV}$. The former requires a $1.9\sigma$ downward shift from its central value to achieve stability, while the more precisely measured latter requires a $5.1\sigma$ shift. Furthermore, the strong coupling constant $\alpha_{S}^{\left( 5 \right)} \left( M_{Z} \right)$ requires a $3.7\sigma$ upward shift from the PDG world average to stabilize the potential. These deviations suggest that the SM scalar potential is metastable, motivating the introduction of an additional scalar field that acquires a non-zero VEV to stabilize the potential. The extended scalar potential considered in this work can develop multiple minima, and it is therefore necessary to verify that the electroweak symmetry breaking (EWSB) minimum corresponds to the global minimum. To this end, we make use of the HEP tool \texttt{Vevacious}~\cite{Camargo-Molina-OLeary:2014} (\texttt{C++} version: \cite{Camargo-Molina:2013qva}).

\subsection{Renormalization-Group-driven perturbative bounds}\label{subsec:RG_driven_perturbativity}

It is important to determine up to which energy scale a given theory remains perturbatively valid. This question can be systematically addressed by analyzing the Renormalization-Group (RG) flow of the coupling constants. Such an RG analysis is essential for the following reasons:
\begingroup
\begin{enumerate}
    \item Perturbativity must be maintained across the entire energy range over which the theory is considered to be valid, not only at the scale where the initial conditions are imposed (e.g., the electroweak scale).
    \item In general, RG-improved perturbativity bounds place stronger constraints on the parameter space than the conventional bounds $g_{i}, y_{i} \leq \sqrt{4 \pi}$ and $\lambda_{i} \leq 4 \pi$, since couplings that appear perturbative at low energies may develop a Landau pole or enter a non-perturbative regime at higher scales.
\end{enumerate}
\endgroup
A more detailed discussion of the RG evolution is provided in Ref.~\cite{CarcamoHernandez:2023wzf}. For comparison, the RG behavior of the SM and the scotogenic model investigated in this work is illustrated in Figure~\ref{fig:RG_analysis}. We recall that the motivation for generating the RH neutrino masses through a Higgs-like mechanism in the Lagrangian of Eq.~\ref{eqn:ren_Lag} is to explicitly include the RH neutrino Yukawa couplings in the RG analysis. As illustrated in Figure~\ref{fig:RG_analysis}, the coupling constants in this framework remain perturbative provided that the initial values of the new parameters are constrained to remain below $0.2$ at the NP scale. Specifically, the quartic couplings must also remain below this threshold; otherwise, a Landau pole is rapidly developed, thereby pushing the theory into a non-perturbative regime. Generating the RH neutrino masses via a Higgs-like mechanism further motivates the imposition of a global $U(1)_{X}$ symmetry to constrain the structure of the allowed Yukawa interactions. Another important feature to address is the RG evolution of the SM quartic coupling. As illustrated in Figure~\ref{fig:RG_analysis}, the SM quartic coupling becomes negative at a scale of approximately $10^{9} \func{GeV}$, implying that the scalar potential becomes unstable and unbounded-from-below at high field values. To ensure vacuum stability, additional scalar interactions must be introduced, as implemented in the present model. Consequently, the SM-like quartic coupling $\lambda_{1}$ remains positive across the entire energy range due to the contributions from these new interactions. These requirements are formally addressed through the derived BFB conditions in Section~\ref{subsec:bfb}. The RGEs employed in this analysis are provided in Appendix~\ref{app:RGEs}.
\begingroup
\begin{figure}[t]
    \centering
    \begin{subfigure}{0.48\textwidth}
        \centering
        \includegraphics[width=\textwidth]{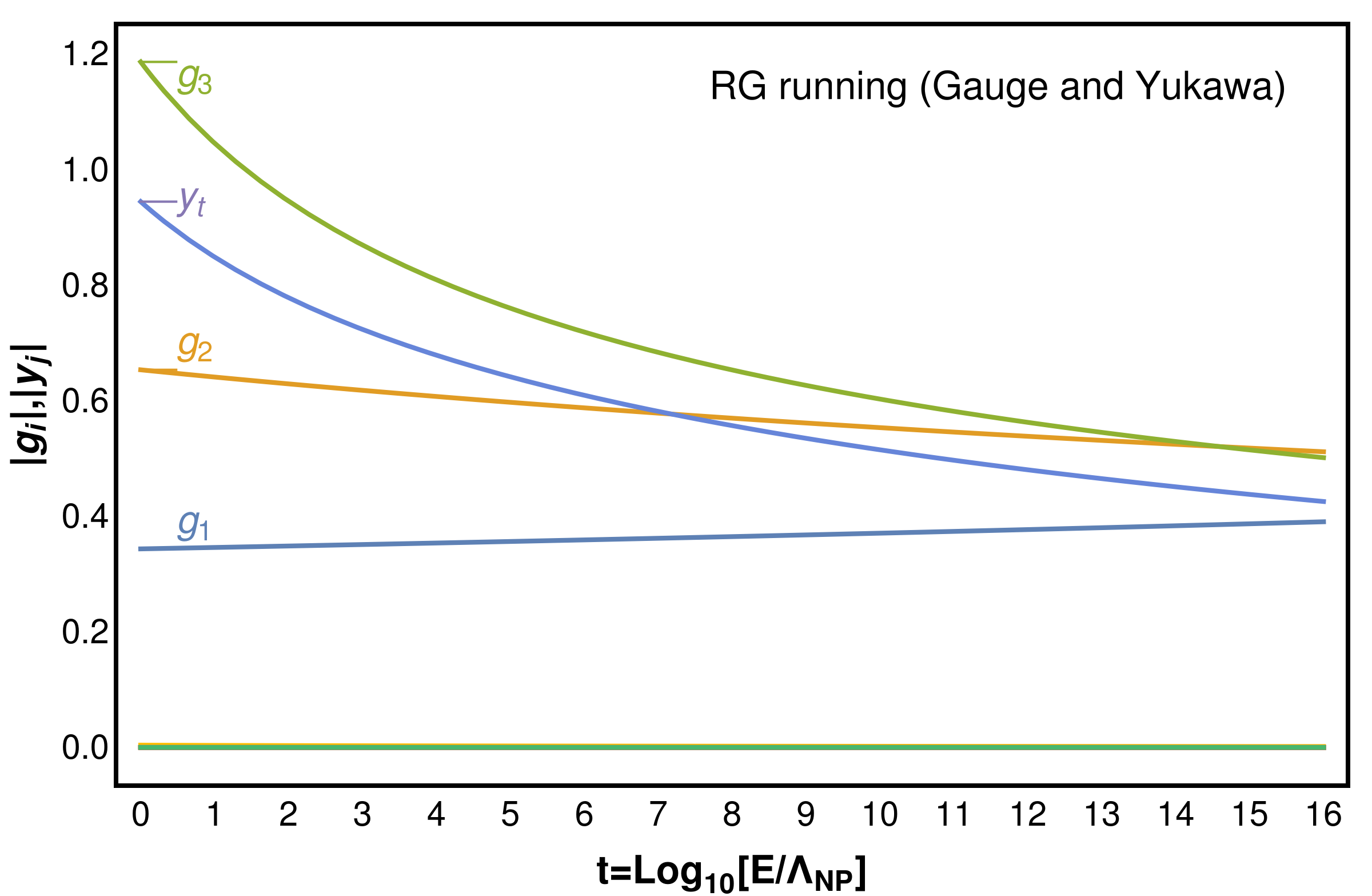}
    \end{subfigure}
    \begin{subfigure}{0.48\textwidth}
        \centering
        \includegraphics[width=\textwidth]{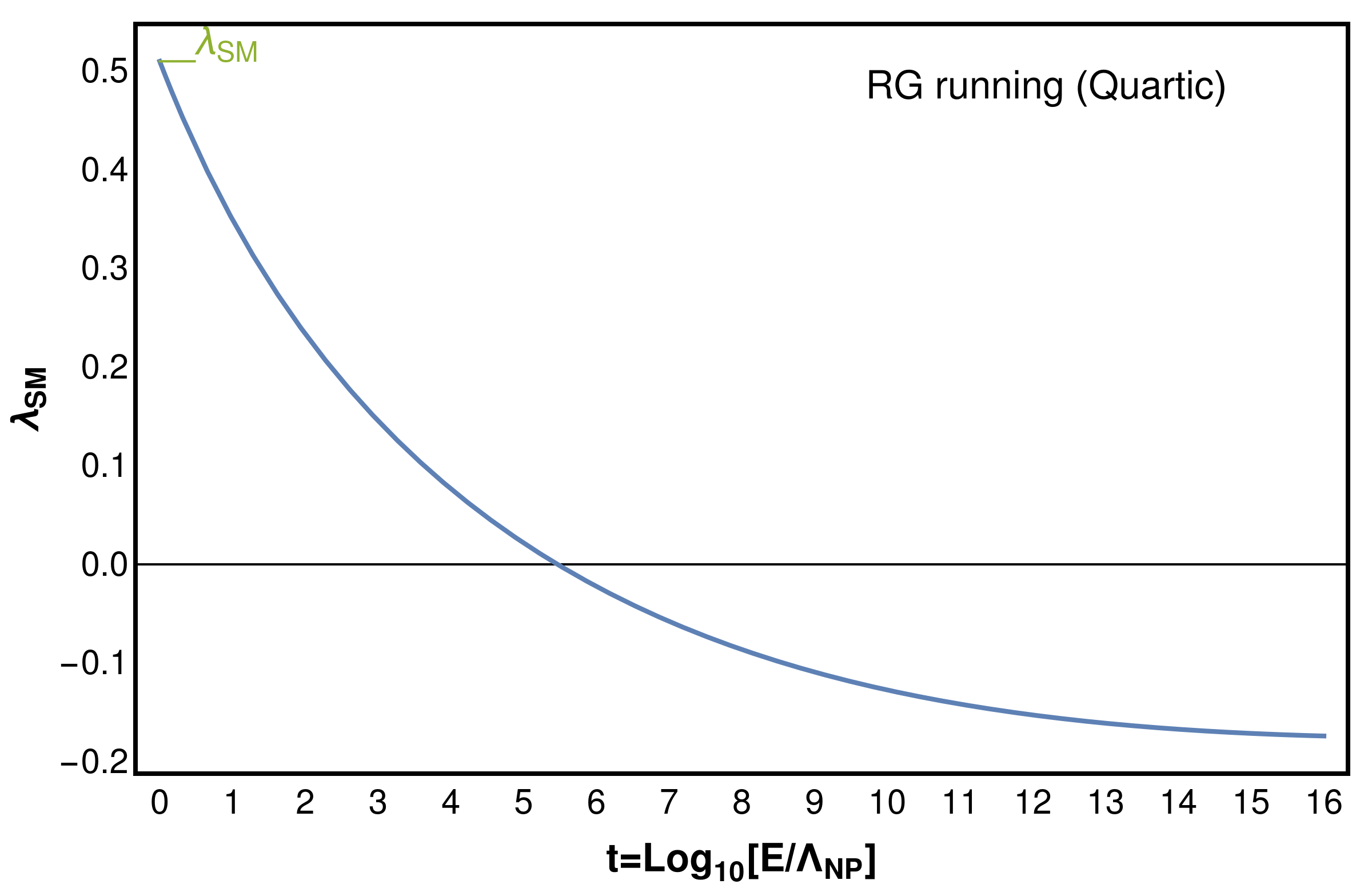}
    \end{subfigure}
    \begin{subfigure}{0.48\textwidth}
        \centering
        \includegraphics[width=\textwidth]{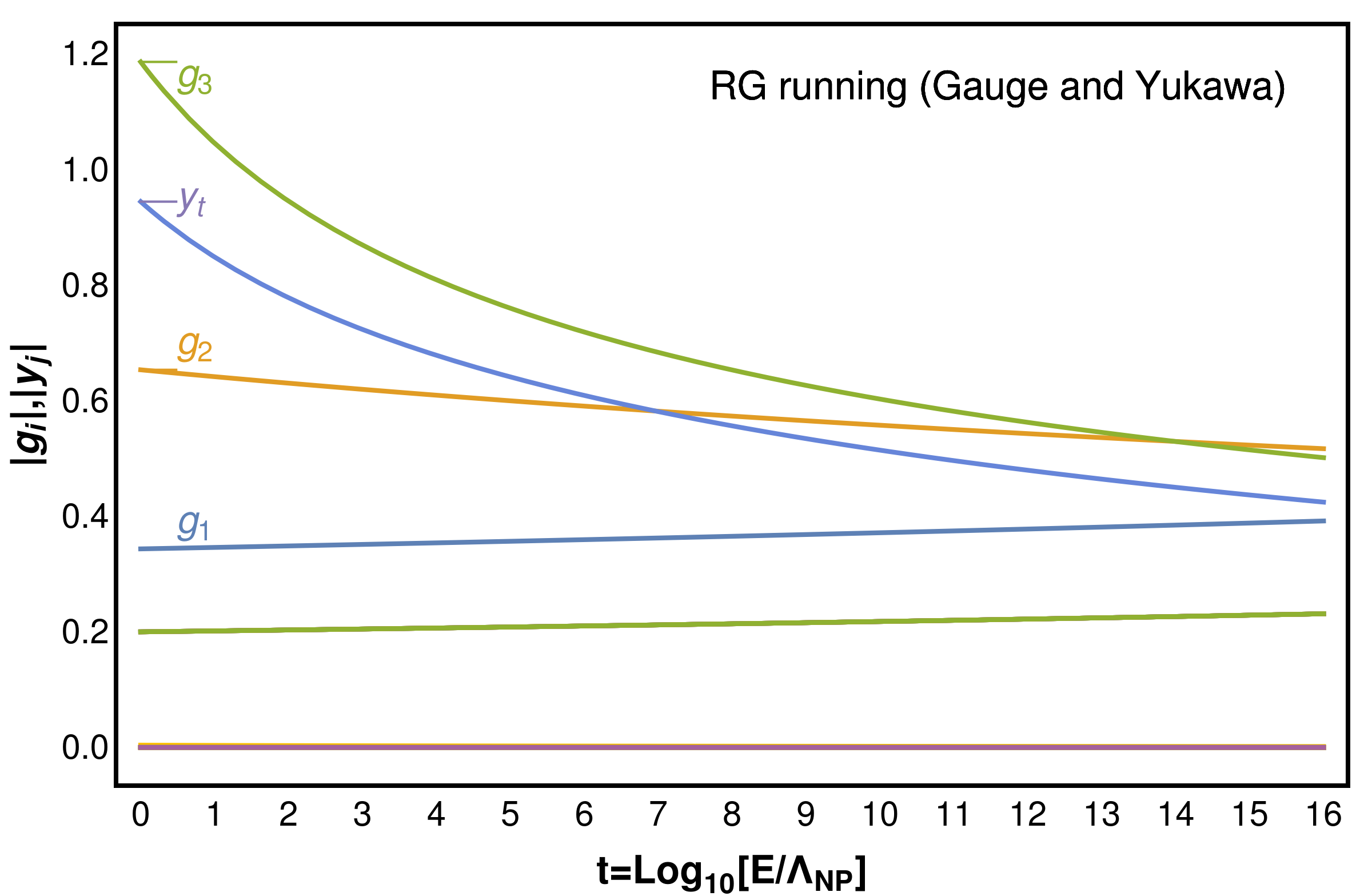}
    \end{subfigure}
    \begin{subfigure}{0.48\textwidth}
        \centering
        \includegraphics[width=\textwidth]{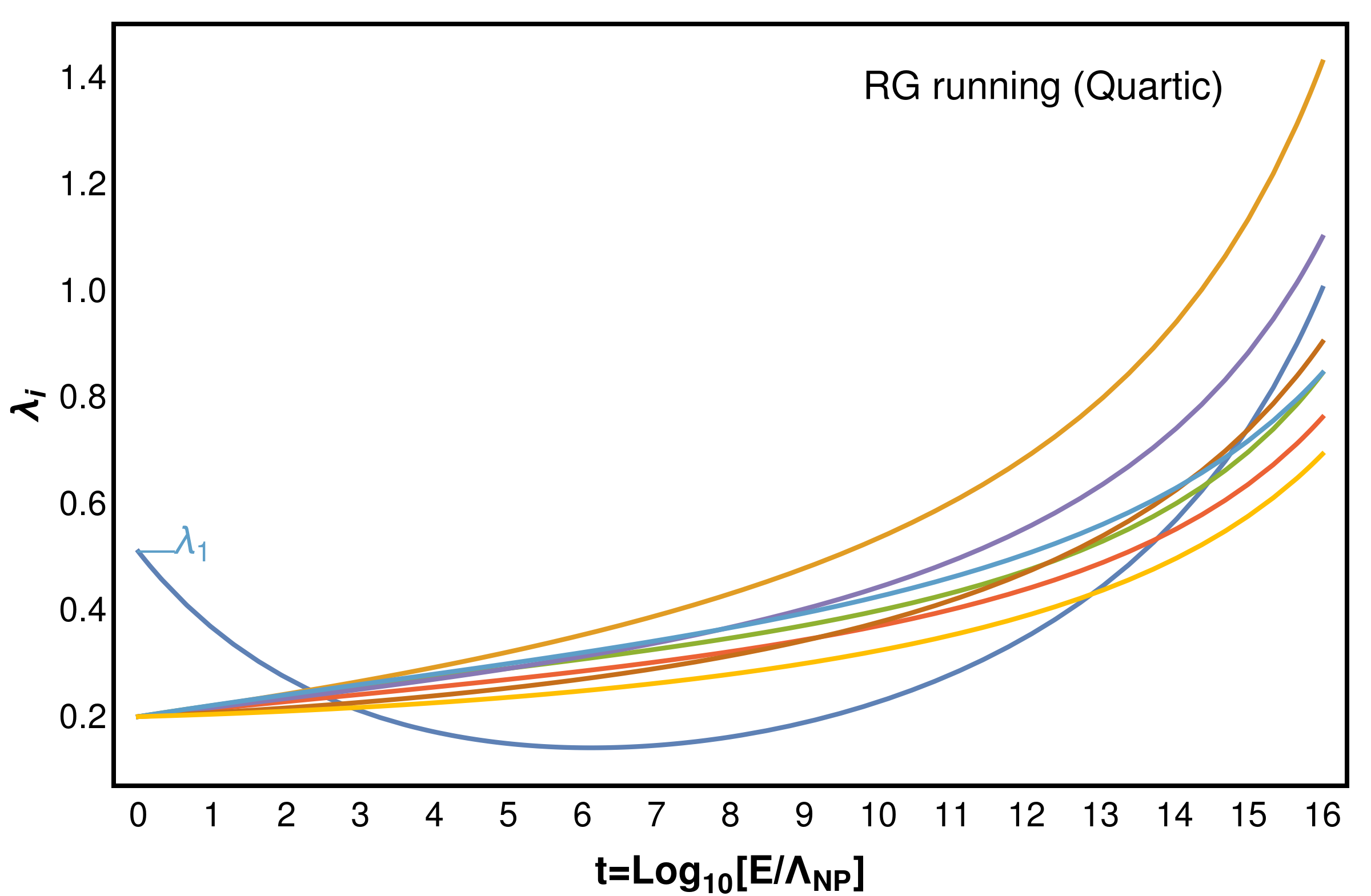}
    \end{subfigure}
    \caption{RG evolution of the SM (upper panels) and the scotogenic model under study (lower panels). The new physics scale $\Lambda_{\func{NP}}$ is set to $1\func{TeV}$ and we consider that this theory is valid from the NP scale to the Planck scale. In the lower-left panel, the Yukawa couplings appearing near around $0.2$ at the initial energy scale correspond to the RH neutrino Yukawa coupling constants $\left( g_{R} \right)$.}
    \label{fig:RG_analysis}
\end{figure}
\endgroup

\section{Flavor and Electroweak observables}\label{sec:flavor_EWobs}

In this section, we primarily discuss flavor and electroweak (EW) observables. Since the scalar sector of the model under study is extended via mixing with an additional scalar to stabilize the SM scalar potential, simply considering NP contributions arising from NP particles is generally insufficient. Instead, we must independently calculate the full contributions within both the SM and BSM frameworks from scratch; the NP contribution is then isolated by subtracting the SM result from the total BSM result. To this end, we perform one-loop calculations for the EW observables of interest using the HEP packages \texttt{FeynArts}~\cite{Hahn:2000kx} and \texttt{FeynCalc}~\cite{Shtabovenko:2020gxv,Shtabovenko:2016sxi}. To ensure a consistent and ultraviolet (UV) finite result, we perform a full renormalization in both the SM and BSM models, specifically employing the on-shell (OS) scheme in the Feynman t'Hooft gauge where $\xi=1$. Furthermore, we adopt an alternative tadpole scheme~\cite{Fleischer:1980ub} rather than the standard tadpole scheme~\cite{Krause:2016gkg},  where all tadpole contributions in the SM vanish due to their counter-terms. The alternative tadpole scheme is preferred because the extended scalar sector receives contributions arising from the mixing with the new scalar; importantly, this scheme ensures that these contributions remain gauge-invariant to all orders of perturbative theory under consistent renormalization conditions. For the sake of consistency, we apply this scheme to both the SM and BSM frameworks throughout our calculations. For flavor physics, the three-body decays $\ell_{\alpha} \rightarrow 3\ell_{\beta}$ are evaluated using \texttt{SPheno}~\cite{Porod:2003um,Porod:2011nf}. Furthermore, we crosscheck our results with \texttt{SPheno}'s output where applicable to validate the reliability of our calculations for observables not supported by the tool, such as the $Z, H \rightarrow \func{invisible}$ decays.

\subsection{Muon anomalous magnetic moment and radiative \texorpdfstring{$\ell_{\alpha} \rightarrow \ell_{\beta} \gamma$}{l_alpha to l_beta gamma} decays}\label{subsec:muong2}

In this section, we discuss the anomalous magnetic moment of the muon and the radiative $\ell_{\alpha} \rightarrow \ell_{\beta} \gamma$ decays. We first briefly review the current status of the muon $g-2$. Several years ago, this observable has exhibited a significant tension between the SM prediction~\cite{Czarnecki:2002nt,Melnikov:2003xd,Aoyama:2012wk,Kurz:2014wya,Davier:2010nc,Gnendiger:2013pva,Colangelo:2014qya,Davier:2017zfy,Masjuan:2017tvw,Colangelo:2017fiz,Hoferichter:2018kwz,Keshavarzi:2018mgv,Colangelo:2018mtw,Hoferichter:2019mqg,Davier:2019can,Keshavarzi:2019abf,Gerardin:2019vio,Bijnens:2019ghy,Colangelo:2019uex,Blum:2019ugy,Aoyama:2019ryr,Aoyama:2020ynm} and the experimental measurements independently reported by the Brookhaven National Laboratory~\cite{Muong-2:2006rrc} and the Fermilab Muon $g-2$ collaboration~\cite{Muong-2:2021ojo,Muong-2:2023cdq}, amounting to a $5.1\sigma$ discrepancy:
\begingroup
\begin{equation}
    \Delta a_{\mu} = a_{\mu}^{\func{exp}} - a_{\mu}^{\func{SM}} = \left( 2.49 \pm 0.48 \right) \times 10^{-9}.
\end{equation}
\endgroup
The most recent experimental measurement of the muon $g-2$ has achieved a precision of $0.127\func{ppm}$~\cite{Muong-2:2025xyk}, enhancing from $0.200\func{ppm}$. Meanwhile, the theoretical prediction based on the 2020 White Paper~\cite{Aoyama:2020ynm} has been subject to a matter of active discussion, owing to a tension between the data-driven dispersive approach and the Lattice QCD result for the Hadronic Vacuum Polarization (HVP) contribution to $a_{\mu}^{\func{SM}}$. This issue has been addressed in the 2025 White Paper, which presents an updated SM prediction and reports the following discrepancy~\cite{Aliberti:2025beg}:
\begingroup
\begin{equation}
    \Delta a_{\mu} = a_{\mu}^{\func{exp}} - a_{\mu}^{\func{SM}} = \left( 0.38 \pm 0.63 \right) \times 10^{-9}.
\end{equation}
\endgroup
This corresponds to a $0.6\sigma$ deviation, indicating no significant tension between the SM prediction and the experimental world average. Therefore, we present the predicted order of muon $g-2$ based on the neutrino oscillation data instead of actively fitting the value. The diagram contributing to the muon $g-2$ and the radiative $\ell_{\alpha} \rightarrow \ell_{\beta} \gamma$ in this model is shown in Figure~\ref{fig:muong2_lalphalbeta}.
\begingroup
\begin{figure}[t]
    \centering
    \includegraphics[width=0.5\textwidth]{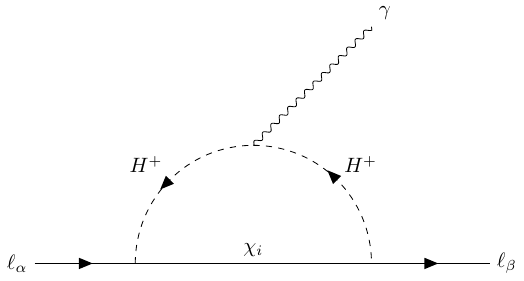}
    \caption{Diagram contributing to the muon anomalous magnetic moment and the charged lepton flavor violation (CLFV) $\ell_{\alpha} \rightarrow \ell_{\beta} \gamma$ decays. Here, $\chi_{i}$ ($i=1,2,3$) denote the physical RH neutrinos.}\label{fig:muong2_lalphalbeta}
\end{figure}
\endgroup
The muon $g-2$ and the radiative decay can be expressed in terms of the coefficient of the dipole operator $c_{R}^{\alpha\beta} \overline{\ell}_{\beta} \sigma_{\mu\nu} P_{R} \ell_{\alpha} F^{\mu\nu} + \func{h.c.}$ from an effective field theory (EFT) perspective~\cite{Alvarez:2023dzz,Darricau:2025vcs}:
\begingroup
\begin{equation}
    c_{R}^{\alpha\beta} = \frac{e}{16 \pi^{2}} \left[ \Gamma_{L}^{\alpha*} \Gamma_{R}^{\beta} \frac{M_{\Psi}}{M_{\Phi}^{2}} \left( f\left( x \right) + Q g\left( x \right) \right) + \frac{ m_{\ell_{\beta}} \Gamma_{L}^{\alpha*} \Gamma_{L}^{\beta} + m_{\ell_{\alpha}} \Gamma_{R}^{\alpha*} \Gamma_{R}^{\beta} }{M_{\Phi}^{2}} \left( \widetilde{f}\left( x \right) + Q \widetilde{g}\left( x \right) \right) \right]
\end{equation}
\endgroup
where $\Gamma_{L,R}$ is the left- and right-handed vertex, respectively, $Q$ is the electric charge of the running fermion in the loop, $x = M_{\Psi}^{2}/M_{\Phi}^{2}$, and $f, g, \widetilde{f}, \widetilde{g}$ are the loop functions defined as:
\begingroup
\begin{equation}
    \begin{split}
        f\left( x \right) &= \frac{x^{2} - 1 - 2 x \ln \left( x \right)}{4 \left( x -1 \right)^{3}}, \\
        g\left( x \right) &= \frac{x - 1 - \ln\left( x \right)}{2 \left( x - 1 \right)^{2}}, \\
        \widetilde{f}\left( x \right) &= \frac{2 x^{3} + 3 x^{2} - 6 x + 1 - 6 x^{2} \ln \left( x \right)}{24 \left( x - 1 \right)^{4}}, \\
        \widetilde{g}\left( x \right) &= \frac{x^{2} - 1 - 2 x \ln \left( x \right)}{8 \left( x - 1 \right)^{3}}.
    \end{split}
\end{equation}
\endgroup
With the coefficient $c_{R}$, the NP contribution to the muon $g-2$ and the branching ratio of the radiative $\ell_{\alpha} \rightarrow \ell_{\beta} \gamma$ decay can be written as:
\begingroup
\begin{equation}
    \begin{split}
        \Delta a_{\mu} &= -4 \frac{m_{\mu}}{e} \func{Re} \left( c_{R}^{\mu\mu} \right), \\
        \func{BR}\left( \ell_{\alpha} \rightarrow \ell_{\beta} \gamma \right) &= \frac{m_{\ell_{\alpha}}^{3}}{4 \pi \Gamma_{\ell_{\alpha}}} \left[ | c_{R}^{\alpha \beta} |^{2} + | c_{R}^{\beta \alpha} |^{2} \right].
    \end{split}
\end{equation}
\endgroup
The analytic muon $g-2$ is derived using \texttt{FeynCalc}~\cite{Shtabovenko:2020gxv,Shtabovenko:2016sxi}:
\begingroup
\begin{equation}
    \Delta a_{\mu} = \sum_{i=1}^{3} \left( -\frac{m_{\mu}^{2}}{16 \pi^{2} m_{H^{+}}^{2}} |y_{2, i}^{L}|^{2} \widetilde{f}^{\prime}\left( x_{i} \right) \right)
\end{equation}
\endgroup
where $y_{2i}^{L}$ is the left-handed vertex for $\chi_{i} - \mu - H^{+}$ and $x_{i} = m_{\chi_{i}}^{2} / m_{H^{+}}^{2}$ and the $\widetilde{f}^{\prime}$ is the loop function defined as:
\begingroup
\begin{equation}
    \widetilde{f}^{\prime}\left( x \right) = 4 \widetilde{f}\left( x \right) = \frac{1}{6 \left( 1 - x \right)^{4}} \left( 2 x^{3} + 3 x^{2} - 6 x +  1 - 6 x^{2} \ln x \right)
\end{equation}
\endgroup
The derived analytic expression for the muon $g-2$ is in agreement with those given in~\cite{CarcamoHernandez:2023wzf,Darricau:2025vcs} in the limit of vanishing right-handed vertex. The current experimental bounds and future sensitivities on the radiative decays $\ell_{\alpha} \rightarrow \ell_{\beta} \gamma$ are summarized in Table~\ref{tab:exp_lalpha_lbeta_gamma}.

\begingroup
\begin{table}[t]
\setlength{\tabcolsep}{5pt}
\centering\renewcommand{\arraystretch}{1.5} 
\begin{tabular}{c|c|c}
\toprule
\toprule
Observable & Current bound & Future sensitivity \\
\midrule
$\func{BR}\left( \mu \rightarrow e \gamma \right)$ & $< 1.5 \times 10^{-13}$ (MEG II~\cite{MEGII:2025gzr}) & $6.0 \times 10^{-14}$ (MEG II~\cite{MEGII:2018kmf}) \\
$\func{BR}\left( \tau \rightarrow e \gamma \right)$ & $< 3.3 \times 10^{-8}$ (BaBar~\cite{BaBar:2009hkt}) & $3.0 \times 10^{-9}$ (Belle II~\cite{Belle-II:2018jsg}) \\
$\func{BR}\left( \tau \rightarrow \mu \gamma \right)$ & $< 4.2 \times 10^{-8}$ (Belle~\cite{Belle:2021ysv}) & $10^{-9}$ (Belle II~\cite{Belle-II:2018jsg}) \\
\midrule
$\func{BR}\left( \mu \rightarrow 3e \right)$ & $< 1.0 \times 10^{-12}$ (SINDRUM~\cite{SINDRUM:1987nra}) & $10^{-15(-16)}$ (Mu3e~\cite{Blondel:2013ia}) \\
$\func{BR}\left( \tau \rightarrow 3e \right)$ & $< 2.7 \times 10^{-8}$ (Belle~\cite{Hayasaka:2010np}) & $5 \times 10^{-10}$ (Belle II~\cite{Belle-II:2018jsg}) \\
$\func{BR}\left( \tau \rightarrow 3\mu \right)$ & $< 1.9 \times 10^{-8}$ (Belle II~\cite{Belle-II:2024sce}) & $5 \times 10^{-10}$ (Belle II~\cite{Belle-II:2018jsg}) \\
 &  & $5 \times 10^{-11}$ (FCC-ee~\cite{FCC:2018byv}) \\
\midrule
$\func{CR}\left( \mu \rightarrow e, N \right)$ & $< 7.0 \times 10^{-13}$ ($\func{Au}$, SINDRUM~\cite{SINDRUM:1987nra}) & $10^{-14}$ (SiC, DeeMe~\cite{Nguyen:2015vkk}) \\
 &  & $2.6 \times 10^{-17}$ ($\func{Al}$, COMET~\cite{Krikler:2015msn,COMET:2018auw,Moritsu:2022lem}) \\
 &  & $8.0 \times 10^{-17}$ ($\func{Al}$, Mu2e~\cite{Mu2e:2014fns} \\
\bottomrule
\bottomrule
\end{tabular}
\caption{Current experimental bounds and future projected sensitivities for the radiative decays $\ell_{\alpha} \rightarrow \ell_{\beta} \gamma$, $\ell_{\alpha} \rightarrow 3\ell_{\beta}$ and the $\mu \rightarrow e$ conversion ratio on nucleus $N$, all at $90\% \func{C.L.}$.}
\label{tab:exp_lalpha_lbeta_gamma}
\end{table}
\endgroup

\subsection{\texorpdfstring{$\ell_{\alpha} \rightarrow 3\ell_{\beta}$}{l_alpha to 3l_beta} decays and \texorpdfstring{$\mu \rightarrow e$}{mu to e} conversion rate}\label{subsec:lito3lj}

We discuss the $\ell_{\alpha} \rightarrow 3\ell_{\beta}$ decays. Unlike the adiative $\ell_{\alpha} \rightarrow \ell_{\beta} \gamma$ decays, the $\ell_{\alpha} \rightarrow 3\ell_{\beta}$ decays receive contributions not only from dipole operators, but also from anapole, penguin, and box diagrams. It is well known that a strong correlation exists between $\ell_{\alpha} \rightarrow 3\ell_{\beta}$ and $\ell_{\alpha} \rightarrow \ell_{\beta} \gamma$ decay rates when their contributions are dominated by dipole operators~\cite{Alvarez:2023dzz,Darricau:2025vcs}. The analytic expression for the $\ell_{\alpha} \rightarrow 3\ell_{\beta}$ decays is given in Equation~\ref{eqn:ana_lalpha_3lbeta}~\cite{Abada:2014kba}:
\begingroup
\begin{equation}
    \begin{split}
        \func{BR}\left( \ell_{\alpha} \rightarrow 3\ell_{\beta} \right) &= \frac{m_{\ell_{\alpha}}^{5}}{512 \pi^{3} \Gamma_{\ell_{\alpha}}} \Bigg[ e^{4} \left( | K_{2}^{L} |^{2} + |K_{2}^{R}|^{2} \right) \left( \frac{16}{3} \ln \frac{m_{\ell_{\alpha}}}{m_{\ell_{\beta}}} - \frac{22}{3} \right) \\
        &+ \frac{1}{24} \left( |A_{LL}^{S}|^{2} + |A_{RR}^{S}|^{2} \right) + \frac{1}{12} \left( |A_{LR}^{S}|^{2} + |A_{RL}^{S}|^{2} \right) \\
        &+ \frac{2}{3} \left( |\hat{A}_{LL}^{V}|^{2} + |\hat{A}_{RR}^{V}|^{2} \right) + \frac{1}{3} \left( |\hat{A}_{LR}^{V}|^{2} + |\hat{A}_{RL}^{V}|^{2} \right) + 6 \left( |\hat{A}_{LL}^{T}|^{2} + |\hat{A}_{RR}^{T}|^{2} \right) \\
        &+ \frac{e^{2}}{3} \left( K_{2}^{L} A_{RL}^{S*} + K_{2}^{R} A_{LR}^{S*} + \func{c.c.} \right) - \frac{2 e^{2}}{3} \left( K_{2}^{L} \hat{A}_{RL}^{V*} + K_{2}^{R} \hat{A}_{LR}^{V*} + \func{c.c.} \right) \\
        &- \frac{4 e^{2}}{3} \left( K_{2}^{L} \hat{A}_{RR}^{V*} + K_{2}^{R} \hat{A}_{LL}^{V*} + \func{c.c.} \right) \\
        &- \frac{1}{2} \left( A_{LL}^{S} A_{LL}^{T*} + A_{RR}^{S} A_{RR}^{T*} + \func{c.c.} \right) - \frac{1}{6} \left( A_{LR}^{S} \hat{A}_{LR}^{V*} + A_{RL}^{S} \hat{A}_{RL}^{V*} + \func{c.c.} \right) \Bigg]
    \end{split}
    \label{eqn:ana_lalpha_3lbeta}
\end{equation}
\endgroup
where
\begingroup
\begin{equation}
    \hat{A}_{XY}^{V} = A_{XY}^{V} + e^{2} K_{1}^{X} \quad \left( X, Y = L, R \right).
\end{equation}
\endgroup
In the above equation, $K_{2}^{L,R}$ denotes the form factor associated with the dipole operator, defined as $K_{2}^{L,R} = 2 c_{R} / m_{\ell_{\alpha}}$~\cite{Darricau:2025vcs}, while $K_{1}^{L,R}$ corresponds to that of the anapole operator. The quantities $A$ represent the four-fermion form factors. Explicit definitions of each form factor are provided in~\cite{Abada:2014kba}. Next, we investigate the $\mu \rightarrow e$ conversion and its analytic expression is given in the Equation~\ref{eqn:CRmuN}~\cite{Abada:2014kba}.
\begingroup
\begin{equation}
    \begin{split}
        \func{CR}\left( \mu \rightarrow e, \func{Nucleus}\left( A, Z \right) \right) &= \frac{p_{e} E_{e} m_{\mu}^{3} G_{F}^{2} \alpha^{3} Z_{\func{eff}}^{4} F_{P}^{2}}{8\pi^{2} Z \Gamma_{\func{capt}}} \\
        &\times \Bigg[ \Big| ( Z+N ) \left( g_{LV}^{\left( 0 \right)} + g_{LS}^{\left( 0 \right)} \right) + ( Z-N ) \left( g_{LV}^{\left( 1 \right)} + g_{LS}^{\left( 1 \right)} \right) \Big|^{2} \\
        &+ \Big| ( Z+N ) \left( g_{RV}^{\left( 0 \right)} + g_{RS}^{\left( 0 \right)} \right) + ( Z-N ) \left( g_{RV}^{\left( 1 \right)} + g_{RS}^{\left( 1 \right)} \right) \Big|^{2} \Bigg]
    \end{split}
    \label{eqn:CRmuN}
\end{equation}
\endgroup
where $p_{e}$ and $E_{e}$ are the momentum and energy of the electron, $G_{F}$ is the Fermi constant, defined as $\sqrt{2} e^{2}/(8 s_{w}^{2} M_{W}^{2})$, $Z_{\func{eff}}$ means the effective atomic charge~\cite{Chiang:1993xz}, $F_{P}$ is the nuclear matrix element, $Z$ and $N$ are the number of protons and neutrons, respectively, and $\Gamma_{\func{capt}}$ is the total muon capture rate. The effective couplings $g_{XK}^{\left(0,1\right)}$ (with $X=L, R$ and $K = S, V$) can be decomposed in terms of the nucleon form factors $G_{K}$:
\begingroup
\begin{equation}
    \begin{split}
        g_{XK}^{\left( 0 \right)} &= \frac{1}{2} \sum_{q=u,d,s} \left( g_{XK\left( q \right)} G_{K}^{\left( q, p \right)} + g_{XK\left( q \right)} G_{K}^{\left( q, n \right)} \right), \\
        g_{XK}^{\left( 1 \right)} &= \frac{1}{2} \sum_{q=u,d,s} \left( g_{XK\left( q \right)} G_{K}^{\left( q, p \right)} - g_{XK\left( q \right)} G_{K}^{\left( q, n \right)} \right), \\
    \end{split}
\end{equation}
\endgroup
The numerical values of the nucleon form factors are~\cite{Kosmas:2001mv}.
\begingroup
\begin{equation}
    \begin{split}
        G_{V}^{\left( u, p \right)} = G_{V}^{\left( d, n \right)} = 2.0, \quad G_{V}^{\left( d, p \right)} = G_{V}^{\left( u, n \right)} = 1.0, \quad
        G_{V}^{\left( s, p \right)} = G_{V}^{\left( s, n \right)} = 0.0, \\
        G_{S}^{\left( u, p \right)} = G_{S}^{\left( d, n \right)} = 5.1, \quad G_{S}^{\left( d, p \right)} = G_{S}^{\left( u, n \right)} = 4.3, \quad
        G_{S}^{\left( s, p \right)} = G_{S}^{\left( s, n \right)} = 2.5. \\
    \end{split}
\end{equation}
\endgroup
The effective couplings $g_{XY\left( q \right)}$ can be written in terms of the form factors arising from the anapole, penguin and box operators.
\begingroup
\begin{equation}
    \begin{split}
        g_{LV\left( q \right)} &= \frac{\sqrt{2}}{G_{F}} \left[ e^{2} Q_{q} \left( K_{1}^{L} - K_{2}^{R} \right) - \frac{1}{2} \left( C_{\ell \ell qq}^{VLL} + C_{\ell \ell qq}^{VLR} \right) \right], \\
        g_{RV\left( q \right)} &= \left. g_{LV\left( q \right)} \right|_{L \rightarrow R}, \\
        g_{LS\left( q \right)} &= -\frac{\sqrt{2}}{G_{F}} \frac{1}{2} \left( C_{\ell\ell qq}^{SLL} + C_{\ell\ell qq}^{SLR} \right), \\
        g_{RS\left( q \right)} &= \left. g_{LS\left( q \right)} \right|_{L \rightarrow R},
    \end{split}
\end{equation}
\endgroup
where $Q_{q}$ denotes the electric charge of the quark, $C_{\ell \ell qq}^{IXK}$ is the form factors arising from the two-lepton and two $d$-quark (or two $u$-quark) interactions, with $X=L,R$ and $K=S, V$. The current experimental bounds and future sensitivities for both decays are summarized in Table~\ref{tab:exp_lalpha_lbeta_gamma}.

\subsection{EW precision observable: oblique parameters}\label{subsec:EW_oblique}

The oblique parameters quantify radiative NP corrections in BSM frameworks. They are parameterized by six quantities: $S, T, U, V, W$ and $X$~\cite{Peskin:1990zt,Peskin:1991sw,Grimus:2008nb}, as defined in Equation~\ref{eqn:oblique_params}:
\begingroup
\begin{equation}
    \begin{split}
        \frac{\alpha}{4 s_{w}^{2} c_{w}^{2}} S &= \frac{A_{ZZ}\left( m_{Z}^{2} \right) - A_{ZZ}\left( 0 \right)}{m_{Z}^{2}} - \left. \frac{\partial A_{\gamma\gamma}\left( p^2 \right)}{\partial p^2} \right|_{p^2=0} - \left. \frac{c_{w}^{2}- s_{w}^{2}}{c_{w} s_{w}} \frac{\partial A_{\gamma Z}\left( p^2 \right)}{\partial p^{2}} \right|_{p^2=0} \\
        \alpha T &= \frac{A_{WW}\left( 0 \right)}{m_{W}^{2}} - \frac{A_{ZZ}\left( 0 \right)}{m_{Z}^{2}} \\
        \frac{\alpha}{4 s_{W}^{2}} U &= \frac{A_{WW}\left( m_{W}^{2} \right) - A_{WW}\left( 0 \right)}{m_{W}^{2}} - c_{W}^{2} \frac{A_{ZZ}\left( m_{Z}^{2} \right) - A_{ZZ}\left( 0 \right)}{m_{Z}^{2}} \\
        &- \left. s_{W}^{2} \frac{\partial A_{\gamma\gamma}\left( p^2 \right)}{\partial p^2} \right|_{p^2=0} - \left. 2 c_{w} s_{w} \frac{\partial A_{\gamma Z}\left( p^2 \right)} {\partial p^2} \right|_{p^2=0} \\
        \alpha V &= \left. \frac{\partial A_{ZZ}\left( p^2 \right)}{\partial p^2} \right|_{p^2=m_{Z}^{2}} - \frac{A_{ZZ}\left( m_{Z}^{2} \right) - A_{ZZ}\left( 0 \right)}{m_{Z}^{2}} \\
        \alpha W &= \left. \frac{\partial A_{WW}\left( p^2 \right)}{\partial p^2} \right|_{p^2=m_{W}^{2}} - \frac{A_{WW}\left( m_{W}^{2} \right) - A_{WW}\left( 0 \right)}{m_{W}^{2}} \\
        \frac{\alpha}{s_{w} c_{w}} X &= \left. \frac{\partial A_{\gamma Z}\left( p^2 \right)}{\partial p^2} \right|_{p^2=0} - \frac{A_{\gamma Z}\left( m_{Z}^{2} \right) - A_{\gamma Z}\left( 0 \right)}{m_{Z}^{2}}
    \end{split}
    \label{eqn:oblique_params}
\end{equation}
\endgroup
where $\alpha$ is the fine-structure constant, $c_{w}$ and $s_{w}$ denote the cosine and sine of the weak mixing angle, respectively, and $A_{VV^{\prime}}$ represents the transverse components of the $V-V^{\prime}$ vacuum polarization tensors:
\begingroup
\begin{equation}
    \Pi_{VV^{\prime}}^{\mu\nu} \left( p \right) = \left( g^{\mu\nu} - \frac{p^{\mu}p^{\nu}}{p^2} \right) A_{VV^{\prime}} \left( p^2 \right) + \frac{p^{\mu}p^{\nu}}{p^2} B_{VV^{\prime}} \left( p^2 \right).
\end{equation}
\endgroup
Considering a NP scale of $1\func{TeV}$, which we also adopt as the renormalization scale $\mu$ for RG evolution, the higher-order oblique parameters $V, W$ and $X$ are kinematically suppressed and remain sub-dominant relative to $S, T$ and $U$ parameters. Consequently, our analysis focuses primarily on the $S, T$ and $U$ parameters, while the $V, W$ and $X$ parameters are utilized as an internal consistency check to verify the UV finiteness of the analytic expressions. The transverse components of the vacuum polarization tensors are detailed in Appendix~\ref{app:SE}. Throughout this work, the observables under investigation are defined as the net contributions from NP, representing the difference between the BSM and SM contributions:
\begingroup
\begin{equation}
    \begin{split}
        \Delta S &= S_{\func{BSM}} - S_{\func{SM}}, \\
        \Delta T &= T_{\func{BSM}} - T_{\func{SM}}, \\
        \Delta U &= U_{\func{BSM}} - U_{\func{SM}}. \\
        \label{eqn:STU_observables}
    \end{split}
\end{equation}
The recently updated experimental constraints on the oblique parameters, within $1\sigma$ confidence intervals, are as follows~\cite{ParticleDataGroup:2024cfk}:
\begingroup
\begin{equation}
    \begin{split}
        \Delta S_{\func{exp}} &= 0.021 \pm 0.096, \\
        \Delta T_{\func{exp}} &= 0.040 \pm 0.120, \\
        \Delta U_{\func{exp}} &= 0.008 \pm 0.092. \\
    \end{split}
\end{equation}
\endgroup
Preliminary assessments of the FCC-ee's projected sensitivity toward the $S$ and $T$ parameters have been recently discussed~\cite{deBlas:2025gyz}.

\subsection{EW precision observable: \texorpdfstring{$Z \rightarrow \overline{\ell}_{\alpha} \ell_{\beta}$}{Z to l_alpha l_beta}}\label{subsec:EW_Ztolilj}

The one-loop vertex correction topologies to the $Z \rightarrow \overline{\ell}_{\alpha} \ell_{\beta}$ in this BSM model are seen in Figure~\ref{fig:Zlilj_VC_top}.
\begingroup
\begin{figure}[t]
    \centering
    \begin{subfigure}{0.24\textwidth}
        \centering
        \includegraphics[width=\textwidth]{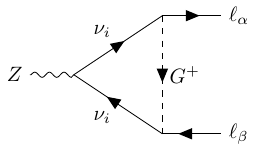}
    \end{subfigure}
    \begin{subfigure}{0.24\textwidth}
        \centering
        \includegraphics[width=\textwidth]{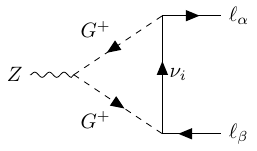}
    \end{subfigure}
    \begin{subfigure}{0.24\textwidth}
        \centering
        \includegraphics[width=\textwidth]{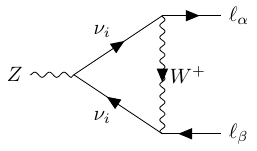}
    \end{subfigure}
    \begin{subfigure}{0.24\textwidth}
        \centering
        \includegraphics[width=\textwidth]{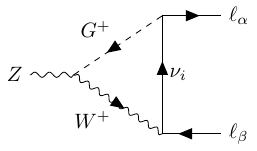}
    \end{subfigure}
    \hfill
    \begin{subfigure}{0.24\textwidth}
        \centering
        \includegraphics[width=\textwidth]{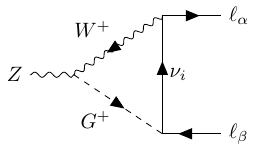}
    \end{subfigure}
    \begin{subfigure}{0.24\textwidth}
        \centering
        \includegraphics[width=\textwidth]{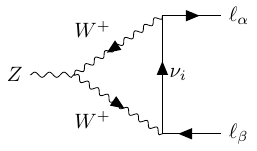}
    \end{subfigure}
    \begin{subfigure}{0.24\textwidth}
        \centering
        \includegraphics[width=\textwidth]{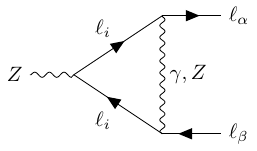}
    \end{subfigure}
    \begin{subfigure}{0.24\textwidth}
        \centering
        \includegraphics[width=\textwidth]{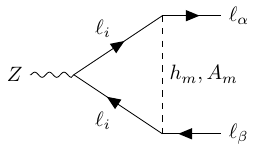}
    \end{subfigure}
    \begin{subfigure}{0.24\textwidth}
        \centering
        \includegraphics[width=\textwidth]{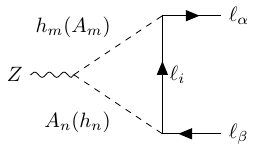}
    \end{subfigure}
    \begin{subfigure}{0.24\textwidth}
        \centering
        \includegraphics[width=\textwidth]{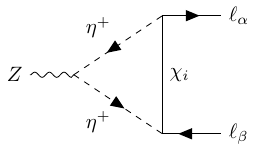}
    \end{subfigure}
    \begin{subfigure}{0.24\textwidth}
        \centering
        \includegraphics[width=\textwidth]{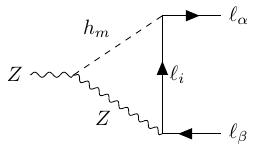}
    \end{subfigure}
    \begin{subfigure}{0.24\textwidth}
        \centering
        \includegraphics[width=\textwidth]{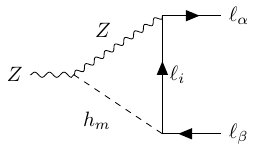}
    \end{subfigure}
    \caption{One-loop Feynman diagrams for the $Z \rightarrow \overline{\ell}_{\alpha} \ell_{\beta}$ vertex corrections within this BSM model. Here, the symbol $\ell$ denotes the SM charged leptons $(e, \mu, \tau)$, while $\nu_{i}$ represent the massless SM Dirac neutrinos. $h_{m}$ and $A_{m}$ are the CP-even and -odd scalars, respectively. The indices $\alpha, \beta, i$ serve as generation indices ranging from $1$ to $3$, whereas the indices $m, n$ label the generations of the new scalars from $1$ to $2$.}
    \label{fig:Zlilj_VC_top}
\end{figure}
\endgroup
Regarding the flavor-violating observables $Z \rightarrow \overline{\ell}_{\alpha} \ell_{\beta}$ where $\alpha \neq \beta$, the contributing topology is the vertex correction mediated by $\chi_{i}$ fields, as shown in Figure~\ref{fig:Zlilj_VC_top}. This is because flavor-changing currents in this framework are exclusively generated by the Majorana particles. For the flavor-conserving current, all topologies shown in Figure~\ref{fig:Zlilj_VC_top} contribute in this BSM model. The SM flavor-conserving contributions are obtained by decoupling all the $\chi$-related contributions and by taking the appropriate limits of the extended scalar sector: specifically, by reducing the scalar fields $h_{i} \rightarrow h_{\func{SM}}$ and the pseudo-scalars $A_{i} \rightarrow G^{0}$, where $G^{0}$ denotes the would-be Goldstone boson associated with the $Z$ boson in the SM. The one-loop amplitude for $Z \rightarrow \overline{\ell}_{\alpha} \ell_{\beta}$ can be decomposed in terms of form factors:
\begingroup
\begin{equation}
    \begin{split}
        \mathcal{A}\left( Z \rightarrow \overline{\ell}_{\alpha} \ell_{\beta} \right) &= F_{SL} \psi\left( q_{1}, m_{\ell_{\alpha}} \right) \left( \gamma \cdot \epsilon\left( p \right) \right) P_{L} \psi\left( -q_{2}, m_{\ell_{\beta}} \right) \\
        &+ F_{SR} \psi\left( q_{1}, m_{\ell_{\alpha}} \right) \left( \gamma \cdot \epsilon\left( p \right) \right) P_{R} \psi\left( -q_{2}, m_{\ell_{\beta}} \right) \\
        &+ F_{L} \psi\left( q_{1}, m_{\ell_{\alpha}} \right) \left( q_{1} \cdot \epsilon\left( p \right) \right) P_{L} \psi\left( -q_{2}, m_{\ell_{\beta}} \right) \\
        &+ F_{R} \psi\left( q_{1}, m_{\ell_{\alpha}} \right) \left( q_{1} \cdot \epsilon\left( p \right) \right) P_{R} \psi\left( -q_{2}, m_{\ell_{\beta}} \right),
    \end{split}
\end{equation}
\endgroup
where $p$ is the four-momentum of the incoming $Z$ boson, $q_{1}$ and $q_{2}$ are the four-momenta of the outgoing leptons $\ell_{\alpha}$ and $\ell_{\beta}$, respectively, and $P_{L,R} = \frac{1}{2} \left( 1 \mp \gamma^{5} \right)$ are the left- and right-handed chiral projection operators. Before addressing the counter-term (CT) contributions, it is necessary to discuss the infrared (IR) divergences arising from the photon-mediated diagram shown in Figure~\ref{fig:Zlilj_VC_top}. In general, the experimental amplitude can be decomposed into its SM and NP contributions as follows:
\begingroup
\begin{equation}
    \mathcal{A}_{\func{exp}} = \mathcal{A}_{\func{SM}} + \mathcal{A}_{\func{NP}}
\end{equation}
\endgroup
If a BSM model does not extend the SM gauge symmetry by additional gauge symmetries, and does not modify the vacuum structure of the SM scalar potential, it is sufficient to identify the NP amplitude $\mathcal{A}_{\func{NP}}$ solely with the contributions from the new particles~\cite{Darricau:2025vcs}. However, the current BSM model modifies the SM scalar sector by introducing a new scalar field that develops its own VEV and mixes with the SM Higgs boson. Consequently, it is necessary to redefine the extraction of NP contributions. In this work, we define the NP contribution as:
\begingroup
\begin{equation}
    \mathcal{A}_{\func{NP}} = \mathcal{A}_{\func{BSM}} - \mathcal{A}_{\func{SM}}
\end{equation}
\endgroup
where $\mathcal{A}_{\func{BSM}}$ represents the total amplitude calculated within this BSM model. In the limit where the extended scalar sector reduces to the SM scalar potential, the redefined NP amplitude consistently reduces to the contribution mediated exclusively by the new BSM particles. In defining the NP contribution, both the BSM and SM aplitudes share identical infrared (IR) structures; consequently, the IR-divergent terms cancel exactly during the subtraction, allowing us to focus on the IR-finite components of each amplitude. However, this cancellation is less straightforward for the $h_{1} \rightarrow \overline{\ell}_{\alpha} \ell_{\beta}$ and $h_{\func{SM}} \rightarrow \overline{\ell}_{\alpha} \ell_{\beta}$ processes. Because the SM-like Higgs $h_{1}$ depends on the CP-even scalar mixing angle, it is required to incorporate the bremsstrahlung effects to ensure an IR-finite result. The counter-term topologies for the flavor-violating $Z \rightarrow \overline{\ell}_{\alpha} \ell_{\beta}$ where $\alpha \neq \beta$ are seen in Figure~\ref{fig:Zlilj_CT_top}.
\begingroup
\begin{figure}[t]
    \centering
    \begin{subfigure}{0.30\textwidth}
        \centering
        \includegraphics[width=\textwidth]{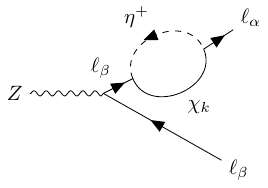}
    \end{subfigure}
    \begin{subfigure}{0.30\textwidth}
        \centering
        \includegraphics[width=\textwidth]{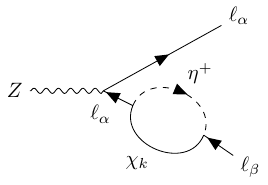}
    \end{subfigure}
    \caption{Flavor-violating $Z \rightarrow \overline{\ell}_{\alpha} \ell_{\beta}$ CT topologies. Here, $\eta^{+}$ is the NP charged scalar and $\chi_{k}$ $\left( k = 1, 2, 3 \right)$ is the RH neutrinos.}
    \label{fig:Zlilj_CT_top}
\end{figure}
\endgroup
For the flavor-conserving counterterms, it is more practical to employ analytic expressions rather than diagrammatic representations. This approach is necessitated by the requirement to renormalize all the parameters within the tree-level Lagrangian relevant to the $Z \rightarrow \overline{\ell}_{\alpha} \ell_{\beta}$ process. The derived analytic expression for the flavor-conserving $Z\rightarrow \overline{\ell}_{\alpha} \ell_{\alpha}$ CT is given in Equation~\ref{eqn:Analytic_Zlili_CT}:
\begingroup
\begin{equation}
\begin{split}
    \frac{\mathcal{L}_{Z \rightarrow \overline{\ell}_{\alpha}\ell_{\alpha}}^{\func{CT}}}{\mathcal{L}_{Z \rightarrow \overline{\ell}_{\alpha}\ell_{\alpha}}^{\func{Tree}}} &= \frac{e}{c_{w} s_{w}} \left( T_{3} - s_{w}^{2} Q_{f} \right) \\
    &\times \left( \frac{1}{2} \delta Z_{ZZ} + \delta Z_{e} - \frac{1}{2} \delta Z_{\ell_{\alpha} \ell_{\alpha}} - \frac{1}{2} \delta \overline{Z}_{\ell_{\alpha} \ell_{\alpha}} + \frac{-T_{3}+\left( T_{3} - Q_{f} \right) \left( s_{w}^{2}/c_{w}^{2} \right)}{T_{3}-s_{w}^{2} Q_{f}}  \frac{\delta s_{w}}{s_{w}} \right) \\
    &+ e Q_{f} \frac{1}{2} \delta Z_{AZ},
\end{split}
\label{eqn:Analytic_Zlili_CT}
\end{equation}
\endgroup
where $c_{w}$ and $s_{w}$ are the cosine and sine of the weak mixing angle, $T_{3}$ is the third component of the isospin doublet, and $Q_{f}$ is the charge of the fermion. The terms $\delta Z_{ZZ}, \delta Z_{e}, \delta Z_{\ell_{\alpha} \ell_{\alpha}},$ and $\delta Z_{AZ}$ denote the $Z$-boson wave function (WF) renormalization constant, the charge renormalization constant, the charged-lepton WF renormalization constant, and the $\gamma - Z$ mixing renormalization constant, respectively. Since the \texttt{FeynArts}~\cite{Hahn:2000kx} model file was generated by \texttt{SARAH}~\cite{Staub:2013tta,Staub:2015kfa}, the flavor-conserving analytic CTs follow the \texttt{SARAH}~\cite{Staub:2013tta,Staub:2015kfa} convention. Consequently, the charge renormalization constant $\delta Z_{e}$ and the charged-lepton WF constants $\delta Z_{\ell \ell}$ differ by a global negative sign compared to the definitions in~\cite{Denner:1991kt}. Once UV and IR finite form factors are given, the branching ratio is given by integrating out the phase space.
\begingroup
\begin{equation}
    \begin{split}
        \func{BR}\left( Z \rightarrow \overline{\ell}_{\alpha} \ell_{\beta} \right) &= \frac{\lambda^{1/2}\left( M_{Z}^{2}, m_{\ell_{\alpha}}^{2}, m_{\ell_{\beta}}^{2} \right)}{192 \pi M_{Z}^{5} \Gamma_{Z}} \Bigg[ \left( | F_{L}^{\alpha\beta} |^{2} + | F_{R}^{\alpha\beta} |^{2} \right) \left( M_{Z}^{2} - m_{\ell_{\alpha}}^{2} - m_{\ell_{\beta}}^{2} \right) \lambda\left( M_{Z}^{2}, m_{\ell_{\alpha}}^{2}, m_{\ell_{\beta}}^{2} \right) \\
        &- 4 \left( | F_{SL}^{\alpha\beta} |^{2} + | F_{SR}^{\alpha\beta} |^{2} \right) \left( \lambda\left( M_{Z}^{2}, m_{\ell_{\alpha}}^{2}, m_{\ell_{\beta}}^{2} \right) - 3 M_{Z}^{2} \left( M_{Z}^{2} - m_{\ell_{\alpha}}^{2} - m_{\ell_{\beta}}^{2} \right) \right) \\
        &- 4 \left( m_{\ell_{\alpha}} m_{\ell_{\beta}} \func{Re}\left( F_{L}^{\alpha\beta} ( F_{R}^{\alpha\beta} )^{*} \right) + m_{\ell_{\beta}} \func{Re}\left( F_{R}^{\alpha\beta} ( F_{SL}^{\alpha\beta} )^{*} + F_{L}^{\alpha\beta} ( F_{SR}^{\alpha\beta} )^{*} \right) \right. \\
        &+ \left. m_{\ell_{\alpha}} \func{Re}\left( F_{R}^{\alpha\beta} ( F_{SR}^{\alpha\beta} )^{*} + F_{L}^{\alpha\beta} ( F_{SL}^{\alpha\beta} )^{*} \right) \right) \lambda\left( M_{Z}^{2}, m_{\ell_{\alpha}}^{2}, m_{\ell_{\beta}}^{2} \right) \\
        &+ 48 M_{Z}^{2} m_{\ell_{\alpha}} m_{\ell_{\beta}} \func{Re} \left( F_{SL}^{\alpha\beta} ( F_{SR}^{\alpha\beta} )^{*} \right) \Bigg]
    \end{split}
\end{equation}
\endgroup
where $\lambda$ is the K\"all\'en function:
\begingroup
\begin{equation}
    \lambda\left( x, y, z \right) \equiv x^{2} + y^{2} + z^{2} - 2 x y - 2 x z - 2 y z.
\end{equation}
\endgroup

The SM predictions and experimental measurements for flavor-conserving $Z \rightarrow \ell_{\alpha}^{+} \ell_{\alpha}^{-}$ decays, as well as the current bounds and future sensitivities for flavor-violating $Z \rightarrow \ell_{\alpha}^{\pm} \ell_{\beta}^{\mp}$ processes, are presented in Table~\ref{tab:experimental_Ztolilj}.

\begingroup
\begin{table}[t]
\setlength{\tabcolsep}{5pt}
\centering\renewcommand{\arraystretch}{1.5} 
\begin{tabular}{c|c|c}
\toprule
\toprule
Observable & Experimental measurements & SM prediction \\
\midrule
$\Gamma\left( Z \rightarrow e^{+} e^{-} \right)$ & $83.91 \pm 0.12 \func{MeV}$ (LEP~\cite{ALEPH:2005ab}) & $83.965 \pm 0.016 \func{MeV}$~\cite{Freitas:2014hra} \\
$\Gamma\left( Z \rightarrow \mu^{+} \mu^{-} \right)$ & $83.99 \pm 0.18 \func{MeV}$ (LEP~\cite{ALEPH:2005ab}) & $83.965 \pm 0.016 \func{MeV}$~\cite{Freitas:2014hra} \\
$\Gamma\left( Z \rightarrow \tau^{+} \tau^{-} \right)$ & $84.08 \pm 0.22 \func{MeV}$ (LEP~\cite{ALEPH:2005ab}) & $83.775 \pm 0.016 \func{MeV}$~\cite{Freitas:2014hra} \\
$\func{BR}\left( Z \rightarrow e^{\pm} \mu^{\mp} \right)$ & $< 4.2 \times 10^{-7}$ (ATLAS~\cite{ATLAS:2014vur}) & $\mathcal{O}\left( 10^{-10} \right)$ (FCC-ee~\cite{FCC:2018byv}) \\
$\func{BR}\left( Z \rightarrow e^{\pm} \tau^{\mp} \right)$ & $< 4.1 \times 10^{-6}$ (ATLAS~\cite{ATLAS:2021bdj}) & $\mathcal{O}\left( 10^{-10} \right)$ (FCC-ee~\cite{FCC:2018byv}) \\
$\func{BR}\left( Z \rightarrow \mu^{\pm} \tau^{\mp} \right)$ & $< 5.3 \times 10^{-6}$ (ATLAS~\cite{ATLAS:2021bdj}) & $\mathcal{O}\left( 10^{-10} \right)$ (FCC-ee~\cite{FCC:2018byv}) \\
\midrule
$R_{\mu e}\left( Z \rightarrow \ell\ell \right)$ & $1.0001 \pm 0.0024$ (PDG~\cite{ParticleDataGroup:2024cfk}) & $1.0$~\cite{Freitas:2014hra} \\
$R_{\tau e}\left( Z \rightarrow \ell\ell \right)$ & $1.0020 \pm 0.0032$ (PDG~\cite{ParticleDataGroup:2024cfk}) & $0.9977$~\cite{Freitas:2014hra} \\
$R_{\tau \mu}\left( Z \rightarrow \ell\ell \right)$ & $1.0010 \pm 0.0026$ (PDG~\cite{ParticleDataGroup:2024cfk}) & $0.9977$~\cite{Freitas:2014hra} \\
\midrule
$\Gamma\left( Z \rightarrow \func{Invisible} \right)$ & $499.0 \pm 1.5 \func{MeV}$ (PDG~\cite{ParticleDataGroup:2024cfk}) & $501.45 \pm 0.05 \func{MeV}$~\cite{Freitas:2014hra} \\
\midrule
$\func{BR}\left( H \rightarrow \mu^{+} \mu^{-} \right)$ & $\left( 2.6 \pm 1.3 \right) \times 10^{-4}$ (PDG~\cite{ParticleDataGroup:2024cfk}) & $\left( 2.17 \pm 0.13 \right) \times 10^{-4}$~\cite{Denner:2011mq} \\
$\func{BR}\left( H \rightarrow \tau^{+} \tau^{-} \right)$ & $\left( 0.06_{-0.007}^{+0.008} \right)$ (PDG~\cite{ParticleDataGroup:2024cfk}) & $0.0624 \pm 0.0035$~\cite{Denner:2011mq} \\
\midrule
$R_{\tau \mu}\left( H \rightarrow \ell\ell \right)$ & $230 \pm 146$ (PDG~\cite{ParticleDataGroup:2024cfk}) & $288$~\cite{LHCHiggsCrossSectionWorkingGroup:2016ypw} \\
\bottomrule
\bottomrule
\end{tabular}
\caption{Comparison between SM predictions and experimental measurements for flavor-conserving (at $68\%$ C.L.) and flavor-violating leptonic $Z/H$ decays. The lepton flavor universality ratio is defined as $R_{\alpha\beta} \equiv \func{BR}( V \rightarrow \ell_{\alpha}^{+} \ell_{\alpha}^{-} ) / \func{BR} ( V \rightarrow \ell_{\beta}^{+} \ell_{\beta}^{-} )$, where $V = Z, H$. Since the theoretical uncertainties for the SM lepton universality ratios are neligible, only the central values are shown.}
\label{tab:experimental_Ztolilj}
\end{table}
\endgroup

\subsection{EW precision observable: \texorpdfstring{$Z \rightarrow \text{Invisible}$}{Z to Invisible}}\label{subsec:EW_Ztoinv}

The one-loop diagrams contributing to the $Z \rightarrow \func{Invisible}$ decay are presented in Figure~\ref{fig:ZInv_top}. 
\begingroup
\begin{figure}[t]
    \centering
    \begin{subfigure}{0.24\textwidth}
        \centering
        \includegraphics[width=\textwidth]{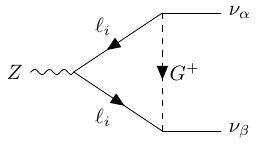}
    \end{subfigure}
    \begin{subfigure}{0.24\textwidth}
        \centering
        \includegraphics[width=\textwidth]{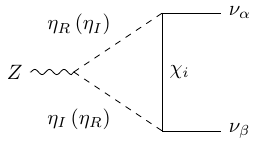}
    \end{subfigure}
    \begin{subfigure}{0.24\textwidth}
        \centering
        \includegraphics[width=\textwidth]{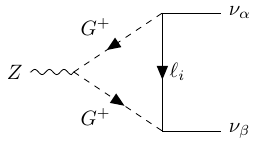}
    \end{subfigure}
    \begin{subfigure}{0.24\textwidth}
        \centering
        \includegraphics[width=\textwidth]{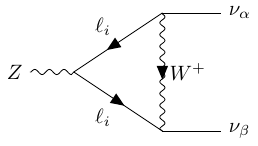}
    \end{subfigure}
    \begin{subfigure}{0.24\textwidth}
        \centering
        \includegraphics[width=\textwidth]{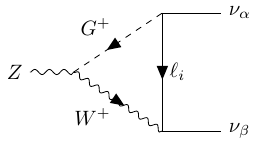}
    \end{subfigure}
    \begin{subfigure}{0.24\textwidth}
        \centering
        \includegraphics[width=\textwidth]{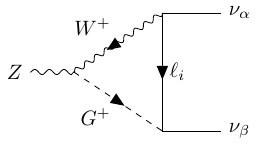}
    \end{subfigure}
    \begin{subfigure}{0.24\textwidth}
        \centering
        \includegraphics[width=\textwidth]{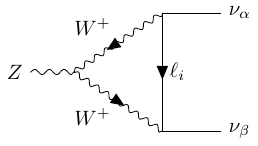}
    \end{subfigure}
    \begin{subfigure}{0.24\textwidth}
        \centering
        \includegraphics[width=\textwidth]{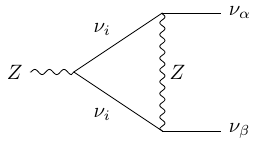}
    \end{subfigure}
    \hspace*{\fill}
    \caption{One-loop Feynman diagrams contributing to the $Z \rightarrow \func{Invisible}$ decay in this BSM model. For each topology featuring charged particles, the corresponding diagram with reversed charge flow is included. The active neutrinos in these diagrams are Majorana particles. The indices $\alpha, \beta$ and $i$ run from $1$ to $3$.}
    \label{fig:ZInv_top}
\end{figure}
\endgroup
The counter-term contributions for the flavor-violating $Z \rightarrow \nu_{\alpha}\nu_{\beta}$ decay are given diagrammatically in Figure~\ref{fig:Zinv_CT_top}.
\begingroup
\begin{figure}[t]
    \centering
    \begin{subfigure}{0.30\textwidth}
        \centering
        \includegraphics[width=\textwidth]{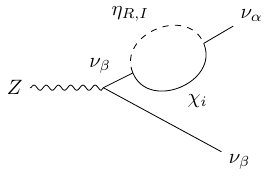}
    \end{subfigure}
    \begin{subfigure}{0.30\textwidth}
        \centering
        \includegraphics[width=\textwidth]{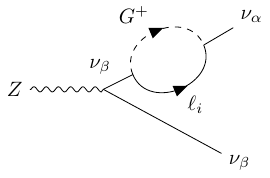}
    \end{subfigure}
    \begin{subfigure}{0.30\textwidth}
        \centering
        \includegraphics[width=\textwidth]{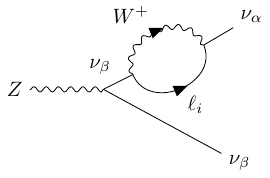}
    \end{subfigure}
    \caption{CT topologies for the flavor-violating $Z \rightarrow \nu_{\alpha}\nu_{\beta}$ process. Analogous to the vertex correction diagrams in Figure~\ref{fig:ZInv_top}, these include corresponding topologies with reversed charge flow, with the exception of diagrams mediated by neutral particles. Furthermore, while radiative corrections also occur on the lower neutrino leg, similar to those in Figure~\ref{fig:Zlilj_CT_top}, they are omitted here for brevity.}
    \label{fig:Zinv_CT_top}
\end{figure}
\endgroup
The counter-term Lagrangian for the flavor-conserving $Z \rightarrow \nu_{\alpha}\nu_{\alpha}$ decay is given in Equation~\ref{eqn:Analytic_Zinv_CT}.
\begingroup
\begin{equation}
\begin{split}
    \frac{\mathcal{L}_{Z \rightarrow \nu_{\alpha}\nu_{\alpha}}^{\func{CT}}}{\mathcal{L}_{Z \rightarrow \nu_{\alpha}\nu_{\alpha}}^{\func{Tree}}} = \frac{e}{c_{w} s_{w}} T_{3} \times \left( \frac{1}{2} \delta Z_{ZZ} + \delta Z_{e} - \frac{c_{w}^{2}-s_{w}^{2}}{c_{w}^{2}} \frac{\delta s_{w}}{s_{w}} - \frac{1}{2} \left( \delta Z_{\nu_{\alpha}\nu_{\alpha}} + \overline{\delta Z}_{\nu_{\alpha}\nu_{\alpha}} \right) \right)
\end{split}
\label{eqn:Analytic_Zinv_CT}
\end{equation}
\endgroup

The $Z \rightarrow \func{Invisible}$ decay in this model is mediated by the active light Majorana neutrinos, whereas in the SM the corresponding decay proceeds through massless left-handed neutrinos. Consequently, the SM contributions can not be simply recovered by taking an appropriate limit of the $Z \rightarrow \func{Invisible}$ decay in the present model. For this reason, it is preferred to present the SM contribution separately. The diagrams contributing to the SM $Z$ invisible decay are shown in Figure~\ref{fig:SM_ZInv_top}.
\begingroup
\begin{figure}[t]
    \centering
    \begin{subfigure}{0.24\textwidth}
        \centering
        \includegraphics[width=\textwidth]{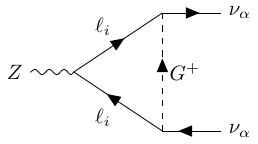}
    \end{subfigure}
    \begin{subfigure}{0.24\textwidth}
        \centering
        \includegraphics[width=\textwidth]{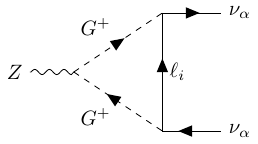}
    \end{subfigure}
    \begin{subfigure}{0.24\textwidth}
        \centering
        \includegraphics[width=\textwidth]{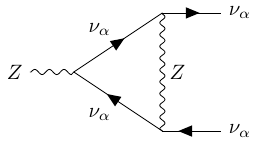}
    \end{subfigure}
    \begin{subfigure}{0.24\textwidth}
        \centering
        \includegraphics[width=\textwidth]{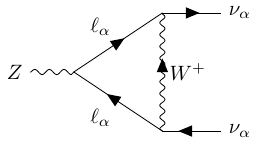}
    \end{subfigure}
    \begin{subfigure}{0.24\textwidth}
        \centering
        \includegraphics[width=\textwidth]{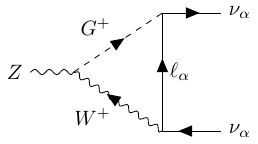}
    \end{subfigure}
    \begin{subfigure}{0.24\textwidth}
        \centering
        \includegraphics[width=\textwidth]{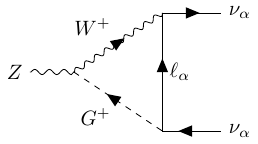}
    \end{subfigure}
    \begin{subfigure}{0.24\textwidth}
        \centering
        \includegraphics[width=\textwidth]{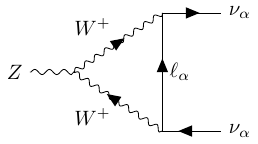}
    \end{subfigure}
    \hspace*{\fill}
    \caption{One-loop Feynman diagrams contributing to the $Z \rightarrow \func{Invisible}$ decay in the SM. The index $\alpha$ runs from $1$ to $3$.}
    \label{fig:SM_ZInv_top}
\end{figure}
\endgroup
Since the final states of the $Z \rightarrow \func{Invisible}$ decay differ between the BSM and SM, the NP decay width must be defined as the difference between the BSM and SM contributions, given by:
\begingroup
\begin{equation}
    \Gamma_{\func{NP}} \equiv \Gamma_{\func{BSM}} - \Gamma_{\func{SM}} = \sum_{i=1}^{3} \sum_{j=1}^{3} \frac{1}{1+\delta_{ij}} \Gamma\left( Z \rightarrow \nu_{i} \nu_{j} \right) - \sum_{i=1}^{3} \Gamma\left( Z \rightarrow \nu_{i,L} \overline{\nu}_{i,L} \right),
\end{equation}
\endgroup
where $\delta_{ij}$ is the Kronecker delta, introduced to account for the symmetry factor arising from identical final states in the BSM contribution. The experimental $Z \rightarrow \func{Invisible}$ decay is given in Table~\ref{tab:experimental_Ztolilj}.

\subsection{EW precision observable: \texorpdfstring{$H \rightarrow \overline{\ell}_{\alpha} \ell_{\beta}$}{H to l_alpha l_beta}}\label{subsec:EW_Htolilj}

We turn to the discussion of the $H \rightarrow \overline{\ell}_{\alpha} \ell_{\beta}$ decays. One important difference, compared to the leptonic decays of $Z$, is that we have two different Higgses, which are the SM-like Higgs $h_{1}$ and the SM Higgs $h$. For the comparison of the leptonic decays of Higgs, we identify the SM-like Higgs $h_{1}$ with the SM Higgs $h$ in the alignment limit, where the couplings of $h_{1}$ reproduce those of the SM. The one-loop vertex correction topologies to the $h_{1} \rightarrow \overline{\ell}_{\alpha} \ell_{\beta}$ decays in the BSM model are given in Figure~\ref{fig:h1lilj_VC_top}. 
\begingroup
\begin{figure}[t]
    \centering
    \begin{subfigure}{0.24\textwidth}
        \centering
        \includegraphics[width=\textwidth]{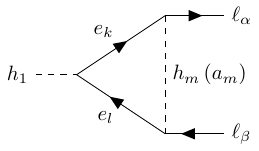}
    \end{subfigure}
    \begin{subfigure}{0.24\textwidth}
        \centering
        \includegraphics[width=\textwidth]{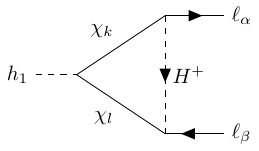}
    \end{subfigure}
    \begin{subfigure}{0.24\textwidth}
        \centering
        \includegraphics[width=\textwidth]{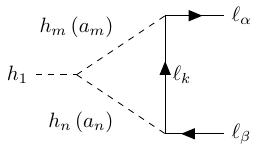}
    \end{subfigure}
    \begin{subfigure}{0.24\textwidth}
        \centering
        \includegraphics[width=\textwidth]{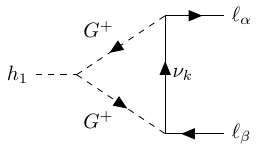}
    \end{subfigure}
    \begin{subfigure}{0.24\textwidth}
        \centering
        \includegraphics[width=\textwidth]{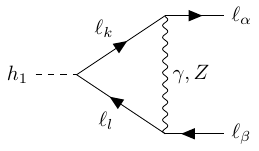}
    \end{subfigure}
    \begin{subfigure}{0.24\textwidth}
        \centering
        \includegraphics[width=\textwidth]{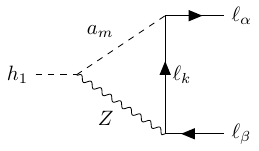}
    \end{subfigure}
    \begin{subfigure}{0.24\textwidth}
        \centering
        \includegraphics[width=\textwidth]{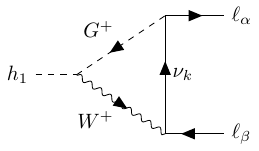}
    \end{subfigure}
    \begin{subfigure}{0.24\textwidth}
        \centering
        \includegraphics[width=\textwidth]{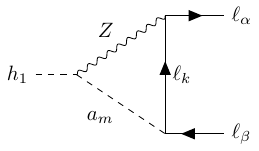}
    \end{subfigure}
    \begin{subfigure}{0.24\textwidth}
        \centering
        \includegraphics[width=\textwidth]{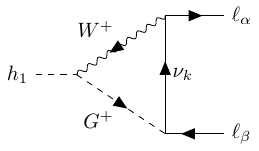}
    \end{subfigure}
    \begin{subfigure}{0.24\textwidth}
        \centering
        \includegraphics[width=\textwidth]{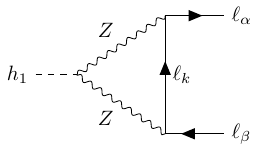}
    \end{subfigure}
    \begin{subfigure}{0.24\textwidth}
        \centering
        \includegraphics[width=\textwidth]{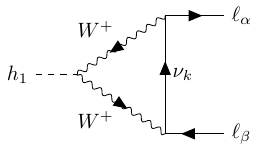}
    \end{subfigure}
    \caption{One-loop Feynman diagrams for the $h_{1} \rightarrow \overline{\ell}_{\alpha} \ell_{\beta}$ vertex corrections within this BSM model. Here, the symbol $\ell$ denotes the SM charged leptons $(e, \mu, \tau)$, while $\nu_{i}$ represent the massless SM Dirac neutrinos. $h_{m}$ and $A_{m}$ are the CP-even and -odd scalars, respectively. The indices $\alpha, \beta, i$ serve as generation indices ranging from $1$ to $3$, whereas the indices $m, n$ label the generations of the new scalars from $1$ to $2$.}
    \label{fig:h1lilj_VC_top}
\end{figure}
\endgroup
The SM contributions to $h \rightarrow \overline{\ell}_{\alpha} \ell_{\alpha}$ are obtained in analogy with the leptonic $Z$ decay by decoupling the $\chi$-related fields and taking the appropriate limits of the extended scalar sector to recover the SM scalar sector. The one-loop amplitude for the $H \rightarrow \overline{\ell}_{\alpha} \ell_{\beta}$ can be decomposed in terms of the form factors:
\begingroup
\begin{equation}
    \begin{split}
        \mathcal{A}\left( H \rightarrow \overline{\ell}_{\alpha} \ell_{\beta} \right) = G_{L} \psi\left( q_{1}, m_{\ell_{\alpha}} \right) P_{L} \psi\left( -q_{2}, m_{\ell_{\beta}} \right) + G_{R} \psi\left( q_{1}, m_{\ell_{\alpha}} \right) P_{R} \psi\left( -q_{2}, m_{\ell_{\beta}} \right).
    \end{split}
\end{equation}
\endgroup
Unlike the $Z \rightarrow \overline{\ell}_{\alpha} \ell_{\beta}$ decays, the processes $H \rightarrow \overline{\ell}_{\alpha} \ell_{\beta}$ with $H = h_{1}, h$ do not share a common IR structure, since the SM-like Higgs $h_{1}$ couples through its scalar mixing matrix. Consequently, the Bremsstrahlung contributions must be evaluated separately for each process in order to obtain an IR-finite result. The IR-divergences arising from the one-loop vertex corrections are cancelled by the corresponding Bremsstrahlung contributions, as follows:
\begingroup
\begin{equation}
    2 \func{Re}\left[ \mathcal{A}_{\func{tree}} \left( h_{1} \rightarrow \ell \ell \right) \left( \mathcal{A}_{\func{OL}} \left( h_{1} \rightarrow \ell \ell \right) \right)^{*} \right] + | \mathcal{A}_{\func{soft}} \left( h_{1} \rightarrow \ell \ell \gamma \right) |^{2} = 0
\end{equation}
\endgroup
where $\mathrm{OL}$ denotes the one-loop contribution, and the amplitude for soft-photon emission is approximated as
\begingroup
\begin{equation}
    | \mathcal{A}_{\func{soft}} \left( h_{1} \rightarrow \ell \ell \gamma \right) |^{2} \approx | \mathcal{A}_{\func{Born}} \left( h_{1} \rightarrow \ell \ell \right) |^{2} \times \left( \frac{c_{20}}{\left( q_{1} \cdot k \right)^{2}} + \frac{c_{11}}{\left( q_{1} \cdot k \right) \left( q_{2} \cdot k \right)} + \frac{c_{02}}{\left( q_{2} \cdot k \right)^{2}} \right).
    \label{eqn:soft_photon_approx}
\end{equation}
\endgroup
Equation~\ref{eqn:soft_photon_approx} confirms the eikonal structure of the soft-photon emission, where $k$ denotes the photon four-momentum and the coefficients as:
\begingroup
\begin{equation}
    \begin{split}
        c_{20} &= -e^{2} m_{\ell_{\alpha}}^{2} = -e^{2} q_{1}^{2}, \\
        c_{11} &= e^{2} \left( m_{H}^{2} -  2 m_{\ell_{\alpha}}^{2} \right) = 2 e^{2} \left( q_{1} \cdot q_{2} \right), \\
        c_{02} &= -e^{2} m_{\ell_{\alpha}}^{2} = -e^{2} q_{2}^{2}.
    \end{split}
\end{equation}
\endgroup
The phase space integral over the soft-photon emission region, where $|\mathbf{k}| < \Delta E$, is evaluated in~\cite{tHooft:1978jhc,Denner:1991kt}, giving rise to the result:
\begingroup
\begin{equation}
    \begin{split}
        I_{ij} &= \int_{\textbf{k} \le \Delta E} \frac{d^{3} k}{2 \omega_{k}} \frac{2 q_{i} q_{j}}{\left( q_{i} \cdot k \right) \left( q_{j} \cdot k \right)} \\
        &= 4\pi \frac{\alpha q_{i} q_{j}}{\left( \alpha q_{i} \right)^{2} - q_{j}^{2}} \Bigg[ \frac{1}{2} \ln \frac{\left( \alpha q_{i} \right)^{2}}{q_{j}^{2}} \ln \frac{4 \Delta E^{2}}{\lambda^{2}} \\
        &+ \left[ \frac{1}{4}\ln^{2}\frac{u_{0}-|\mathbf{u}|}{u_{0}+|\mathbf{u}|} + \func{Li}_{2} \left( 1 - \frac{u_{0} + |\mathbf{u}|}{v} \right) + \func{Li}_{2} \left( 1 - \frac{u_{0} - |\mathbf{u}|}{v} \right) \right]_{u=q_{j}}^{u=\alpha q_{i}} \Bigg]
    \end{split}
\end{equation}
\endgroup
where $\lambda$ is a fictitious photon mass to regularize the IR-singularities, $\func{Li}_{2}$ is the Spence dilogarithm, and 
\begingroup
\begin{equation}
    v = \frac{\left( \alpha q_{i} \right)^{2} - q_{j}^{2}}{2 \left( \alpha q_{i0} - q_{j0} \right)}
\end{equation}
\endgroup
and $\alpha$ is defined as:
\begingroup
\begin{equation}
    \alpha^{2} q_{i}^{2} - 2 \alpha q_{i} q_{j} + q_{j}^{2} = 0, \quad \frac{\alpha q_{i0}-q_{j0}}{q_{j0}} > 0.
\end{equation}
\endgroup
Given the UV- and IR-finite form factors, the branching ratio of $H \rightarrow \overline{\ell}_{\alpha} \ell_{\beta}$ is given by:
\begingroup
\begin{equation}
    \begin{split}
        \func{BR}\left( H \rightarrow \overline{\ell}_{\alpha} \ell_{\beta} \right) &= \frac{\lambda^{1/2}\left( m_{H}^{2}, m_{\ell_{\alpha}}^{2}, m_{\ell_{\beta}}^{2} \right)}{16 \pi m_{H}^{3} \Gamma_{H}} \Big[ \left( | G_{L}^{\alpha\beta} |^{2} + | G_{R}^{\alpha\beta} |^{2} \right) \left( m_{H}^{2} - m_{\ell_{\alpha}}^{2} - m_{\ell_{\beta}}^{2} \right) \\
        &- 4 m_{\ell_{\alpha}} m_{\ell_{\beta}} \func{Re} \left( G_{L}^{\alpha\beta} ( G_{R}^{\alpha\beta} )^{*} \right) \Big]
    \end{split}
\end{equation}
\endgroup
The experimental bound on $H \rightarrow \overline{\ell}_{\alpha} \ell_{\beta}$ is given in Table~\ref{tab:experimental_Ztolilj}.

\subsection{EW precision observable: \texorpdfstring{$H \rightarrow \text{Invisible}$}{H to Invisible}}\label{subsec:EW_Htoinv}

The final observable under consideration is the $H \rightarrow \func{Invisible}$ decay, with the contributing diagrams shown in Figure~\ref{fig:h1inv_VC_top}.
\begingroup
\begin{figure}[t]
    \centering
    \begin{subfigure}{0.24\textwidth}
        \centering
        \includegraphics[width=\textwidth]{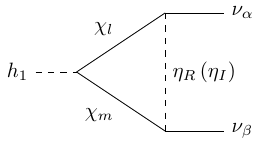}
    \end{subfigure}
    \begin{subfigure}{0.24\textwidth}
        \centering
        \includegraphics[width=\textwidth]{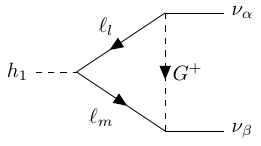}
    \end{subfigure}
    \begin{subfigure}{0.24\textwidth}
        \centering
        \includegraphics[width=\textwidth]{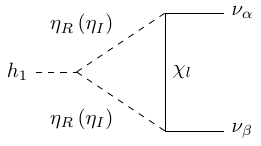}
    \end{subfigure}
    \begin{subfigure}{0.24\textwidth}
        \centering
        \includegraphics[width=\textwidth]{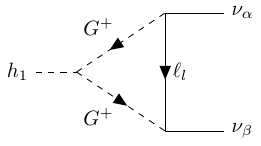}
    \end{subfigure}
    \begin{subfigure}{0.24\textwidth}
        \centering
        \includegraphics[width=\textwidth]{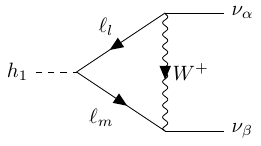}
    \end{subfigure}
    \begin{subfigure}{0.24\textwidth}
        \centering
        \includegraphics[width=\textwidth]{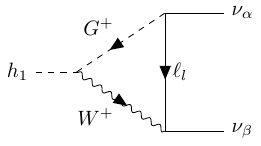}
    \end{subfigure}
    \begin{subfigure}{0.24\textwidth}
        \centering
        \includegraphics[width=\textwidth]{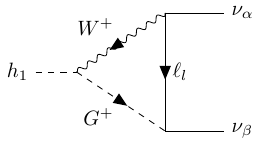}
    \end{subfigure}
    \begin{subfigure}{0.24\textwidth}
        \centering
        \includegraphics[width=\textwidth]{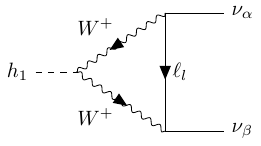}
    \end{subfigure}
    \caption{One-loop Feynman diagrams contributing to the $h_{1} \rightarrow \func{Invisible}$ vertex corrections within this BSM model. The indices $\alpha, \beta, l, m$ run from $1$ to $3$. As in the case of the $Z \rightarrow \func{Invisible}$ decay, each diagram featuring charged particles is accompanied by its counterpart with reversed charge flow.}
    \label{fig:h1inv_VC_top}
\end{figure}
\endgroup
Since the SM contribution to this process vanishes exactly, it is a purely NP effect. Furthermore, the absence of a tree-level $H-\nu-\nu$ vertex implies that no CTs are required, rendering the vertex correction UV finite. The total $H \rightarrow \func{Invisible}$ decay amplitude is then obtained by summing over all partial contributions indexed by $\left( \alpha, \beta \right)$.

\section{Numerical analysis and discussions}\label{sec:numerical_analysis}

In this section, we discuss the scanning methodology and flavor observables of interest. A notable feature of the present work is that our analysis is not driven by large Yukawa couplings, which can easily develop a Landau pole at relatively low energy scales, as discussed in Section~\ref{subsec:RG_driven_perturbativity}. As a result, many flavor observables will be suppressed by several orders of magnitude; therefore, we will simply note their predicted values in passing, while focusing on more phenomenologically accessible observables. This analysis pursues two primary objectives: to identify the region of parameter space in which the difference between the normal and inverted hierarchies is maximized, and to assess whether the results of the numerical scan are consistent with current experimental bounds. Each of these questions is addressed in turn in the sections that follow.

\subsection{Scanning methodology}\label{subsec:scanning_methodology}

This section outlines the scanning methodology adopted in the present work. The model is implemented in \texttt{SARAH}~\cite{Staub:2013tta,Staub:2015kfa}, which is used to generate the \texttt{SPheno}~\cite{Porod:2003um,Porod:2011nf} spectrum calculator and the \texttt{FeynArts}~\cite{Hahn:2000kx} model file employed for the analytic derivation of the flavor observables discussed in Section~\ref{sec:flavor_EWobs}. Analytic derivations are performed with~\texttt{FeynCalc}~\cite{Shtabovenko:2020gxv,Shtabovenko:2016sxi}, while the numerical evaluation of Passarino-Veltman functions is carried out using \texttt{LoopTools}~\cite{Hahn:1998yk}. Vacuumm stability is assessed with \texttt{Vevacious}~\cite{Camargo-Molina-OLeary:2014}, and the scalar constraints from both the SM and BSM sectors are tested with \texttt{HiggsTool}~\cite{Bahl:2022igd}. Relic density and spin-independent cross sections are computed using \texttt{micrOMEGAs}~\cite{Alguero:2023zol}. Throughout the scan, only parameter points whose vacuum state is confirmed to be stable are retained, provided they also pass all the scalar constraints. Considering the high dimensionality of the parameter space, a random scan (random walk) proves computationally inefficient; we therefore employ a Markov Chain Monte Carlo (MCMC) approach. The MCMC procedure adopted in this work proceeds as follows:
\begingroup
\begin{enumerate}
    \item Generate an initial point in the parameter space yielding a finite log likelihood value, defined as
    \begin{equation}
        \ln \mathcal{L} = \sum_{i} \ln \mathcal{L}_{i} = -\sum_{i} \frac{\left( \mathcal{O}_{i}^{\mathrm{pred}} - \mathcal{O}_{i}^{\mathrm{exp}} \right)^{2}}{2\sigma_{i}^{2}},
    \end{equation}
    where $\mathcal{O}_{i}^{\mathrm{pred}}$ and $\mathcal{O}_{i}^{\mathrm{exp}}$ denote the predicted and experimental values of the $i$-th observable, respectively, and $\sigma_{i}$ is the corresponding $1\sigma$ uncertainty.
    \item Propose a candidate point sampled from a Gaussian distribution centered on the current point.
    \item Evaluate the $\chi^{2}$ value at the candidate point and determine whether to accept or reject it according to the Metropolis--Hastings criterion.
    \item Repeat steps $2$--$4$ until the chain reaches sufficient convergence.
\end{enumerate}
\endgroup
The primary observables driving the MCMC algorithm are the SM Higgs boson mass at one-loop level and the relic density, whose experimental values are taken as $m_{h} = 125.20 \pm 0.11 \func{GeV}$~\cite{ParticleDataGroup:2024cfk} and $\Omega h^{2} = 0.120 \pm 0.012$ of Equation~\ref{eqn:relicrenew}, respectively. The ranges of the input parameters are listed in Table~\ref{tab:range_input_parameters}. A hierarchy between the diagonal and off-diagonal $g_{R}$ coupling constants is imposed within these ranges to ensure that the physical right-handed neutrino masses remain sufficiently large. In addition, it was found that when the DM candidate mass falls below $100 \func{GeV}$, the numerical evaluation in \texttt{micrOMEGAs} becomes computationally expensive due to the opening of numerous SM decay channels. To mitigate this, a lower bound of $m_{\func{DM}} > 120 \func{GeV}$ is imposed as an additional scan constraint. After the scan, all points for which the Higgs boson mass and relic density fall within $5\sigma$ of their respective experimental values are retained.
\begingroup
\begin{table}[t]
\setlength{\tabcolsep}{5pt}
\centering\renewcommand{\arraystretch}{1.5} 
\begin{tabular}{c|c|c|c}
\toprule
\toprule
Parameter & Range & Parameter & Range \\
\midrule
$\lambda_{1}$ & $\left[ 0.40, 0.60 \right]$ & $v_{S}$ & $\left[ 2 \times 10^{3}, 2 \times 10^{4} \right]$ \\
$\lambda_{2,\cdot,8}$ & $\left[ 10^{-5}, 0.2 \right]$ & $g_{R,ii}$ & $\left[ 0.05, 0.2 \right]$ \\
$\mu_{\eta}^{2}$ & $\left[ 10^{4}, 10^{6} \right]$ & $g_{R,ij}$ & $\left[ 10^{-5}, 10^{-2} \right]$ \\
$\mu_{\func{sb}}^{2}$ & $\left[ 10^{4}, 10^{6} \right]$ & $c_{R,ij}$ & $\left[ -1, 1 \right]$ \\
\bottomrule
\bottomrule
\end{tabular}
\caption{Ranges of the input parameters. Here $i,j=1,2,3$ and the parameters $c_{R,ij}$ are taken to be real in the numerical scan.}
\label{tab:range_input_parameters}
\end{table}
\endgroup


\subsection{Neutrino normal hierarchy and inverse hierarchy}

We begin with a discussion of the scanned neutrino masses for both normal and inverted hierarchies, as summarized in Table~\ref{tab:numass_scanned}.
\begingroup
\begin{table}[t]
\setlength{\tabcolsep}{5pt}
\centering\renewcommand{\arraystretch}{1.5} 
\begin{tabular}{c|cc|cc}
\toprule
\toprule
 & \multicolumn{2}{c}{Normal hierarchy} & \multicolumn{2}{c}{Inverted hierarchy} \\ \cmidrule(lr){2-3} \cmidrule(lr){4-5}
 & Minimum value & Maximum value & Minimum value & Maximum value \\
 \midrule
 $m_{\nu_{1}}$ & $1.0008 \times 10^{-19}$ & $2.1409 \times 10^{-11}$ & $4.9376 \times 10^{-11}$ & $4.9649 \times 10^{-11}$ \\
 $m_{\nu_{2}}$ & $8.3196 \times 10^{-12}$ & $2.3129 \times 10^{-11}$ & $5.0076 \times 10^{-11}$ & $5.0399 \times 10^{-11}$ \\
 $m_{\nu_{3}}$ & $5.0332 \times 10^{-11}$ & $5.5482 \times 10^{-11}$ & $1.0005 \times 10^{-19}$ & $5.1159 \times 10^{-13}$ \\
 \midrule
 $\sum_{i=1}^{3} m_{\nu_{i}}$ & $5.8698 \times 10^{-11}$ & $9.9913 \times 10^{-11}$ & $9.9453 \times 10^{-11}$ & $1.0000 \times 10^{-10}$ \\
\bottomrule
\bottomrule
\end{tabular}
\caption{Minimum and maximum values of the neutrino masses obtained from the parameter scan, for both normal hierarchy and inverted hierarchy. All retained points satisfy the neutrino mass-squared difference constraints at the $3\sigma$ level.} 
\label{tab:numass_scanned}
\end{table}
\endgroup
As discussed in the Introduction, the cosmological upper bound on the sum of neutrino masses~\cite{Planck:2018vyg}, $\sum_{i=1}^{3} m_{\nu_{i}} \leq 10^{-10}\func{GeV}$, is imposed throughout the numerical scan. More recently, this bound has been further tightened by the DESI BAO measurement to $\sum_{i=1}^{3} m_{\nu_{i}} < 6.42 \times 10^{-11} \func{GeV}$~\cite{Elbers:2025vlz}. Should this result be confirmed by other experiments, the inverted hierarchy scenario would be disfavoured, as its minimum neutrino mass sum cannot satisfy the updated bound. Furthermore, a significant portion of the parameter points in the normal hierarchy would also be excluded, though a small viable region of parameter space would remain consistent with this bound. 

\subsection{Dark matter analysis}\label{subsec:DM_analysis}

In this section, we discuss the phenomenology of the DM, focusing on the scanned relic density and spin independent cross section.
\begingroup
\begin{figure}[t]
    \centering
    \includegraphics[width=0.96\textwidth]{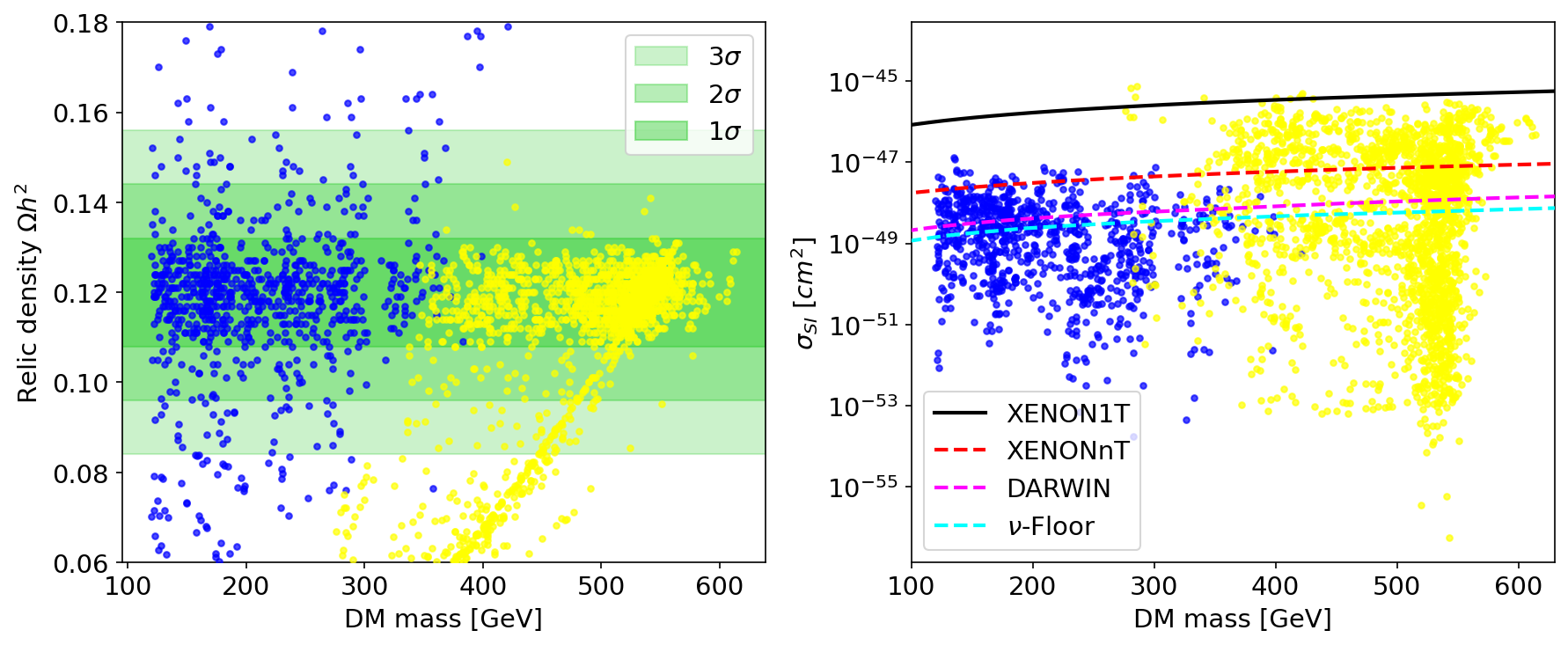}
    \caption{Scanned DM mass versus relic density (left panel) and spin-independent proton cross section (right panel). In both panels, blue points correspond to the fermionic DM candidate $\chi_{1}$, while yellow points correspond to the CP-odd scalar DM candidate $\eta_{I}$. In the right panel, the current experimental upper bound from XENON1T~\cite{XENON:2018voc} and future sensitivities from XENONnT~\cite{XENON:2020kmp}, DARWIN~\cite{DARWIN:2016hyl}, and the neutrino floor ($\nu$-floor)~\cite{OHare:2021utq} are also shown.}
    \label{fig:DMmass_relic_xsec}
\end{figure}
\endgroup
As discussed, this model features three DM candidates: the CP-even scalar $\eta_{R}$, the CP-odd scalar $\eta_{I}$, and the lightest RH fermion $\chi_{1}$. Throughout the scan, it is found that the mass of the CP-even scalar $\eta_{R}$ is nearly degenerate with that of the CP-odd scalar $\eta_{I}$, but is consistently slightly heavier. Consequently, $\eta_{R}$ does not appear as a viable scalar DM candidate, leaving two DM candidates, $\eta_{I}$ and $\chi_{1}$, as shown in Figure~\ref{fig:DMmass_relic_xsec}. Interestingly, the mass ranges of both DM candidates are strongly constrained by the relic density requirement. Among all the surviving points, the fermionic DN candidate accounts for nearly $29\%$, while the scalar DM candidate accounts for $71\%$, of the numerical scan. The fermionic DM mass is found to line in the range $120-350 \func{GeV}$, while the CP-odd scalar mass is found in the range $350-600 \func{GeV}$. The spin-independent proton cross sections are shown in the right panel, along with the current experimental upper bound from XENON1T~\cite{XENON:2018voc} and projected sensitivities from XENONnT~\cite{XENON:2020kmp}, DARWIN~\cite{DARWIN:2016hyl}, and the neutrino floor ($\nu$-floor)~\cite{OHare:2021utq}. As shown in the right panel of Figure~\ref{fig:DMmass_relic_xsec}, the majority of the parameter points lie below even the $\nu$-floor sensitivity, indicating that the spin-independent cross section places no significant constraint on the parameter space of this model.

\subsection{Flavor observable analsysis}\label{subsec:flavor_obs_analysis}

In this section, we discuss the flavor observables examined in our analysis. Notably, no region of parameter space was found in which a distinction between the normal and inverted neutrino mass hierarchies is manifest. Moreover, all lepton flavor-violating decays are controlled by the coupling constant $g_{X}$, which is tightly constrained by neutrino oscillation data through the one-loop generation of neutrino masses. As a result, their branching ratios are highly suppressed. We therefore present the order-of-magnitude estimates of the muon $g-2$ contribution and the relevant flavor-violating observables, all of which remain many orders of magnitude below current experimental bounds.
\begingroup
\begin{equation}
    \begin{split}
    \mathcal{O}\left( \Delta a_{\mu} \right) &= 10^{-14}, \\
    \mathcal{O}\left( \func{BR}\left( \mu \rightarrow e \gamma \right) \right) &= [ 10^{-35}, 10^{-26} ], \\
    \mathcal{O}\left( \func{BR}\left( \tau \rightarrow e \gamma \right) \right) &= [ 10^{-35} , 10^{-26} ], \\
    \mathcal{O}\left( \func{BR}\left( \tau \rightarrow \mu \gamma \right) \right) &= [ 10^{-33} , 10^{-25} ], \\
    \mathcal{O}\left( \func{CR}\left( \mu \rightarrow e, \func{Al} \right) \right) &= [ 10^{-37} , 10^{-28} ], \\
    \mathcal{O}\left( \func{BR}\left( \mu \rightarrow 3e \right) \right) &= [ 10^{-36}, 10^{-28} ], \\
    \mathcal{O}\left( \func{BR}\left( \tau \rightarrow 3e \right) \right) &= [ 10^{-36}, 10^{-28} ], \\
    \mathcal{O}\left( \func{BR}\left( \tau \rightarrow 3\mu \right) \right) &= [ 10^{-35}, 10^{-27} ], \\
    \mathcal{O}\left( \func{BR}\left( Z \rightarrow e^{\pm} \mu^{\mp} \right) \right) &= [ 10^{-40}, 10^{-31} ], \\
    \mathcal{O}\left( \func{BR}\left( Z \rightarrow e^{\pm} \tau^{\mp} \right) \right) &= [ 10^{-38}, 10^{-30} ], \\
    \mathcal{O}\left( \func{BR}\left( Z \rightarrow \mu^{\pm} \tau^{\mp} \right) \right) &= [ 10^{-37}, 10^{-29} ], \\
    \mathcal{O}\left( \func{BR}\left( H \rightarrow e^{\pm} \mu^{\mp} \right) \right) &= [ 10^{-46}, 10^{-33} ], \\
    \mathcal{O}\left( \func{BR}\left( H \rightarrow e^{\pm} \tau^{\mp} \right) \right) &= [ 10^{-43}, 10^{-30} ], \\
    \mathcal{O}\left( \func{BR}\left( H \rightarrow \mu^{\pm} \tau^{\mp} \right) \right) &= [ 10^{-44}, 10^{-29} ].
    \end{split}
    \label{eqn:muong2_FVobs_scanned}
\end{equation}
\endgroup
Beyond the highly suppressed muon $g-2$ and flavor-violating observables, a few other observables call for discussion, starting with the oblique parameters shown in Figure~\ref{fig:STU}.
\begingroup
\begin{figure}[t]
    \centering
    \includegraphics[width=0.96\textwidth]{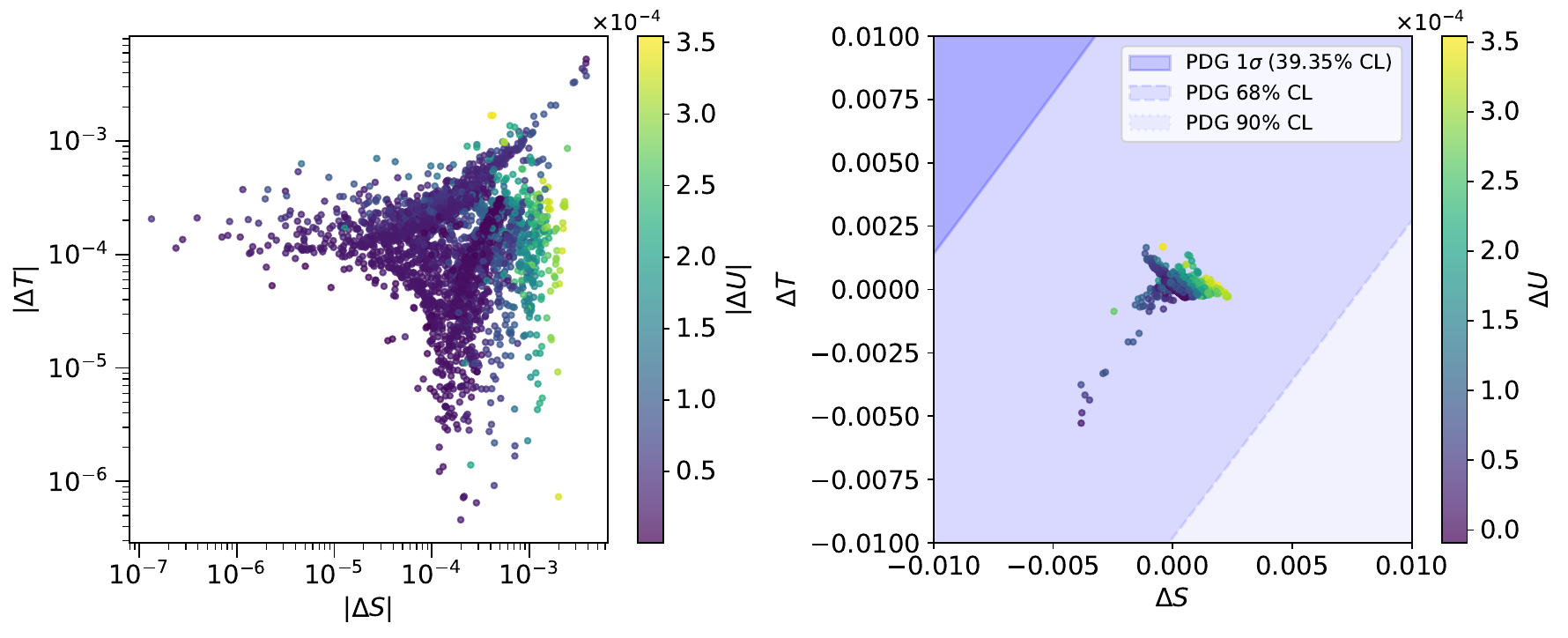}
    \caption{Scanned results of the oblique parameters. The left panel takes the absolute values of the $S$ and $T$ parameters and express them in logarithmic scale. The right panel displays them on a linear scale alongside the experimental bounds.}
    \label{fig:STU}
\end{figure}
\endgroup
The oblique parameters are less suppressed compared to the highly suppressed flavor observables. We investigated the source of this relative enhancement and found the SM Higgs mass to be the main driving factor. As discussed in Section~\ref{subsec:EW_oblique}, the oblique parameters are determined by the difference between the BSM and SM contributions. In the BSM contribution, the SM-like Higgs mass $m_{h_{1}}$ is scanned within its $5\sigma$ allowed range, and we confirmed that the dominant contribution arises from the difference between the SM-like Higgs and SM Higgs contributions. Ohter new physics contributions were also examined but found to be relatively small in comparison. As a result, the oblique parameters are the most accessible to near-future experiments, as they are quite sensitive to deviations in the SM Higgs mass. Next, we discuss lepton flavor universality, whose experimental bounds are strongly constrained as shown in Table~\ref{tab:experimental_Ztolilj}, and the result is given in Figure~\ref{fig:RZtm_RHtm}.
\begingroup
\begin{figure}[t]
    \centering
    \includegraphics[width=0.96\textwidth]{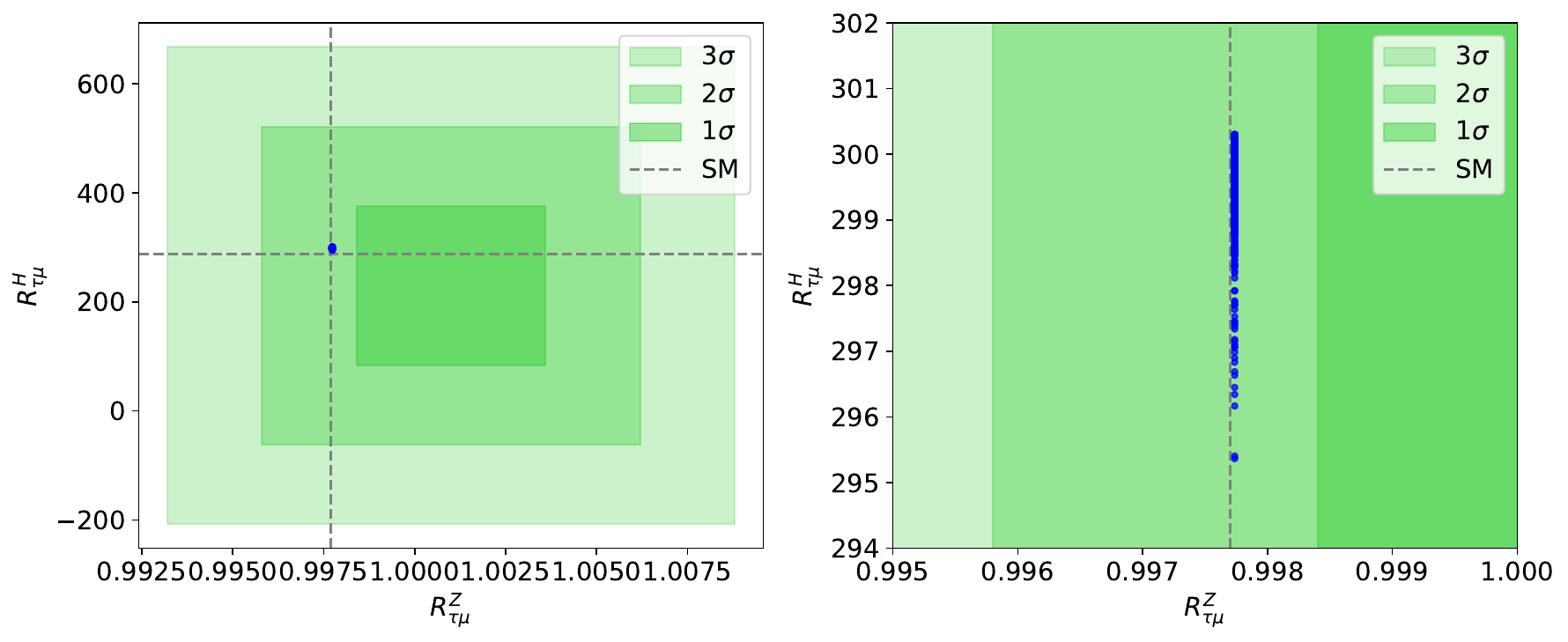}
    \caption{Scanned results of the lepton flavor universality. The dashed line with "SM" represents the SM prediction given in Table~\ref{tab:experimental_Ztolilj}. The left panel shows all scanned points alongside the experimental bounds at $3\sigma$. The right panel provides a zoomed-in view.}
    \label{fig:RZtm_RHtm}
\end{figure}
\endgroup
Since the experimental measurement of $H \rightarrow e^{+}e^{-}$ is unavailable due to the highly suppressed branching ratio from the light electron mass, we focus on the lepton flavor universality between the muon and tau. All scanned points are extremely well aligned with the SM prediction, falling within $2\sigma$ of the experimental bounds. This indicates that the one-loop corrections are well under control, confirming that our scan does not violate the well-established precision constraints on the $Z$ and $H$ bosons. A further notable feature of these results is the IR contribution to the $H \rightarrow \ell^{+}\ell^{-}$ decays. For $H \rightarrow \tau^{+}\tau^{-}$, the IR contribution is smaller by roughly one to two orders of magnitude than the sum of vertex correction and counter-term contributions. For $H \rightarrow \mu^{+}\mu^{-}$, the IR contribution is smaller by roughly one order of magnitude compared to the summed contribution. For $H \rightarrow e^{+}e^{-}$, the IR contribution is of comparable magnitude to the summed contribution. These results confirm that for heavier leptons, the IR contributions remain relatively small, whereas for lighter leptons, they become as significant as the other contributions combined.
\begingroup
\begin{figure}[t]
    \centering
    \includegraphics[width=0.96\textwidth]{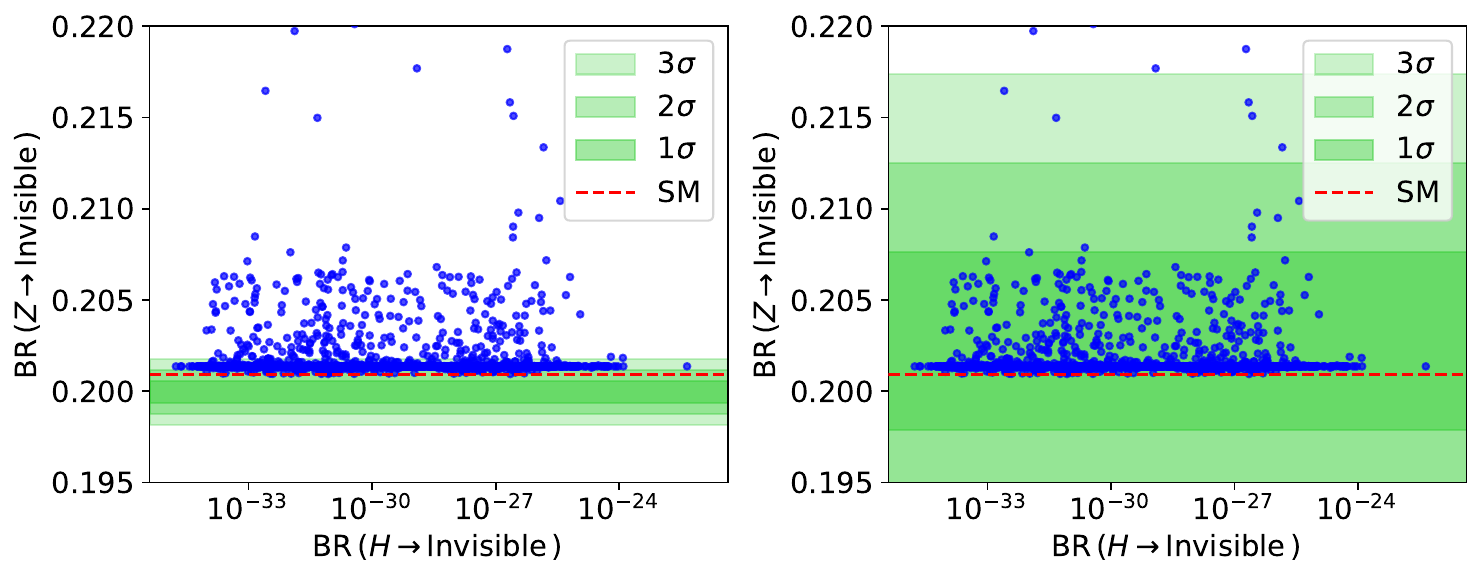}
    \caption{Scanned results of $H \rightarrow \func{Invisible}$ versus $Z \rightarrow \func{Invisible}$. In both panels, the dashed line denotes the SM prediction for the $Z \rightarrow \func{Invisible}$ branching ratio given in Table~\ref{tab:experimental_Ztolilj}. The left panel is based on the world avergae experimental bound, whereas the right panel is based on the recent ATLAS experimental bound~\cite{ATLAS:2023ynf}}
    \label{fig:Hinv_Zinv}. 
\end{figure}
\endgroup
The last observable we discuss is the $Z \rightarrow \func{Invisible}$ decay, with the scanned results shown in Figure~\ref{fig:Hinv_Zinv}. We confirm that most of the numerical points are consistent with the world-average experimental bound at the $3\sigma$ level. We find that the NP contribution to the $Z \rightarrow \func{Invisible}$ decay is mainly driven by the diagonal contributions, giving rise to a numerical enhancement of $0.0004$, whereas the off-diagonal contributions remain relatively suppressed, of order $\mathcal{O}\left( 10^{-12} \right)$. We also investigated the points that deviate significantly from the world-average bound, and found that these deviations arise from a numerically unstable region in which the mass of the CP-even or CP-odd scalar is nearly degenerated with that of the lightest right-handed neutrino $\chi_{1}$, in the range $100-600 \func{GeV}$. The recent $Z \rightarrow \func{Invisible}$ measurement reported by the ATLAS collaboration features a broader $1\sigma$ uncertainty~\cite{ATLAS:2023ynf}
\begingroup
\begin{equation}
    \Gamma\left( Z \rightarrow \func{Invisible} \right) = 506 \pm 2 (\func{stat.}) \pm 12(\func{syst.}) \func{MeV},
\end{equation}
\endgroup
corresponding to $\func{BR}\left( Z \rightarrow \func{Invisible} \right) = 0.202765 \pm 0.004875$ using the experimental $Z$ boson decay width $\Gamma_{Z} = 2.4955 \func{GeV}$ from PDG~\cite{ParticleDataGroup:2024cfk}. This result is favored by our numerical findings at the $3\sigma$ level. 

\section{Conclusion}\label{sec:conclusion}

The SM has explained many phenomena with remarkable precision, yet several key observations remain unexplained within its framework - most notably experimentally measured nonzero neutrino masses and the existence of DM. Additionally, the SM quartic coupling constant turns negative at high energies around $10^{9} \func{GeV}$, signaling an instability of the electroweak vacuum. To address these limitations, we extend the original scotogenic model by introducing an additional complex singlet scalar and a $U(1)^{\prime}$ global symmetry. Within this BSM framework, we discuss the diagonalization of the flavor and scalar sectors, the one-loop neutrino mass generation mechanism for both normal and inverted hierarchies along with the relevant experimental bounds, and potential DM candidates together with their relic density constraints, further refined by incorporating EW corrections.
\\~\\
We investigate all constraints arising from the extended scalar sector, namely the BFB conditions, vacuum stability conditions, and perturbativity bounds induced by RG running of each parameter. For the BFB conditions, we derive both necessary and sufficient conditions by requiring $V \rightarrow \infty$ for all possible field directions in the large field limit. For vacuum stability, we briefly discuss that the SM vacuum is considered metastable within the current experimental uncertainties on the top quark mass and the strong coupling constant $\alpha_{S}$. To identify regions of parameter space with a stable vacuum, we employ the public tool \texttt{Vevacious}. Finally, we investigate the RG-induced perturbativity bounds. We first motivate their necessary: perturbativity must be preserved over the entire energy range from the initial NP scale to the Planck scale, and RG-induced bounds typically place strong constraints on the couplings, making it essential to impose an upper threshold before the theory enters a Landau pole or non-perturbative regime. With this motivation, we confirm that the coupling constant $g_{X}$ and quartic coupling constant $\lambda_{i}$ with $i = 1, \cdots, 8$ must each remain below $0.2$, as determined by RG evolution of the parameters in both the SM and the present BSM model.
\\~\\
We next discuss several flavor and precision observables. Since this BSM model modifies the vacuum structure of the SM, accounting solely for NP contributions from NP particles is generally insufficient. Instead, we separately compute each observable in both the SM and BSM models, and take their difference as the net NP contribution. To ensure a UV-finite result, we perform a full renormalization in both models, employing the on-shell (OS) scheme in the Feynman t'Hooft gauge $(\xi = 1)$, as well as the alternative tadpole scheme. For the muon $g-2$ and the radiative decays $\ell_{\alpha} \rightarrow \ell_{\beta} \gamma$, we first briefly review the current status of the $\Delta a_{\mu}$ tension, and show that both observables can be expressed in terms of the dipole operator coefficient $c_{R}$, subject to the latest experimental bounds. We then discuss the $\ell_{\alpha} \rightarrow 3\ell_{\beta}$ decays and the $\mu \rightarrow e$ conversion rate, presenting their analytic expressions alongside the corresponding experimental bounds. Next, we discuss the precision observables, namely the oblique parameters and the decays $Z \rightarrow \overline{\ell}_{\alpha} \ell_{\beta}$, $Z \rightarrow \func{Invisible}$, $H \rightarrow \overline{\ell}_{\alpha} \ell_{\beta}$ and $H \rightarrow \func{Invisible}$, providing analytic expressions and experimental bounds for each. For the $H \rightarrow \overline{\ell}_{\alpha} \ell_{\beta}$ decay in particular, we explicitly examine the IR structure and its cancellation via the Bremsstrahlung effects.
\\~\\
Finally, we perform a numerical scan. We first discuss the scanned neutrino mass spectra for both normal and inverted hierarchies. For the normal hierarchy, the sum of active neutrino masses ranges from $5.8698 \times 10^{-11}$ to $9.9913 \times 10^{-11} \func{GeV}$, whereas for the inverted hierarchy it ranges from $9.9453 \times 10^{-11}$ to $1.0000 \times 10^{-11} \func{GeV}$. The scanned inverted hierarchy spectrum would be ruled out by the recent DESI BAO result, $\sum_{i=1}^{3} m_{\nu_{i}} < 6.42 \times 10^{-11} \func{GeV}$, if the result is confirmed by other experiments. Furthermore, the same bound excludes a large portion of the normal hierarchy parameter space as well, leaving only a small viable region.
\\
We next discuss the potential DM candidates, their relic densities, and their spin-independent proton cross sections. We find that the CP-even scalar $\eta_{R}$ is consistently slightly heavier than the CP-odd scalar $\eta_{I}$, leaving two viable DM candidates: $\eta_{I}$ and $\chi_{1}$. The fermionic DM candidate $\chi_{1}$ has mass ranging from $120-350 \func{GeV}$, while the CP-odd scalar candidate $\eta_{I}$ ranges from $350-600 \func{GeV}$. Within the allowed parameter space, the fermionic and scalar DM candidates account for approximately $29\%$ and $71\%$ of the surviving points. We further examine their mass spectra against the spin-independent proton cross sections, comparing with the current experimental sensitivity from XENON1T and the projected sensitivities from XENONnT, DARWIN, and the $\nu$-floor, finding that none of these constraints place significant restrictions on the parameter space.
\\
We find that no region of parameter space exhibits a manifest distinction between the normal and inverted neutrino mass hierarchies. Moreover, all flavor-violating observables, including the muon $g-2$, are suppressed by many orders of magnitude relative to their current experimental bounds. Beyond these highly suppressed flavor observables, several results call for discussion. The oblique parameters are comparatively less suppressed, and the dominant contribution is traced to the mass splitting between the SM-like Higgs $m_{h_{1}}$ and the SM Higgs $m_{h}$. Other NP contributions are also examined but foundn to be small in comparison, confirming that the oblique parameters are particularly sensitive to deviations in the SM Higgs mass. We further investigate lepton flavor universality via the ratio of $Z/H \rightarrow \ell\ell$ decay rates, and confirm that our numerical results are consistent with the well-established experimental bounds on leptonic $Z$ and $H$ decays. A notable feature emerging from the leptonic decays is the IR contribution to $H \rightarrow \overline{\ell}_{\alpha} \ell_{\beta}$: while this IR contribution remains relatively small for heavier leptons, it becomes comparable in magnitude to all other contributions combined for lighter leptons. 
\\
Finally, we discuss the $Z \rightarrow \func{Invisible}$ decay. The numerical scan reveals that the NP contribution to this decay is mainly driven by the diagonal contributions, giving rise to a numerical enhancement of $0.0004$, which is well aligned with the world-average experimental bound at the $3\sigma$ level. We also investigated the points that deviate significantly from the world average, finding that they originate from a numerically unstable region in which the mass of the CP-even or CP-odd scalar is nearly degenerate with that of the lightest RH neutrino $\chi_{1}$, in the range $100-600\func{GeV}$. We further compared our results with the recent ATLAS measurement of the decay, finding that our numerical predictions are favored at the $3\sigma$ level.

\section*{Acknowledgements}

HL and SKK are supported by the National Research Foundation of Korea under Grant NRF-2023R1A2C100609111.

\appendix

\section{Casas-Ibarra parameterization}\label{subapp:CI}

The neutrino Yukawa coupling constant $g_{X}$ encodes the neutrino oscillation data. They can be expressed using the Casas-Ibarra parameterization~\cite{Casas:2001sr}:
\begingroup
\begin{equation}
    g_{X} = U_{L}^{\dagger} D_{L}^{-1/2} U_{R} D_{\nu}^{1/2} U_{\func{PMNS}}^{T}
\end{equation}
\endgroup
where $U_{L}$ and $D_{L}$ are the unitary mixing matrix and the diagonalized mass matrix of the one-loop matrix $M_{L}$ given in Equation~\ref{eqn:oneloop_ML}, respectively:
\begingroup
\begin{equation}
    D_{L} = U_{L}^{T} M_{L} U_{L},
\end{equation}
\endgroup
and $D_{\nu}$ is the diagonalized mass matrix of $M_{\nu}$ given in Equation~\ref{eqn:oneloop_Mnu} and $U_{\func{PMNS}}$ is the unitary Pontecorvo-Maki-Nakagawa-Sakata mixing matrix. Furthermore, the parameters entering $M_{L}$ and $M_{\nu}$ do not uniquely determine the coupling matrix $g_{X}$~\cite{Alvarez:2023dzz}. The remaining degrees of freedom are therefore encoded in the matrix $U_{R}$, which can be parameterized as:
\begingroup
\begin{equation}
    U_{R} = U_{R,23} U_{R,13} U_{R,12} =
    \begin{pmatrix}
        1 & 0 & 0 \\
        0 &  c_{R,23} & s_{R,23} \\
        0 & -s_{R,23} & c_{R_23}
    \end{pmatrix}
    \begin{pmatrix}
        c_{R,13} & 0 & s_{R,13} \\
        0 &  1 & 0 \\
        -s_{R,13} & 0 & c_{R,13}
    \end{pmatrix}
    \begin{pmatrix}
         c_{R,12} & s_{R,12} & 0 \\
        -s_{R,12} &  c_{R,12} & 0 \\
        0 & 0 & 1
    \end{pmatrix}
\end{equation}
\endgroup
where $c_{R,ij} = \cos\theta_{ij}$ with complex angle $\theta_{ij}$ and $s_{R,ij} = \sqrt{1-c_{R,ij}^{2}}$.

\section{Self-energies}\label{app:SE}

In this Section, we collect all the self-energies (SEs) used to derive the analytic expressions for the observables.

\subsection{Photon self-energy}\label{subapp:AASE}

The diagrams contributing to the photon SE in this work are seen in Figure~\ref{fig:AASE}:
\begingroup
\begin{figure}[t]
    \centering
    \begin{subfigure}{0.24\textwidth}
        \centering
        \includegraphics[width=\textwidth]{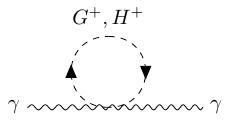}
    \end{subfigure}
    \begin{subfigure}{0.24\textwidth}
        \centering
        \includegraphics[width=\textwidth]{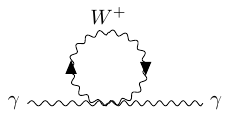}
    \end{subfigure}
    \begin{subfigure}{0.24\textwidth}
        \centering
        \includegraphics[width=\textwidth]{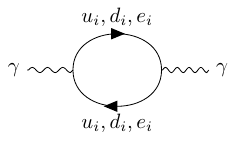}
    \end{subfigure}
    \begin{subfigure}{0.24\textwidth}
        \centering
        \includegraphics[width=\textwidth]{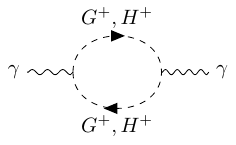}
    \end{subfigure}
    \begin{subfigure}{0.24\textwidth}
        \centering
        \includegraphics[width=\textwidth]{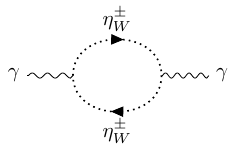}
    \end{subfigure}
    \begin{subfigure}{0.24\textwidth}
        \centering
        \includegraphics[width=\textwidth]{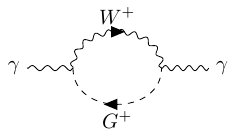}
    \end{subfigure}
    \begin{subfigure}{0.24\textwidth}
        \centering
        \includegraphics[width=\textwidth]{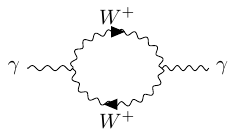}
    \end{subfigure}
    \caption{Feynman diagrams contributing to the photon SE in the scotogenic model. For the sixth diagram, an equivalent diagram with reversed charge flow is included. The charged Higgs $H^{+}$ contributions are absent in the SM photon SE.}
    \label{fig:AASE}
\end{figure}
\endgroup
The counter-term for the photon self-energy in the scotogenic model is given in Equation~\ref{eqn:dZAA}:
\begingroup
{
\allowdisplaybreaks
\begin{align}
    \delta Z_{AA} = -\left. \frac{\partial \Sigma_{T}^{AA} \left( p^{2} \right)}{\partial p^{2}} \right|_{p^{2} = 0} \nonumber \\
    = \frac{e^2 (s_w^2 - 1)}{144 c_w^2 (D-1) \pi^{2}}
    \Bigg[
    &\phantom{{}+{}} N_c \sum_{i=1}^{3} \Big(
    (2D - 4)\, B_0(0, m_{d_i}^2, m_{d_i}^2)
    + 8\, m_{d_i}^2\, DB_0(0, m_{d_i}^2, m_{d_i}^2)
    \Big) \nonumber \\
    &+ N_c \sum_{i=1}^{3} \Big(
    (8D - 16)\, B_0(0, m_{u_i}^2, m_{u_i}^2)
    + 32\, m_{u_i}^2\, DB_0(0, m_{u_i}^2, m_{u_i}^2)
    \Big) \nonumber \\
    &+ \sum_{i=1}^{3} \Big(
    (18D - 36)\, B_0(0, m_{e_i}^2, m_{e_i}^2)
    + 72\, m_{e_i}^2\, DB_0(0, m_{e_i}^2, m_{e_i}^2)
    \Big) \nonumber \\
    &+ (27 - 27D)\, B_0(0, M_W^2, M_W^2)
    + (36 - 36D)\, M_W^2\, DB_0(0, M_W^2, M_W^2) \nonumber \\
    &+ 9\, B_0(0, M_{H^\pm}^2, M_{H^\pm}^2)
    - 36\, M_{H^\pm}^2\, DB_0(0, M_{H^\pm}^2, M_{H^\pm}^2) 
    \Bigg]
    \label{eqn:dZAA}
\end{align}
}
\endgroup
where the $B_{0}$ and $DB_{0}$ are the Passarino-Veltman functions. The charged Higgs contributions are absent in the SM photon SE.

\subsection{Photon-\texorpdfstring{$Z$}{Z} self-energy}\label{subapp:AZSE}

The diagrams contributing to the photon-$Z$ SE in this work are seen in Figure~\ref{fig:AZSE}:
\begingroup
\begin{figure}[t]
    \centering
    \begin{subfigure}{0.24\textwidth}
        \centering
        \includegraphics[width=\textwidth]{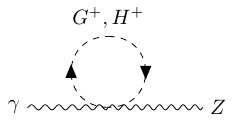}
    \end{subfigure}
    \begin{subfigure}{0.24\textwidth}
        \centering
        \includegraphics[width=\textwidth]{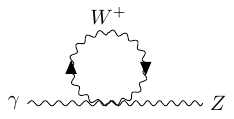}
    \end{subfigure}
    \begin{subfigure}{0.24\textwidth}
        \centering
        \includegraphics[width=\textwidth]{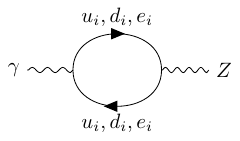}
    \end{subfigure}
    \begin{subfigure}{0.24\textwidth}
        \centering
        \includegraphics[width=\textwidth]{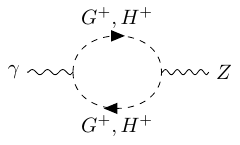}
    \end{subfigure}
    \begin{subfigure}{0.24\textwidth}
        \centering
        \includegraphics[width=\textwidth]{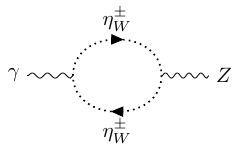}
    \end{subfigure}
    \begin{subfigure}{0.24\textwidth}
        \centering
        \includegraphics[width=\textwidth]{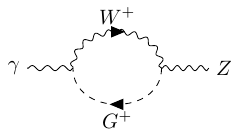}
    \end{subfigure}
    \begin{subfigure}{0.24\textwidth}
        \centering
        \includegraphics[width=\textwidth]{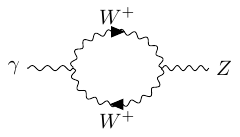}
    \end{subfigure}
    \caption{Feynman diagrams contributing to the photon SE in the scotogenic model. For the sixth diagram, an equivalent diagram with reversed charge flow is included. The charged Higgs $H^{+}$ contributions are absent in the SM photon SE.}
    \label{fig:AZSE}
\end{figure}
\endgroup
The counter-terms for the photon-$Z$ self-energy in the scotogenic model is given in Equation~\ref{eqn:dZAZ}:
\begingroup
{\allowdisplaybreaks
\begin{align}
    \delta Z_{ZA} &= \frac{2}{M_{Z}^{2}} \Sigma_{T}^{AZ} \left( 0 \right) = -\frac{e^{2} c_{w} \func{B}_{0}\left( 0, M_{W}^{2}, M_{W}^{2} \right)}{4 \pi^{2} s_{w}}, \nonumber \\
    \delta Z_{AZ} &= -2 \func{Re} \frac{\Sigma_{T}^{AZ}\left( M_{Z}^{2} \right)}{M_{Z}^{2}} = \frac{e^2}{144\, c_{w} (1-D) M_{Z}^2 \pi^2 s_{w}} \Bigg[ \nonumber \\
    &+ 36 M_{H^{\pm}}^2 \left(1 - 2 s_{w}^2\right) B_0\!\left(0, M_{H^{\pm}}^2, M_{H^{\pm}}^2\right) \nonumber \\
    &- 36 M_{W}^2 \left(3 - 2D + 2(D-1) s_{w}^2\right) B_0\!\left(0, M_{W}^2, M_{W}^2\right) \nonumber \\
    &+ 9 \left(4 M_{H^{\pm}}^2 - M_{Z}^2\right)\left(2 s_{w}^2 - 1\right) B_0\!\left(M_{Z}^2, M_{H^{\pm}}^2, M_{H^{\pm}}^2\right) \nonumber \\
    &+ 9 \Big( 4 M_{W}^2 \left(4 - 3D + 3(D-1) s_{w}^2\right) \nonumber \\
    &\qquad + M_{Z}^2 \left(5 - 6D + 6(D-1) s_{w}^2\right) - (D-1) e^2 v^2 \Big) B_0\!\left(M_{Z}^2, M_{W}^2, M_{W}^2\right) \nonumber \\
    &\qquad + 36 \left(4 s_{w}^2 - 1\right) \sum_{i=1}^{3} m_{e_{i}}^2\, B_0\!\left(0, m_{e_{i}}^2, m_{e_{i}}^2\right) \nonumber \\
    &- 9 \left(4 s_{w}^2 - 1\right) \sum_{i=1}^{3} \left((D-2) M_{Z}^2 + 4 m_{e_{i}}^2\right) B_0\!\left(M_{Z}^2, m_{e_{i}}^2, m_{e_{i}}^2\right) \nonumber \\
    &+ 4 \left(4 s_{w}^2 - 3\right) N_c \sum_{i=1}^{3} m_{d_{i}}^2\, B_0\!\left(0, m_{d_{i}}^2, m_{d_{i}}^2\right) \nonumber \\
    &+ 8 \left(8 s_{w}^2 - 3\right) N_c \sum_{i=1}^{3} m_{u_{i}}^2\, B_0\!\left(0, m_{u_{i}}^2, m_{u_{i}}^2\right) \nonumber \\
    &- \left(4 s_{w}^2 - 3\right) N_c \sum_{i=1}^{3} \left((D-2) M_{Z}^2 + 4 m_{d_{i}}^2\right) B_0\!\left(M_{Z}^2, m_{d_{i}}^2, m_{d_{i}}^2\right) \nonumber \\
    &- 2 \left(8 s_{w}^2 - 3\right) N_c \sum_{i=1}^{3} \left((D-2) M_{Z}^2 + 4 m_{u_{i}}^2\right) B_0\!\left(M_{Z}^2, m_{u_{i}}^2, m_{u_{i}}^2\right) \Bigg]    
    \label{eqn:dZAZ}
\end{align}
}
\endgroup
where the charged Higgs contributions are absent in the SM photon-$Z$ SE.

\subsection{\texorpdfstring{$Z$}{Z} self-energy}\label{subapp:ZZSE}

From $Z$ SE, the tadpole contributions start to appear. Therefore, we split them for better visibility. The diagrams contributing to the $Z$ tadpoles and $Z$ SEs are seen in Figures~\ref{fig:ZZSE_tadpole} and \ref{fig:ZZSE_SE}.
\begingroup
\begin{figure}[t]
    \centering
    \begin{subfigure}{0.24\textwidth}
        \centering
        \includegraphics[width=\textwidth]{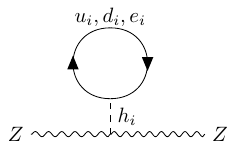}
    \end{subfigure}
    \begin{subfigure}{0.24\textwidth}
        \centering
        \includegraphics[width=\textwidth]{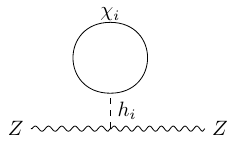}
    \end{subfigure}
    \begin{subfigure}{0.24\textwidth}
        \centering
        \includegraphics[width=\textwidth]{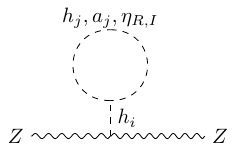}
    \end{subfigure}
    \begin{subfigure}{0.24\textwidth}
        \centering
        \includegraphics[width=\textwidth]{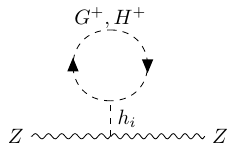}
    \end{subfigure}
    \begin{subfigure}{0.24\textwidth}
        \centering
        \includegraphics[width=\textwidth]{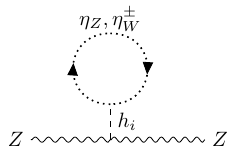}
    \end{subfigure}
    \begin{subfigure}{0.24\textwidth}
        \centering
        \includegraphics[width=\textwidth]{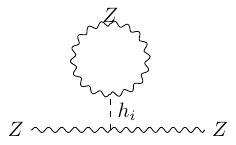}
    \end{subfigure}
    \begin{subfigure}{0.24\textwidth}
        \centering
        \includegraphics[width=\textwidth]{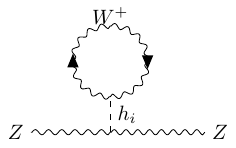}
    \end{subfigure}
    \caption{Feynman diagrams contributing to the $Z$ boson tadpole in the scotogenic model, where $i,j=1,2$. In the SM limit, $h_{i}$ and $a_{i}$ reduce to the SM Higgs boson $h$ and the would-be Goldstone boson $a$ of the $Z$ boson, respectively, while the contributions involving the $\chi, \eta_{R,I}$ and $H^{+}$ fields are absent.}
    \label{fig:ZZSE_tadpole}
\end{figure}
\endgroup
\begingroup
\begin{figure}[t]
    \centering
    \begin{subfigure}{0.24\textwidth}
        \centering
        \includegraphics[width=\textwidth]{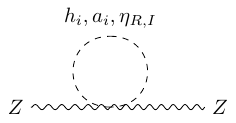}
    \end{subfigure}
    \begin{subfigure}{0.24\textwidth}
        \centering
        \includegraphics[width=\textwidth]{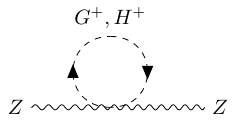}
    \end{subfigure}
    \begin{subfigure}{0.24\textwidth}
        \centering
        \includegraphics[width=\textwidth]{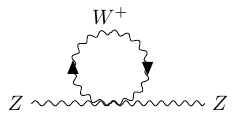}
    \end{subfigure}
    \begin{subfigure}{0.24\textwidth}
        \centering
        \includegraphics[width=\textwidth]{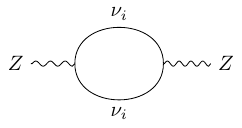}
    \end{subfigure}
    \begin{subfigure}{0.24\textwidth}
        \centering
        \includegraphics[width=\textwidth]{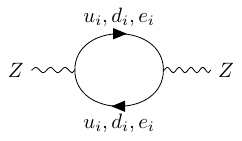}
    \end{subfigure}
    \begin{subfigure}{0.24\textwidth}
        \centering
        \includegraphics[width=\textwidth]{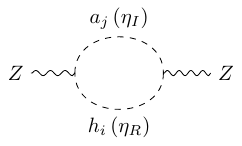}
    \end{subfigure}
    \begin{subfigure}{0.24\textwidth}
        \centering
        \includegraphics[width=\textwidth]{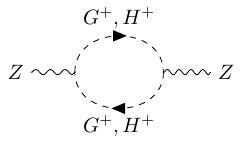}
    \end{subfigure}
    \begin{subfigure}{0.24\textwidth}
        \centering
        \includegraphics[width=\textwidth]{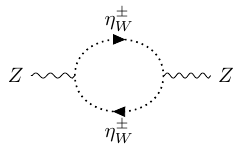}
    \end{subfigure}
    \begin{subfigure}{0.24\textwidth}
        \centering
        \includegraphics[width=\textwidth]{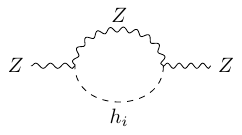}
    \end{subfigure}
    \begin{subfigure}{0.24\textwidth}
        \centering
        \includegraphics[width=\textwidth]{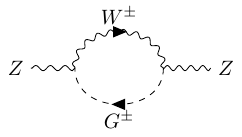}
    \end{subfigure}
    \begin{subfigure}{0.24\textwidth}
        \centering
        \includegraphics[width=\textwidth]{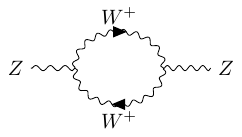}
    \end{subfigure}
    \caption{Feynman diagrams contributing to the $Z$ boson tadpole in the scotogenic model, where $i,j=1,2$. In the SM limit, $h_{i}$ and $a_{i}$ reduce to the SM Higgs boson $h$ and the would-be Goldstone boson $a$ of the $Z$ boson, respectively, while the contributions involving the $\chi$ and $H^{+}$ fields are absent.}
    \label{fig:ZZSE_SE}
\end{figure}
\endgroup
With the explicit tadpole contributions, the SE in the alternative tadpole scheme is redefined as~\cite{Krause:2016gkg}:
\begingroup
\begin{equation}
    \Sigma_{T}^{ZZ}\left( p^{2} \right) \equiv \Sigma_{T}^{ZZ}\left( p^{2} \right) + \Sigma_{T}^{ZZ,\func{Tad}}\left( p^{2} \right)
\end{equation}
\endgroup
This definition applies to all other SEs which feature explicit tadpole contributions. Then, the mass and field CT terms for the $Z$ SE are given in Equations~\ref{eqn:dMZ2} and \ref{eqn:dZZZ}.
\begingroup
{\allowdisplaybreaks
\begin{align}
\delta M_{Z}^{2} &= \func{Re} \Sigma_{T}^{ZZ} \left( M_{Z}^{2} \right) \nonumber \\
&= \frac{1}{64 c_{w}^2 (-1+D) M_{Z}^2 \pi ^2 s_{w}^2}
e^2 (M_{\eta_{R}}-M_{\eta_{I}}-M_{Z}) (M_{\eta_{R}}+M_{\eta_{I}}-M_{Z}) \nonumber \\
&\times (M_{\eta_{R}}-M_{\eta_{I}}+M_{Z}) (M_{\eta_{R}}+M_{\eta_{I}}+M_{Z}) B_0\Big(M_{Z}^2,M_{\eta_{R}}^2,M_{\eta_{I}}^2\Big) \nonumber \\ 
&- \frac{1}{64 c_{w}^2 (-1+D) \pi ^2 s_{w}^2}
\Big(4 M^{}_{H^{+}}-M_{Z}^2\Big) \Big(e-2 e s_{w}^2\Big)^2 \
B_0\Big(M_{Z}^2,M^{2}_{H^{+}},M^{2}_{H^{+}}\Big) \nonumber \\ 
&+ \frac{1}{64 c_{w}^2 (-1+D) \pi ^2 s_{w}^2}
e^2 M_{Z}^2 \Big(-3+4 s_{w}^2\Big) \Big(-15+12 D+(20-16 D) s_{w}^2+4 \
(-1+D) s_{w}^4\Big) \nonumber \\ &\times B_0\Big(M_{Z}^2,M_{W}^2,M_{W}^2\Big) \nonumber \\ 
&+ \frac{1}{64 c_{w}^2 (-1+D) \pi ^2 s_{w}^2}
3 (-2+D) e^2 M_{Z}^2 B_0\Big(M_{Z}^2,0,0\Big) \nonumber \\ 
&+ \sum_{k=1}^3\frac{1}{64 c_{w}^2 (-1+D) \pi ^2 s_{w}^2}
e^2 B_0\Big(M_{Z}^2,m_{e_k}^2,m_{e_k}^2\Big) \Big((-2+D) M_{Z}^2 \Big(1-4 s_{w}^2+8 s_{w}^4\Big) \nonumber \\
&-2 \Big(-3+D+8 s_{w}^2-16 s_{w}^4\Big) m_{e_k}^2\Big) \nonumber \\ 
&+ \sum_{k=1}^3\frac{1}{576 c_{w}^2 (-1+D) \pi ^2 s_{w}^2}
e^2 N_{c} B_0\Big(M_{Z}^2,m_{d_k}^2,m_{d_k}^2\Big) \Big((-2+D) M_{Z}^2 \Big(9-12 s_{w}^2+8 s_{w}^4\Big) \nonumber \\
&+\Big(54-18 D-48 s_{w}^2+32 s_{w}^4\Big) m_{d_k}^2\Big) \nonumber \\ 
&+ \sum_{k=1}^3\frac{1}{576 c_{w}^2 (-1+D) \pi ^2 s_{w}^2}
e^2 N_{c} B_0\Big(M_{Z}^2,m_{u_k}^2,m_{u_k}^2\Big) \Big((-2+D) M_{Z}^2 \Big(9-24 s_{w}^2+32 s_{w}^4\Big) \nonumber \\
&-2 \Big(-27+9 D+48 s_{w}^2-64 s_{w}^4\Big) m_{u_k}^2\Big) \nonumber \\ 
&+ \sum_{i=1}^2\frac{1}{32 c_{w}^2 (-2+D) (-1+D) M_{Z}^2 \pi ^2 s_{w}^2}
-e^2 B_0\Big(0,M_{h_i}^2,M_{h_i}^2\Big) M_{h_i}^2 \Big(-(-1+D) M_{Z}^2+M_{h_i}^2\Big) U^{H,2}_{i1} \nonumber \\ 
&+ \sum_{i=1}^2\frac{1}{64 c_{w}^2 (-1+D) M_{Z}^2 \pi ^2 s_{w}^2}
e^2 B_0\Big(M_{Z}^2,M_{Z}^2,M_{h_i}^2\Big) \Big(4 (-1+D) M_{Z}^4-4 M_{Z}^2 M_{h_i}^2+M_{h_i}^4\Big) U^{H,2}_{i1} \nonumber \\ 
&+ \sum_{k=1}^3\frac{1}{16 c_{w}^2 (-2+D) (-1+D) \pi ^2 s_{w}^2 M_{h_i}^2}
e^2 B_0\Big(0,m_{e_k}^2,m_{e_k}^2\Big) m_{e_k}^2 \Big(-(-2+D) \Big(1-4 s_{w}^2+8 s_{w}^4\Big) M_{h_i}^2 \nonumber \\
&+\sum_{i=1}^2 4 (-1+D) m_{e_k}^2 U^{H,2}_{i1}\Big) \nonumber \\ 
&+ \sum_{k=1}^3\frac{1}{144 c_{w}^2 (-2+D) (-1+D) \pi ^2 s_{w}^2 M_{h_i}^2}
e^2 N_{c} B_0\Big(0,m_{u_k}^2,m_{u_k}^2\Big) m_{u_k}^2 \Big(-(-2+D) \nonumber \\
&\times \Big(9-24 s_{w}^2+32 s_{w}^4\Big) M_{h_i}^2+\sum_{i=1}^2 36 (-1+D) m_{u_k}^2 U^{H,2}_{i1}\Big) \nonumber \\ 
&+ \sum_{k=1}^3\frac{1}{144 c_{w}^2 (-2+D) (-1+D) \pi ^2 s_{w}^2 M_{h_i}^2}
e^2 N_{c} B_0\Big(0,m_{d_k}^2,m_{d_k}^2\Big) m_{d_k}^2 \Big(-(-2+D) \nonumber \\
&\times \Big(9-12 s_{w}^2+8 s_{w}^4\Big) M_{h_i}^2+\sum_{i=1}^2 36 (-1+D) m_{d_k}^2 U^{D}_{L,k,k} U^{H,2}_{i1}\Big) \nonumber \\ 
&+ \sum_{i=1}^2 \Big(\sum_{k=1}^3\frac{1}{4 c_{w} (-2+D) \pi ^2 s_{w} v_{S} M_{h_i}^2}
e M_{Z} B_0\Big(0,M_{\chi_k}^2,M_{\chi_k}^2\Big) M_{\chi_k}^4 U^{H}_{i1} U^{H}_{i2} \nonumber \\ 
&- \sum_{i=1}^2\frac{1}{32 c_{w} (-2+D) \pi ^2 s_{w} M_{h_i}^2}
M_{A}^2 M_{Z} B_0\Big(0,M_{A}^2,M_{A}^2\Big) U^{H}_{i1} (4 c_{w} \lambda_{7} M_{Z} s_{w} U^{H}_{i1}+e \lambda_{6} v_{S} U^{H}_{i2}) \nonumber \\ 
&+ \frac{1}{32 c_{w}^2 (-2+D) (-1+D) M_{Z}^2 \pi ^2 s_{w}^2 M_{h_i}^2}
M_{\eta_{I}}^2 B_0\Big(0,M_{\eta_{I}}^2,M_{\eta_{I}}^2\Big) \Big(e^2 M_{\eta_{R}}^2 M_{h_i}^2-e^2 M_{\eta_{I}}^2 M_{h_i}^2 \nonumber \\
&-2 e^2 M_{Z}^2 M_{h_i}^2+D e^2 M_{Z}^2 M_{h_i}^2+\sum_{i=1}^2 -2 (-1+D) (2 \lambda_{3}+2 \lambda_{4}-\lambda_{5}) M_{Z}^4 s_{w}^2 U^{H,2}_{i1} \nonumber \\
&+\sum_{i=1}^2 2 (-1+D) (2 \lambda_{3}+2 \lambda_{4}-\lambda_{5}) M_{Z}^4 s_{w}^4 U^{H,2}_{i1}+\sum_{i=1}^2 -2 c_{w} (-1+D) e \lambda_{8} M_{Z}^3 s_{w} v_{S} U^{H}_{i1} U^{H}_{i2}\Big) \nonumber \\ 
&+ \frac{1}{32 c_{w}^2 (-2+D) (-1+D) M_{Z}^2 \pi ^2 s_{w}^2 M_{h_i}^2}
-M_{\eta_{R}}^2 B_0\Big(0,M_{\eta_{R}}^2,M_{\eta_{R}}^2\Big) \Big(e^2 M_{\eta_{R}}^2 M_{h_i}^2-e^2 M_{\eta_{I}}^2 M_{h_i}^2 \nonumber \\
&+2 e^2 M_{Z}^2 M_{h_i}^2-D e^2 M_{Z}^2 M_{h_i}^2+\sum_{i=1}^2 2 (-1+D) (2 \lambda_{3}+2 \lambda_{4}+\lambda_{5}) M_{Z}^4 s_{w}^2 U^{H,2}_{i1} \nonumber \\
&+\sum_{i=1}^2 -2 (-1+D) (2 \lambda_{3}+2 \lambda_{4}+\lambda_{5}) M_{Z}^4 s_{w}^4 U^{H,2}_{i1}+\sum_{i=1}^2 2 c_{w} (-1+D) e \lambda_{8} M_{Z}^3 s_{w} v_{S} U^{H}_{i1} U^{H}_{i2}\Big) \nonumber \\ 
&+ \frac{1}{16 c_{w}^2 (-2+D) (-1+D) \pi ^2 s_{w}^2 M_{h_i}^2}
M^{}_{H^{+}} B_0\Big(0,M^{}_{H^{+}},M^{}_{H^{+}}\Big) \Big((-2+D) e^2 \Big(1-2 s_{w}^2\Big)^2 M_{h_i}^2\nonumber \\
&+\sum_{i=1}^2 2 (-1+D) M_{Z} s_{w} U^{H}_{i1} \Big(2 \lambda_{3} M_{Z} s_{w} \Big(-1+s_{w}^2\Big) U^{H}_{i1}-c_{w} e \lambda_{8} v_{S} U^{H}_{i2}\Big)\Big) \nonumber \\ 
&+ \frac{1}{32 c_{w}^2 (-2+D) (-1+D) \pi ^2 s_{w}^2 M_{h_i}^2}
-B_0\Big(0,M_{Z}^2,M_{Z}^2\Big) \Big(e^2 M_{Z}^2 M_{h_i}^2 \Big(1-D+\sum_{i=1}^2 2 U^{H,2}_{i1}\Big) \nonumber \\
&+\sum_{i=1}^2 -e^2 M_{h_i}^4 U^{H,2}_{i1}+\sum_{i=1}^2 2 (-1+D) M_{Z}^3 U^{H}_{i1} \Big(M_{Z} \Big((-1+D) e^2+\lambda_{1} s_{w}^2-\lambda_{1} s_{w}^4\Big) U^{H}_{i1}+c_{w} e \lambda_{7} s_{w} v_{S} U^{H}_{i2}\Big)\Big) \nonumber \\ 
&+ \frac{1}{16 (-2+D) (-1+D) \pi ^2 s_{w}^2 M_{h_i}^2}
M_{Z}^2 B_0\Big(0,M_{W}^2,M_{W}^2\Big) \Big(e^2 \Big(14-15 D+4 D^2 \nonumber \\
&-4 \Big(6-7 D+2 D^2\Big) s_{w}^2+4 \Big(2-3 D+D^2\Big) s_{w}^4\Big) M_{h_i}^2 \nonumber \\
&+\sum_{i=1}^2 2 (-1+D) M_{Z} U^{H}_{i1} \Big(M_{Z} \Big(-1+s_{w}^2\Big) \Big((-1+D) e^2+\lambda_{1} s_{w}^2\Big) U^{H}_{i1}-c_{w} e \lambda_{7} s_{w} v_{S} U^{H}_{i2}\Big)\Big) \nonumber \\ 
&+ \sum_{i=1}^2 \Big(\sum_{j=1}^2\frac{1}{32 c_{w} (-2+D) \pi ^2 s_{w} M_{h_i}^2}
-M_{Z} B_0\Big(0,M_{h_j}^2,M_{h_j}^2\Big) M_{h_j}^2 U^{H}_{i1} \nonumber \\
&\times \Big(U^{H}_{i2} \Big(2 e \lambda_{7} v_{S} U^{H,2}_{j1}+8 c_{w} \lambda_{7} M_{Z} s_{w} U^{H}_{j1} U^{H}_{j2}+3 e \lambda_{6} v_{S} U^{H,2}_{j2}\Big) \nonumber \\
&+2 U^{H}_{i1} \Big(2 e \lambda_{7} v_{S} U^{H}_{j1} U^{H}_{j2}+c_{w} M_{Z} s_{w} \Big(3 \lambda_{1} U^{H,2}_{j1}+2 \lambda_{7} U^{H,2}_{j2}\Big)\Big)\Big) 
\label{eqn:dMZ2}
\end{align}
}
\endgroup

\begingroup
{\allowdisplaybreaks
\begin{align}
\delta Z_{ZZ} &= -\func{Re} \left. \frac{\partial \Sigma_{T}^{ZZ} \left( p^{2} \right)}{\partial p^{2}} \right|_{p^{2}=M_{Z}^{2}} \nonumber \\
&= \frac{1}{32 c_{w}^2 (-2+D) (-1+D) M_{Z}^4 \pi ^2 s_{w}^2}
e^2 M_{\eta_{R}}^2 \Big(-M_{\eta_{R}}^2+M_{\eta_{I}}^2\Big) B_0\Big(0,M_{\eta_{R}}^2,M_{\eta_{R}}^2\Big) \nonumber \\ 
&+ \frac{1}{32 c_{w}^2 (-2+D) (-1+D) M_{Z}^4 \pi ^2 s_{w}^2}
e^2 M_{\eta_{I}}^2 \Big(M_{\eta_{R}}^2-M_{\eta_{I}}^2\Big) B_0\Big(0,M_{\eta_{I}}^2,M_{\eta_{I}}^2\Big) \nonumber \\ 
&+ \frac{1}{64 c_{w}^2 (-1+D) M_{Z}^4 \pi ^2 s_{w}^2}
e^2 \Big(M_{\eta_{R}}^4-2 M_{\eta_{R}}^2 M_{\eta_{I}}^2+M_{\eta_{I}}^4-M_{Z}^4\Big) B_0\Big(M_{Z}^2,M_{\eta_{R}}^2,M_{\eta_{I}}^2\Big) \nonumber \\ 
&+ \frac{1}{64 c_{w}^2 (-1+D) \pi ^2 s_{w}^2}
-\Big(e-2 e s_{w}^2\Big)^2 B_0\Big(M_{Z}^2,M^{2}_{H^{+}},M^{2}_{H^{+}}\Big) \nonumber \\ 
&+ \frac{1}{64 c_{w}^2 (-1+D) \pi ^2 s_{w}^2}
e^2 \Big(3 c_{w}^4 (-3+4 D)+2 c_{w}^2 s_{w}^2-s_{w}^4\Big) B_0\Big(M_{Z}^2,M_{W}^2,M_{W}^2\Big) \nonumber \\ 
&+ \frac{1}{64 c_{w}^2 (-1+D) \pi ^2 s_{w}^2}
-3 (-2+D) e^2 B_0\Big(M_{Z}^2,0,0\Big) \nonumber \\ 
&+ \sum_{k=1}^3\frac{1}{64 c_{w}^2 (-1+D) \pi ^2 s_{w}^2}
-(-2+D) e^2 \Big(1-4 s_{w}^2+8 s_{w}^4\Big) B_0\Big(M_{Z}^2,m_{e_k}^2,m_{e_k}^2\Big) \nonumber \\ 
&+ \sum_{k=1}^3\frac{1}{576 c_{w}^2 (-1+D) \pi ^2 s_{w}^2}
-(-2+D) e^2 N_{c} \Big(9-12 s_{w}^2+8 s_{w}^4\Big) B_0\Big(M_{Z}^2,m_{d_k}^2,m_{d_k}^2\Big) \nonumber \\ 
&+ \sum_{k=1}^3\frac{1}{576 c_{w}^2 (-1+D) \pi ^2 s_{w}^2}
-(-2+D) e^2 N_{c} \Big(9-24 s_{w}^2+32 s_{w}^4\Big) B_0\Big(M_{Z}^2,m_{u_k}^2,m_{u_k}^2\Big) \nonumber \\ 
&+ \sum_{i=1}^2\frac{1}{32 c_{w}^2 (-2+D) (-1+D) M_{Z}^2 \pi ^2 s_{w}^2}
-e^2 B_0\Big(0,M_{Z}^2,M_{Z}^2\Big) \Big(M_{Z}^2-M_{h_i}^2\Big) U^{H,2}_{i1} \nonumber \\ 
&+ \sum_{i=1}^2\frac{1}{32 c_{w}^2 (-2+D) (-1+D) M_{Z}^4 \pi ^2 s_{w}^2}
e^2 B_0\Big(0,M_{h_i}^2,M_{h_i}^2\Big) M_{h_i}^2 \Big(M_{Z}^2-M_{h_i}^2\Big) U^{H,2}_{i1} \nonumber \\ 
&+ \sum_{i=1}^2\frac{1}{64 c_{w}^2 (-1+D) M_{Z}^4 \pi ^2 s_{w}^2}
e^2 B_0\Big(M_{Z}^2,M_{Z}^2,M_{h_i}^2\Big) M_{h_i}^2 \Big(-2 M_{Z}^2+M_{h_i}^2\Big) U^{H,2}_{i1} \nonumber \\ 
&- \frac{1}{576 c_{w}^2 (-1+D) M_{Z}^2 \pi ^2 s_{w}^2}
e^2 \Big(-54 M_{Z}^4 DB_{0}\Big(M_{Z}^2,0,0\Big)+27 D M_{Z}^4 DB_{0}\Big(M_{Z}^2,0,0\Big) \nonumber \\
&+9 \Big(M_{\eta_{R}}^4+\Big(M_{\eta_{I}}^2-M_{Z}^2\Big)^2-2 M_{\eta_{R}}^2 \Big(M_{\eta_{I}}^2+M_{Z}^2\Big)\Big) DB_{0}\Big(M_{Z}^2,M_{\eta_{R}}^2,M_{\eta_{I}}^2\Big) \nonumber \\
&+9 \Big(-4 M^{}_{H^{+}}+M_{Z}^2\Big) \Big(M_{Z}-2 M_{Z} s_{w}^2\Big)^2 DB_{0}\Big(M_{Z}^2,M^{}_{H^{+}},M^{}_{H^{+}}\Big)+405 M_{Z}^4 DB_{0}\Big(M_{Z}^2,M_{W}^2,M_{W}^2\Big) \nonumber \\
&-324 D M_{Z}^4 DB_{0}\Big(M_{Z}^2,M_{W}^2,M_{W}^2\Big)+144 (-1+D) M_{Z}^4 s_{w}^6 DB_{0}\Big(M_{Z}^2,M_{W}^2,M_{W}^2\Big) \nonumber \\
&+\sum_{k=1}^3 -18 M_{Z}^4 N_{c} DB_{0}\Big(M_{Z}^2,m_{d_k}^2,m_{d_k}^2\Big)+\sum_{k=1}^3 9 D M_{Z}^4 N_{c} DB_{0}\Big(M_{Z}^2,m_{d_k}^2,m_{d_k}^2\Big)+\sum_{k=1}^3 \nonumber \\
&-18 M_{Z}^4 DB_{0}\Big(M_{Z}^2,m_{e_k}^2,m_{e_k}^2\Big)+\sum_{k=1}^3 9 D M_{Z}^4 DB_{0}\Big(M_{Z}^2,m_{e_k}^2,m_{e_k}^2\Big) \nonumber \\
&+\sum_{k=1}^3 -18 M_{Z}^4 N_{c} DB_{0}\Big(M_{Z}^2,m_{u_k}^2,m_{u_k}^2\Big)+\sum_{k=1}^3 9 D M_{Z}^4 N_{c} DB_{0}\Big(M_{Z}^2,m_{u_k}^2,m_{u_k}^2\Big) \nonumber \\
&+\sum_{k=1}^3 54 M_{Z}^2 N_{c} DB_{0}\Big(M_{Z}^2,m_{d_k}^2,m_{d_k}^2\Big) m_{d_k}^2+\sum_{k=1}^3 -18 D M_{Z}^2 N_{c} DB_{0}\Big(M_{Z}^2,m_{d_k}^2,m_{d_k}^2\Big) m_{d_k}^2 \nonumber \\
&+\sum_{k=1}^3 54 M_{Z}^2 DB_{0}\Big(M_{Z}^2,m_{e_k}^2,m_{e_k}^2\Big) m_{e_k}^2+\sum_{k=1}^3 -18 D M_{Z}^2 DB_{0}\Big(M_{Z}^2,m_{e_k}^2,m_{e_k}^2\Big) m_{e_k}^2 \nonumber \\
&+\sum_{k=1}^3 54 M_{Z}^2 N_{c} DB_{0}\Big(M_{Z}^2,m_{u_k}^2,m_{u_k}^2\Big) m_{u_k}^2+\sum_{k=1}^3 -18 D M_{Z}^2 N_{c} DB_{0}\Big(M_{Z}^2,m_{u_k}^2,m_{u_k}^2\Big) m_{u_k}^2 \nonumber \\
&+12 M_{Z}^2 s_{w}^2 \Big(18 (-5+4 D) M_{Z}^2 DB_{0}\Big(M_{Z}^2,M_{W}^2,M_{W}^2\Big)+\sum_{k=1}^3 \Big(-3 DB_{0}\Big(M_{Z}^2,m_{e_k}^2,m_{e_k}^2\Big) 
\nonumber \\
&\times \Big((-2+D) M_{Z}^2+4 m_{e_k}^2\Big)+N_{c} \Big(-DB_{0}\Big(M_{Z}^2,m_{d_k}^2,m_{d_k}^2\Big) \Big((-2+D) M_{Z}^2+4 m_{d_k}^2\Big) \nonumber \\
&-2 DB_{0}\Big(M_{Z}^2,m_{u_k}^2,m_{u_k}^2\Big) \Big((-2+D) M_{Z}^2+4 m_{u_k}^2\Big)\Big)\Big)\Big)-4 M_{Z}^2 s_{w}^4 \Big(9 (-23+19 D) M_{Z}^2 \nonumber \\
&\times DB_{0}\Big(M_{Z}^2,M_{W}^2,M_{W}^2\Big)+\sum_{k=1}^3 -2 \Big(9 DB_{0}\Big(M_{Z}^2,m_{e_k}^2,m_{e_k}^2\Big) \Big((-2+D) M_{Z}^2+4 m_{e_k}^2\Big)\nonumber \\
&+N_{c} \Big(DB_{0}\Big(M_{Z}^2,m_{d_k}^2,m_{d_k}^2\Big) \Big((-2+D) M_{Z}^2+4 m_{d_k}^2\Big)+4 DB_{0}\Big(M_{Z}^2,m_{u_k}^2,m_{u_k}^2\Big) \nonumber \\
&\times \Big((-2+D) M_{Z}^2+4 m_{u_k}^2\Big)\Big)\Big)\Big)+\sum_{i=1}^2 -36 M_{Z}^4 DB_{0}\Big(M_{Z}^2,M_{Z}^2,M_{h_i}^2\Big) U^{H,2}_{i1} \nonumber \\
&+\sum_{i=1}^2 36 D M_{Z}^4 DB_{0}\Big(M_{Z}^2,M_{Z}^2,M_{h_i}^2\Big) U^{H,2}_{i1}+\sum_{i=1}^2 -36 M_{Z}^2 DB_{0}\Big(M_{Z}^2,M_{Z}^2,M_{h_i}^2\Big) M_{h_i}^2 U^{H,2}_{i1}\nonumber \\
&+\sum_{i=1}^2 9 DB_{0}\Big(M_{Z}^2,M_{Z}^2,M_{h_i}^2\Big) M_{h_i}^4 U^{H,2}_{i1}\Big) 
\label{eqn:dZZZ}
\end{align}
}
\endgroup

\subsection{\texorpdfstring{$W$}{W} self-energy}\label{subapp:WWSE}

The diagrams contributing to the $W$ tadpole and SE are seen in Figure~\ref{fig:WWSE_tadpole} and Figure~\ref{fig:WWSE_SE}.
\begingroup
\begin{figure}[t]
    \centering
    \begin{subfigure}{0.24\textwidth}
        \centering
        \includegraphics[width=\textwidth]{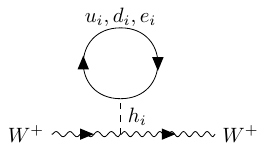}
    \end{subfigure}
    \begin{subfigure}{0.24\textwidth}
        \centering
        \includegraphics[width=\textwidth]{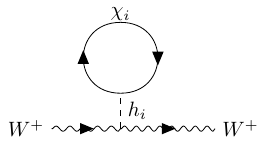}
    \end{subfigure}
    \begin{subfigure}{0.24\textwidth}
        \centering
        \includegraphics[width=\textwidth]{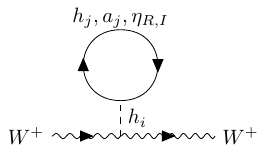}
    \end{subfigure}
    \begin{subfigure}{0.24\textwidth}
        \centering
        \includegraphics[width=\textwidth]{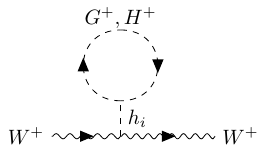}
    \end{subfigure}
    \begin{subfigure}{0.24\textwidth}
        \centering
        \includegraphics[width=\textwidth]{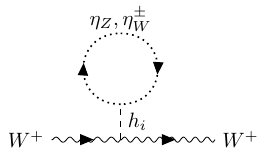}
    \end{subfigure}
    \begin{subfigure}{0.24\textwidth}
        \centering
        \includegraphics[width=\textwidth]{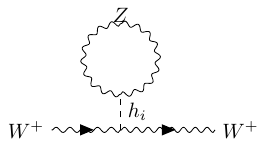}
    \end{subfigure}
    \begin{subfigure}{0.24\textwidth}
        \centering
        \includegraphics[width=\textwidth]{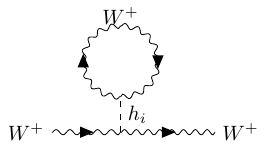}
    \end{subfigure}
    \caption{Feynman diagrams contributing to the $Z$ boson tadpole in the scotogenic model, where $i,j=1,2$. In the SM limit, $h_{i}$ and $a_{i}$ reduce to the SM Higgs boson $h$ and the would-be Goldstone boson $a$ of the $Z$ boson, respectively, while the contributions involving the $\chi, \eta_{R,I}$ and $H^{+}$ fields are absent.}
    \label{fig:WWSE_tadpole}
\end{figure}
\endgroup
\begingroup
\begin{figure}[t]
    \centering
    \begin{subfigure}{0.24\textwidth}
        \centering
        \includegraphics[width=\textwidth]{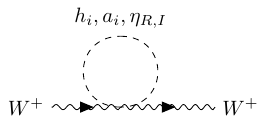}
    \end{subfigure}
    \begin{subfigure}{0.24\textwidth}
        \centering
        \includegraphics[width=\textwidth]{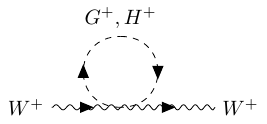}
    \end{subfigure}
    \begin{subfigure}{0.24\textwidth}
        \centering
        \includegraphics[width=\textwidth]{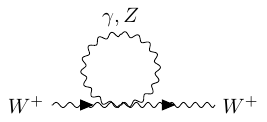}
    \end{subfigure}
    \begin{subfigure}{0.24\textwidth}
        \centering
        \includegraphics[width=\textwidth]{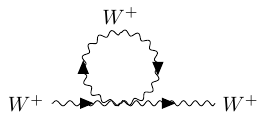}
    \end{subfigure}
    \begin{subfigure}{0.24\textwidth}
        \centering
        \includegraphics[width=\textwidth]{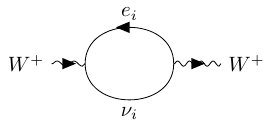}
    \end{subfigure}
    \begin{subfigure}{0.24\textwidth}
        \centering
        \includegraphics[width=\textwidth]{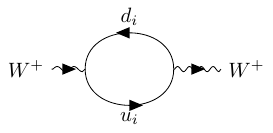}
    \end{subfigure}
    \begin{subfigure}{0.24\textwidth}
        \centering
        \includegraphics[width=\textwidth]{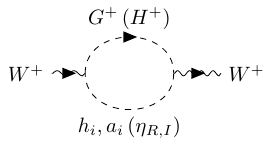}
    \end{subfigure}
    \begin{subfigure}{0.24\textwidth}
        \centering
        \includegraphics[width=\textwidth]{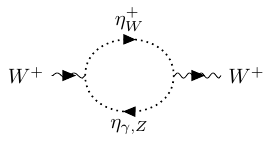}
    \end{subfigure}
    \begin{subfigure}{0.24\textwidth}
        \centering
        \includegraphics[width=\textwidth]{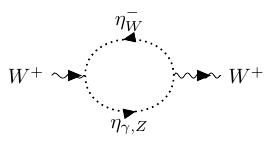}
    \end{subfigure}
    \begin{subfigure}{0.24\textwidth}
        \centering
        \includegraphics[width=\textwidth]{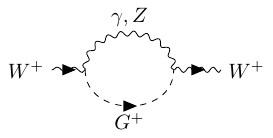}
    \end{subfigure}
    \begin{subfigure}{0.24\textwidth}
        \centering
        \includegraphics[width=\textwidth]{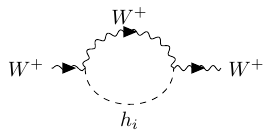}
    \end{subfigure}
    \begin{subfigure}{0.24\textwidth}
        \centering
        \includegraphics[width=\textwidth]{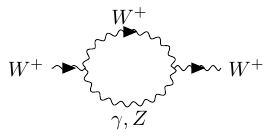}
    \end{subfigure}
    \caption{Feynman diagrams contributing to the $Z$ boson tadpole in the scotogenic model, where $i,j=1,2$. In the SM limit, $h_{i}$ and $a_{i}$ reduce to the SM Higgs boson $h$ and the would-be Goldstone boson $a$ of the $Z$ boson, respectively, while the contributions involving the $\chi, \eta_{R,I}$ and $H^{+}$ fields are absent.}
    \label{fig:WWSE_SE}
\end{figure}
\endgroup
The mass and field CT terms for the $W$ SEs are given in Equations~\ref{eqn:dMW2} and \ref{eqn:dZWW}:
\begingroup
{\allowdisplaybreaks
\begin{align}
\delta M_{W}^{2} &= \func{Re} \Sigma_{T}^{WW} \left( M_{W}^{2} \right) \nonumber \\
&= -\frac{1}{4 \pi ^2}
e^2 M_{W}^2 B_0\Big(M_{W}^2,0,M_{W}^2\Big) \nonumber \\ 
&+ \frac{1}{64 c_{w}^2 (-1+D) M_{Z}^2 \pi ^2 s_{w}^2}
e^2 (M_{\eta_{R}}-M_{H^{+}}-c_{w} M_{Z}) (M_{\eta_{R}}+M_{H^{+}}-c_{w} M_{Z}) \nonumber \\ 
&\times (M_{\eta_{R}}-M_{H^{+}}+c_{w} M_{Z}) (M_{\eta_{R}}+M_{H^{+}}+c_{w} M_{Z}) B_0\Big(M_{W}^2,M_{\eta_{R}}^2,M^{}_{H^{+}}\Big) \nonumber \\ 
&+ \frac{1}{64 c_{w}^2 (-1+D) M_{Z}^2 \pi ^2 s_{w}^2}
e^2 (M_{H^{+}}-M_{\eta_{I}}-c_{w} M_{Z}) (M_{H^{+}}+M_{\eta_{I}}-c_{w} M_{Z}) \nonumber \\
&\times (M_{H^{+}}-M_{\eta_{I}}+c_{w} M_{Z}) (M_{H^{+}}+M_{\eta_{I}}+c_{w} M_{Z}) B_0\Big(M_{W}^2,M^{}_{H^{+}},M_{\eta_{I}}^2\Big) \nonumber \\ 
&+ \frac{1}{64 c_{w}^2 (-1+D) \pi ^2 s_{w}^2}
e^2 M_{Z}^2 \Big(-3+4 s_{w}^2\Big) \Big(-15+12 D+(20-16 D) s_{w}^2+4 (-1+D) s_{w}^4\Big)  \nonumber \\
&\times B_0\Big(M_{W}^2,M_{W}^2,M_{Z}^2\Big) \nonumber \\ 
&+ \sum_{k=1}^3\frac{1}{32 c_{w}^2 (-1+D) M_{Z}^2 \pi ^2 s_{w}^2}
e^2 B_0\Big(M_{W}^2,0,m_{e_k}^2\Big) \Big(M_{W}^2-m_{e_k}^2\Big) \Big((-2+D) M_{W}^2+m_{e_k}^2\Big) \nonumber \\ 
&+ \sum_{k=1}^3 \Big(\sum_{l=1}^3\frac{1}{16 c_{w}^2 (-2+D) (-1+D) M_{Z}^2 \pi ^2 s_{w}^2}
e^2 N_{c} B_0\Big(0,m_{d_l}^2,m_{d_l}^2\Big) m_{d_l}^2 \nonumber \\
&\times \Big(-(-2+D) M_{W}^2+m_{d_l}^2-m_{u_k}^2\Big) U^{D,2}_{L,l,k} \nonumber \\ 
&+ \sum_{k=1}^3 \Big(\sum_{l=1}^3\frac{1}{32 c_{w}^2 (-1+D) M_{Z}^2 \pi ^2 s_{w}^2}
e^2 N_{c} B_0\Big(M_{W}^2,m_{d_l}^2,m_{u_k}^2\Big) \Big(-m_{d_l}^4+\Big(M_{W}^2-m_{u_k}^2\Big) \nonumber \\
&\times \Big((-2+D) M_{W}^2+m_{u_k}^2\Big)+m_{d_l}^2 \Big(-(-3+D) M_{W}^2+2 m_{u_k}^2\Big)\Big) U^{D,2}_{L,l,k} \nonumber \\ 
&- \sum_{i=1}^2\frac{1}{32 c_{w}^2 (-2+D) (-1+D) M_{Z}^2 \pi ^2 s_{w}^2}
e^2 B_0\Big(0,M_{h_i}^2,M_{h_i}^2\Big) M_{h_i}^2 \Big(-(-1+D) M_{W}^2+M_{h_i}^2\Big) U^{H,2}_{i1} \nonumber \\ 
&+ \sum_{i=1}^2\frac{1}{64 c_{w}^2 (-1+D) M_{Z}^2 \pi ^2 s_{w}^2}
e^2 B_0\Big(M_{W}^2,M_{W}^2,M_{h_i}^2\Big) \Big(4 c_{w}^4 (-1+D) M_{Z}^4-4 M_{W}^2 M_{h_i}^2+M_{h_i}^4\Big) U^{H,2}_{i1} \nonumber \\ 
&+ \sum_{i=1}^2 \Big(\sum_{k=1}^3\frac{1}{4 (-2+D) \pi ^2 s_{w}^2 M_{h_i}^2}
e^2 N_{c} B_0\Big(0,m_{d_k}^2,m_{d_k}^2\Big) m_{d_k}^4 U^{D}_{L,k,k} U^{H,2}_{i1} \nonumber \\ 
&+ \sum_{k=1}^3\frac{1}{16 c_{w}^2 (-2+D) (-1+D) M_{Z}^2 \pi ^2 s_{w}^2 M_{h_i}^2}
e^2 B_0\Big(0,m_{e_k}^2,m_{e_k}^2\Big) m_{e_k}^2 \Big(\Big(-(-2+D) M_{W}^2+m_{e_k}^2\Big) M_{h_i}^2 \nonumber \\
&+\sum_{i=1}^2 4 (-1+D) M_{W}^2 m_{e_k}^2 U^{H,2}_{i1}\Big) \nonumber \\ 
&+ \sum_{k=1}^3\frac{1}{16 c_{w}^2 (-2+D) (-1+D) M_{Z}^2 \pi ^2 s_{w}^2 M_{h_i}^2}
e^2 N_{c} B_0\Big(0,m_{u_k}^2,m_{u_k}^2\Big) m_{u_k}^2 \nonumber \\
&\times \Big(\sum_{l=1}^3 \Big((-2+D) M_{Z}^2 \Big(-1+s_{w}^2\Big)-m_{d_l}^2+m_{u_k}^2\Big) M_{h_i}^2 U^{D,2}_{L,l,k}+\sum_{i=1}^2 4 (-1+D) M_{W}^2 m_{u_k}^2 U^{H,2}_{i1}\Big) \nonumber \\ 
&+ \sum_{i=1}^2 \Big(\sum_{k=1}^3\frac{1}{4 (-2+D) \pi ^2 s_{w} v_{S} M_{h_i}^2}
c_{w} e M_{Z} B_0\Big(0,M_{\chi_k}^2,M_{\chi_k}^2\Big) M_{\chi_k}^4 U^{H}_{i1} U^{H}_{i2} \nonumber \\ 
&- \sum_{i=1}^2\frac{1}{32 (-2+D) \pi ^2 s_{w} M_{h_i}^2}
c_{w} M_{A}^2 M_{Z} B_0\Big(0,M_{A}^2,M_{A}^2\Big) U^{H}_{i1} (4 c_{w} \lambda_{7} M_{Z} s_{w} U^{H}_{i1}+e \lambda_{6} v_{S} U^{H}_{i2}) \nonumber \\ 
&- \frac{1}{32 c_{w}^2 (-2+D) (-1+D) M_{Z}^2 \pi ^2 s_{w}^2 M_{h_i}^2}
M_{H^{+}}^2 B_0\Big(0,M_{H^{+}}^2,M_{H^{+}}^2\Big) \Big(-e^2 \Big(M_{\eta_{R}}^2-2 M_{H^{+}}^2+M_{\eta_{I}}^2 \nonumber \\
&-4 M_{Z}^2+2 D M_{Z}^2-2 (-2+D) M_{Z}^2 s_{w}^2\Big) M_{h_i}^2+\sum_{i=1}^2 4 (-1+D) M_{Z}^3 s_{w} \Big(-1+s_{w}^2\Big) \nonumber \\
&\times U^{H}_{i1} \Big(2 \lambda_{3} M_{Z} s_{w} \Big(-1+s_{w}^2\Big) U^{H}_{i1}-c_{w} e \lambda_{8} v_{S} U^{H}_{i2}\Big)\Big) \nonumber \\ 
&- \frac{1}{32 c_{w}^2 (-2+D) (-1+D) M_{Z}^2 \pi ^2 s_{w}^2 M_{h_i}^2}
M_{\eta_{I}}^2 B_0\Big(0,M_{\eta_{I}}^2,M_{\eta_{I}}^2\Big) \Big(-e^2 \Big(M_{H^{+}}^2-M_{\eta_{I}}^2 \nonumber \\
&-(-2+D) M_{Z}^2 \Big(-1+s_{w}^2\Big)\Big) M_{h_i}^2+\sum_{i=1}^2 2 (-1+D) M_{Z}^3 s_{w} \Big(-1+s_{w}^2\Big) U^{H}_{i1} \nonumber \\
&\times \Big((2 \lambda_{3}+2 \lambda_{4}-\lambda_{5}) M_{Z} s_{w} \Big(-1+s_{w}^2\Big) U^{H}_{i1}-c_{w} e \lambda_{8} v_{S} U^{H}_{i2}\Big)\Big) \nonumber \\ 
&- \frac{1}{32 c_{w}^2 (-2+D) (-1+D) M_{Z}^2 \pi ^2 s_{w}^2 M_{h_i}^2}
M_{\eta_{R}}^2 B_0\Big(0,M_{\eta_{R}}^2,M_{\eta_{R}}^2\Big) \Big(e^2 \Big(M_{\eta_{R}}^2-M_{H^{+}}^2\nonumber \\
&+(-2+D) M_{Z}^2 \Big(-1+s_{w}^2\Big)\Big) M_{h_i}^2+\sum_{i=1}^2 2 (-1+D) M_{Z}^3 s_{w} \Big(-1+s_{w}^2\Big) U^{H}_{i1} \nonumber \\
&\times \Big((2 \lambda_{3}+2 \lambda_{4}+\lambda_{5}) M_{Z} s_{w} \Big(-1+s_{w}^2\Big) U^{H}_{i1}-c_{w} e \lambda_{8} v_{S} U^{H}_{i2}\Big)\Big) \nonumber \\ 
\label{eqn:dMW2}
\end{align}
}
\endgroup

\begingroup
{\allowdisplaybreaks
\begin{align}
\delta Z_{WW} &= -\func{Re} \left. \frac{\partial \Sigma_{T}^{WW}\left( p^{2} \right)}{\partial p^{2}} \right|_{p^{2} = M_{W}^{2}} \nonumber \\
&= \frac{1}{32 c_{w}^4 (-2+D) (-1+D) M_{Z}^4 \pi ^2 s_{w}^2}
e^2 M_{\eta_{R}}^2 \Big(-M_{\eta_{R}}^2+M_{H^{+}}^2\Big) B_0\Big(0,M_{\eta_{R}}^2,M_{\eta_{R}}^2\Big) \nonumber \\ 
&+ \frac{1}{32 c_{w}^4 (-2+D) (-1+D) M_{Z}^4 \pi ^2 s_{w}^2}
e^2 M_{H^{+}}^2 \Big(M_{\eta_{R}}^2-2 M_{H^{+}}^2+M_{\eta_{I}}^2\Big) B_0\Big(0,M_{H^{+}}^2,M_{H^{+}}^2\Big) \nonumber \\ 
&+ \frac{1}{32 c_{w}^4 (-2+D) (-1+D) M_{Z}^4 \pi ^2 s_{w}^2}
e^2 M_{\eta_{I}}^2 \Big(M_{H^{+}}^2-M_{\eta_{I}}^2\Big) B_0\Big(0,M_{\eta_{I}}^2,M_{\eta_{I}}^2\Big) \nonumber \\ 
&+ \frac{1}{32 c_{w}^4 (-2+D) (-1+D) \pi ^2}
e^2 \Big(7-4 D+4 (-2+D) s_{w}^2\Big) B_0\Big(0,M_{Z}^2,M_{Z}^2\Big) \nonumber \\ 
&+ \frac{1}{4 \pi ^2}
e^2 B_0\Big(M_{W}^2,0,M_{W}^2\Big) \nonumber \\ 
&+ \frac{1}{64 c_{w}^4 (-1+D) M_{Z}^4 \pi ^2 s_{w}^2}
-e^2 \Big(M_{\eta_{R}}^2-M_{H^{+}}^2+M_{W}^2\Big) \Big(-M_{\eta_{R}}^2+M_{H^{+}}^2+M_{W}^2\Big) B_0\Big(M_{W}^2,M_{\eta_{R}}^2,M_{H^{+}}^2\Big) \nonumber \\ 
&+ \frac{1}{64 c_{w}^4 (-1+D) M_{Z}^4 \pi ^2 s_{w}^2}
-e^2 \Big(M_{H^{+}}^2-M_{\eta_{I}}^2+M_{W}^2\Big) \Big(-M_{H^{+}}^2+M_{\eta_{I}}^2+M_{W}^2\Big) B_0\Big(M_{W}^2,M_{H^{+}}^2,M_{\eta_{I}}^2\Big) \nonumber \\ 
&+ \frac{1}{64 c_{w}^4 (-1+D) \pi ^2 s_{w}^2}
e^2 \Big(-9+12 D+(26-36 D) s_{w}^2+8 (-4+5 D) s_{w}^4-16 (-1+D) s_{w}^6\Big) \nonumber \\
&\times B_0\Big(M_{W}^2,M_{W}^2,M_{Z}^2\Big) \nonumber \\ 
&+ \sum_{k=1}^3\frac{1}{16 c_{w}^4 (-2+D) (-1+D) M_{Z}^4 \pi ^2 s_{w}^2}
e^2 B_0\Big(0,m_{e_k}^2,m_{e_k}^2\Big) m_{e_k}^4 \nonumber \\ 
&- \sum_{k=1}^3\frac{1}{32 c_{w}^4 (-1+D) M_{Z}^4 \pi ^2 s_{w}^2}
e^2 B_0\Big(M_{W}^2,0,m_{e_k}^2\Big) \Big(c_{w}^4 (-2+D) M_{Z}^4+m_{e_k}^4\Big) \nonumber \\ 
&+ \sum_{k=1}^3 \Big(\sum_{l=1}^3\frac{1}{16 c_{w}^4 (-2+D) (-1+D) M_{Z}^4 \pi ^2 s_{w}^2}
e^2 N_{c} B_0\Big(0,m_{d_l}^2,m_{d_l}^2\Big) m_{d_l}^2 \Big(m_{d_l}^2-m_{u_k}^2\Big) U^{D,2}_{L,l,k} \nonumber \\ 
&+ \sum_{k=1}^3 \Big(\sum_{l=1}^3\frac{1}{16 c_{w}^4 (-2+D) (-1+D) M_{Z}^4 \pi ^2 s_{w}^2}
e^2 N_{c} B_0\Big(0,m_{u_k}^2,m_{u_k}^2\Big) m_{u_k}^2 \Big(-m_{d_l}^2+m_{u_k}^2\Big) U^{D,2}_{L,l,k} \nonumber \\ 
&- \sum_{k=1}^3 \Big(\sum_{l=1}^3\frac{1}{32 c_{w}^4 (-1+D) M_{Z}^4 \pi ^2 s_{w}^2}
e^2 N_{c} B_0\Big(M_{W}^2,m_{d_l}^2,m_{u_k}^2\Big) \Big(c_{w}^4 (-2+D) M_{Z}^4+m_{d_l}^4\nonumber \\
&-2 m_{d_l}^2 m_{u_k}^2+m_{u_k}^4\Big) U^{D,2}_{L,l,k} \nonumber \\ 
&+ \sum_{i=1}^2\frac{1}{32 c_{w}^4 (-2+D) (-1+D) M_{Z}^4 \pi ^2 s_{w}^2}
e^2 B_0\Big(0,M_{h_i}^2,M_{h_i}^2\Big) M_{h_i}^2 \Big(M_{W}^2-M_{h_i}^2\Big) U^{H,2}_{i1} \nonumber \\ 
&+ \sum_{i=1}^2\frac{1}{64 c_{w}^4 (-1+D) M_{Z}^4 \pi ^2 s_{w}^2}
e^2 B_0\Big(M_{W}^2,M_{W}^2,M_{h_i}^2\Big) M_{h_i}^2 \Big(-2 M_{W}^2+M_{h_i}^2\Big) U^{H,2}_{i1} \nonumber \\ 
&+ \frac{1}{32 c_{w}^2 (-2+D) (-1+D) M_{Z}^2 \pi ^2 s_{w}^2}
e^2 B_0\Big(0,M_{W}^2,M_{W}^2\Big) \Big(M_{Z}^2 s_{w}^2+\sum_{i=1}^2 \Big(M_{Z}^2 \Big(-1+s_{w}^2\Big)+M_{h_i}^2\Big) U^{H,2}_{i1}\Big) \nonumber \\ 
&+ \frac{1}{64 c_{w}^2 (-1+D) M_{Z}^2 \pi ^2 s_{w}^2}
e^2 \Big(16 (-1+D) M_{Z}^4 s_{w}^2 \Big(-1+s_{w}^2\Big)^2 DB_{0}\Big(M_{W}^2,0,M_{W}^2\Big)\nonumber \\
&-\Big(M_{\eta_{R}}^4+\Big(M_{H^{+}}^2-M_{W}^2\Big)^2-2 M_{\eta_{R}}^2 \Big(M_{H^{+}}^2+M_{W}^2\Big)\Big) DB_{0}\Big(M_{W}^2,M_{\eta_{R}}^2,M_{H^{+}}^2\Big)\nonumber \\
&-M_{H^{+}}^4 DB_{0}\Big(M_{W}^2,M_{H^{+}}^2,M_{\eta_{I}}^2\Big)+2 M_{H^{+}}^2 M_{\eta_{I}}^2 DB_{0}\Big(M_{W}^2,M_{H^{+}}^2,M_{\eta_{I}}^2\Big)-M_{\eta_{I}}^4 DB_{0}\Big(M_{W}^2,M_{H^{+}}^2,M_{\eta_{I}}^2\Big)\nonumber \\
&+2 M_{H^{+}}^2 M_{W}^2 DB_{0}\Big(M_{W}^2,M_{H^{+}}^2,M_{\eta_{I}}^2\Big)+2 M_{\eta_{I}}^2 M_{W}^2 DB_{0}\Big(M_{W}^2,M_{H^{+}}^2,M_{\eta_{I}}^2\Big)\nonumber \\
&-c_{w}^4 M_{Z}^4 DB_{0}\Big(M_{W}^2,M_{H^{+}}^2,M_{\eta_{I}}^2\Big)-M_{Z}^4 DB_{0}\Big(M_{W}^2,M_{W}^2,M_{Z}^2\Big)+12 c_{w}^2 M_{Z}^4 DB_{0}\Big(M_{W}^2,M_{W}^2,M_{Z}^2\Big)\nonumber \\
&-36 c_{w}^4 M_{Z}^4 DB_{0}\Big(M_{W}^2,M_{W}^2,M_{Z}^2\Big)-20 c_{w}^6 M_{Z}^4 DB_{0}\Big(M_{W}^2,M_{W}^2,M_{Z}^2\Big)-4 c_{w}^2 D M_{Z}^4 DB_{0}\Big(M_{W}^2,M_{W}^2,M_{Z}^2\Big)\nonumber \\
&+20 c_{w}^4 D M_{Z}^4 DB_{0}\Big(M_{W}^2,M_{W}^2,M_{Z}^2\Big)+20 c_{w}^6 D M_{Z}^4 DB_{0}\Big(M_{W}^2,M_{W}^2,M_{Z}^2\Big)\nonumber \\
&+4 c_{w}^2 M_{Z}^4 s_{w}^4 DB_{0}\Big(M_{W}^2,M_{W}^2,M_{Z}^2\Big)-4 c_{w}^2 D M_{Z}^4 s_{w}^4 DB_{0}\Big(M_{W}^2,M_{W}^2,M_{Z}^2\Big)\nonumber \\
&+\sum_{k=1}^3 4 c_{w}^4 M_{Z}^4 DB_{0}\Big(M_{W}^2,0,m_{e_k}^2\Big)+\sum_{k=1}^3 -2 c_{w}^4 D M_{Z}^4 DB_{0}\Big(M_{W}^2,0,m_{e_k}^2\Big)\nonumber \\
&+\sum_{k=1}^3 -6 M_{W}^2 DB_{0}\Big(M_{W}^2,0,m_{e_k}^2\Big) m_{e_k}^2+\sum_{k=1}^3 2 D M_{W}^2 DB_{0}\Big(M_{W}^2,0,m_{e_k}^2\Big) m_{e_k}^2\nonumber \\
&+\sum_{k=1}^3 2 DB_{0}\Big(M_{W}^2,0,m_{e_k}^2\Big) m_{e_k}^4+\sum_{k=1}^3 \Big(\sum_{l=1}^3 4 c_{w}^4 M_{Z}^4 N_{c} DB_{0}\Big(M_{W}^2,m_{d_l}^2,m_{u_k}^2\Big) U^{D,2}_{L,l,k}\Big)\nonumber \\
&+\sum_{k=1}^3 \Big(\sum_{l=1}^3 -2 c_{w}^4 D M_{Z}^4 N_{c} DB_{0}\Big(M_{W}^2,m_{d_l}^2,m_{u_k}^2\Big) U^{D,2}_{L,l,k}\Big)\nonumber \\
&+\sum_{k=1}^3 \Big(\sum_{l=1}^3 2 D M_{W}^2 N_{c} DB_{0}\Big(M_{W}^2,m_{d_l}^2,m_{u_k}^2\Big) m_{d_l}^2 U^{D,2}_{L,l,k}\Big)\nonumber \\
&+\sum_{k=1}^3 \Big(\sum_{l=1}^3 -6 M_{Z}^2 N_{c} DB_{0}\Big(M_{W}^2,m_{d_l}^2,m_{u_k}^2\Big) m_{d_l}^2 U^{D,2}_{L,l,k}\Big)\nonumber \\
&+\sum_{k=1}^3 \Big(\sum_{l=1}^3 6 M_{Z}^2 N_{c} s_{w}^2 DB_{0}\Big(M_{W}^2,m_{d_l}^2,m_{u_k}^2\Big) m_{d_l}^2 U^{D,2}_{L,l,k}\Big)\nonumber \\
&+\sum_{k=1}^3 \Big(\sum_{l=1}^3 2 N_{c} DB_{0}\Big(M_{W}^2,m_{d_l}^2,m_{u_k}^2\Big) m_{d_l}^4 U^{D,2}_{L,l,k}\Big)\nonumber \\
&+\sum_{k=1}^3 \Big(\sum_{l=1}^3 2 D M_{W}^2 N_{c} DB_{0}\Big(M_{W}^2,m_{d_l}^2,m_{u_k}^2\Big) m_{u_k}^2 U^{D,2}_{L,l,k}\Big)\nonumber \\
&+\sum_{k=1}^3 \Big(\sum_{l=1}^3 -6 M_{Z}^2 N_{c} DB_{0}\Big(M_{W}^2,m_{d_l}^2,m_{u_k}^2\Big) m_{u_k}^2 U^{D,2}_{L,l,k}\Big)\nonumber \\
&+\sum_{k=1}^3 \Big(\sum_{l=1}^3 6 M_{Z}^2 N_{c} s_{w}^2 DB_{0}\Big(M_{W}^2,m_{d_l}^2,m_{u_k}^2\Big) m_{u_k}^2 U^{D,2}_{L,l,k}\Big)\nonumber \\
&+\sum_{k=1}^3 \Big(\sum_{l=1}^3 -4 N_{c} DB_{0}\Big(M_{W}^2,m_{d_l}^2,m_{u_k}^2\Big) m_{d_l}^2 m_{u_k}^2 U^{D,2}_{L,l,k}\Big)\nonumber \\
&+\sum_{k=1}^3 \Big(\sum_{l=1}^3 2 N_{c} DB_{0}\Big(M_{W}^2,m_{d_l}^2,m_{u_k}^2\Big) m_{u_k}^4 U^{D,2}_{L,l,k}\Big)+\sum_{i=1}^2 4 c_{w}^4 M_{Z}^4 DB_{0}\Big(M_{W}^2,M_{W}^2,M_{h_i}^2\Big) U^{H,2}_{i1}\nonumber \\
&+\sum_{i=1}^2 -4 c_{w}^4 D M_{Z}^4 DB_{0}\Big(M_{W}^2,M_{W}^2,M_{h_i}^2\Big) U^{H,2}_{i1}+\sum_{i=1}^2 4 M_{W}^2 DB_{0}\Big(M_{W}^2,M_{W}^2,M_{h_i}^2\Big) M_{h_i}^2 U^{H,2}_{i1}\nonumber \\
&+\sum_{i=1}^2 -DB_{0}\Big(M_{W}^2,M_{W}^2,M_{h_i}^2\Big) M_{h_i}^4 U^{H,2}_{i1}\Big) 
\label{eqn:dZWW}
\end{align}
}
\endgroup

\subsection{Scalar self-energy}\label{subapp:hhSE}

The diagrams contributing to the scalar tadpole and SE are seen in Figure~\ref{fig:hhSE_tadpole} and Figure~\ref{fig:hhSE_SE}.
\begingroup
\begin{figure}[t]
    \centering
    \begin{subfigure}{0.24\textwidth}
        \centering
        \includegraphics[width=\textwidth]{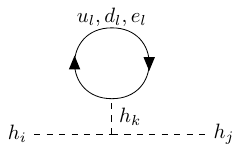}
    \end{subfigure}
    \begin{subfigure}{0.24\textwidth}
        \centering
        \includegraphics[width=\textwidth]{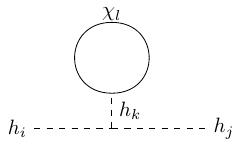}
    \end{subfigure}
    \begin{subfigure}{0.24\textwidth}
        \centering
        \includegraphics[width=\textwidth]{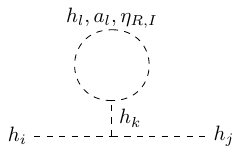}
    \end{subfigure}
    \begin{subfigure}{0.24\textwidth}
        \centering
        \includegraphics[width=\textwidth]{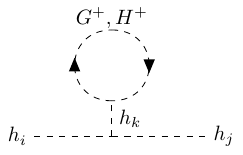}
    \end{subfigure}
    \begin{subfigure}{0.24\textwidth}
        \centering
        \includegraphics[width=\textwidth]{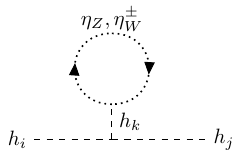}
    \end{subfigure}
    \begin{subfigure}{0.24\textwidth}
        \centering
        \includegraphics[width=\textwidth]{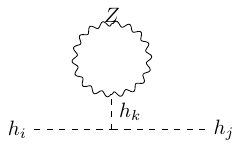}
    \end{subfigure}
    \begin{subfigure}{0.24\textwidth}
        \centering
        \includegraphics[width=\textwidth]{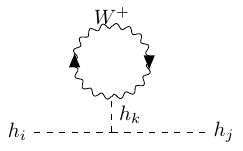}
    \end{subfigure}
    \caption{Feynman diagrams contributing to the scalar tadpole in the scotogenic model, where $i,j=1,2$. In the SM limit, $h_{i}$ and $a_{i}$ reduce to the SM Higgs boson $h$ and the would-be Goldstone boson $a$ of the $Z$ boson, respectively, while the contributions involving the $\chi, \eta_{R,I}$ and $H^{+}$ fields are absent.}
    \label{fig:hhSE_tadpole}
\end{figure}
\endgroup
\begingroup
\begin{figure}[t]
    \centering
    \begin{subfigure}{0.24\textwidth}
        \centering
        \includegraphics[width=\textwidth]{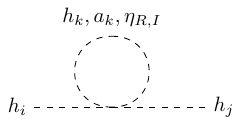}
    \end{subfigure}
    \begin{subfigure}{0.24\textwidth}
        \centering
        \includegraphics[width=\textwidth]{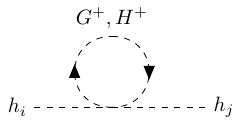}
    \end{subfigure}
    \begin{subfigure}{0.24\textwidth}
        \centering
        \includegraphics[width=\textwidth]{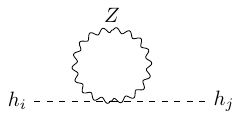}
    \end{subfigure}
    \begin{subfigure}{0.24\textwidth}
        \centering
        \includegraphics[width=\textwidth]{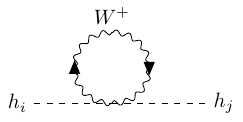}
    \end{subfigure}
    \begin{subfigure}{0.24\textwidth}
        \centering
        \includegraphics[width=\textwidth]{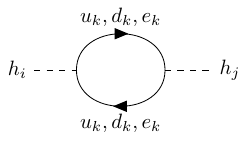}
    \end{subfigure}
    \begin{subfigure}{0.24\textwidth}
        \centering
        \includegraphics[width=\textwidth]{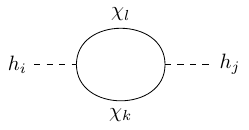}
    \end{subfigure}
    \begin{subfigure}{0.24\textwidth}
        \centering
        \includegraphics[width=\textwidth]{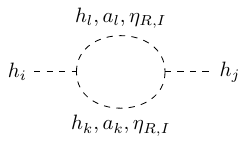}
    \end{subfigure}
    \begin{subfigure}{0.24\textwidth}
        \centering
        \includegraphics[width=\textwidth]{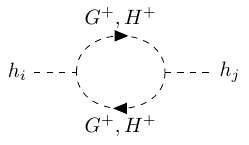}
    \end{subfigure}
    \begin{subfigure}{0.24\textwidth}
        \centering
        \includegraphics[width=\textwidth]{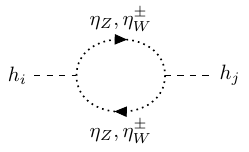}
    \end{subfigure}
    \begin{subfigure}{0.24\textwidth}
        \centering
        \includegraphics[width=\textwidth]{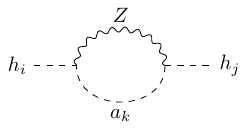}
    \end{subfigure}
    \begin{subfigure}{0.24\textwidth}
        \centering
        \includegraphics[width=\textwidth]{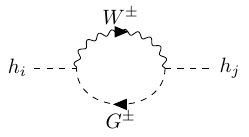}
    \end{subfigure}
    \begin{subfigure}{0.24\textwidth}
        \centering
        \includegraphics[width=\textwidth]{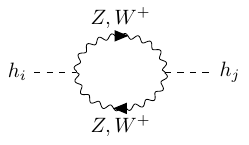}
    \end{subfigure}
    \caption{Feynman diagrams contributing to the scalar SE in the scotogenic model, where $i,j=1,2$. In the SM limit, $h_{i}$ and $a_{i}$ reduce to the SM Higgs boson $h$ and the would-be Goldstone boson $a$ of the $Z$ boson, respectively, while the contributions involving the $\chi, \eta_{R,I}$ and $H^{+}$ fields are absent.}
    \label{fig:hhSE_SE}
\end{figure}
\endgroup
Since the scalar sector is extended, the mass and field CTs are defined as follows. In what follows, we present the final analytic expressions, omitting the lengthy intermediate expressions for brevity.
\begingroup
{\allowdisplaybreaks
\begin{align}
    \delta Z_{h_{1}h_{1}} &= -\func{Re} \left[ \frac{\partial \Sigma_{h_{1}h_{1}}\left( p^{2} \right)}{\partial p^{2}} \right]_{p^{2}=m_{h_{1}}^{2}}, \\
    \delta Z_{h_{1}h_{2}} &= \frac{2}{m_{h_{1}}^{2} - m_{h_{2}}^{2}} \func{Re}\left[ \Sigma_{h_{1}h_{2}} \left( m_{h_{2}}^{2} \right) \right], \\
    \delta Z_{h_{2}h_{1}} &= \frac{2}{m_{h_{2}}^{2} - m_{h_{1}}^{2}} \func{Re}\left[ \Sigma_{h_{1}h_{2}} \left( m_{h_{1}}^{2} \right) \right], \\
    \delta Z_{h_{2}h_{2}} &= -\func{Re} \left[ \frac{\partial \Sigma_{h_{2}h_{2}}\left( p^{2} \right)}{\partial p^{2}} \right]_{p^{2}=m_{h_{2}}^{2}}, \\
    \delta M_{h_{1}}^{2} &= \func{Re}\left[ \Sigma_{h_{1}h_{1}} \left( M_{h_{1}}^{2} \right) \right], \\
    \delta M_{h_{2}}^{2} &= \func{Re}\left[ \Sigma_{h_{2}h_{2}} \left( M_{h_{2}}^{2} \right) \right].
\end{align}
}
\endgroup

\subsection{Charged lepton self-energy}\label{subapp:llSE}

The diagrams contributing to the chaged lepton tadpole and SE are seen in Figure~\ref{fig:llSE_tadpole} and Figure~\ref{fig:llSE_SE}.
\begingroup
\begin{figure}[t]
    \centering
    \begin{subfigure}{0.24\textwidth}
        \centering
        \includegraphics[width=\textwidth]{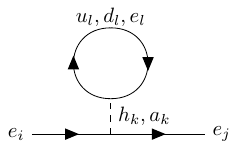}
    \end{subfigure}
    \begin{subfigure}{0.24\textwidth}
        \centering
        \includegraphics[width=\textwidth]{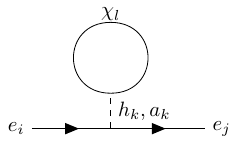}
    \end{subfigure}
    \begin{subfigure}{0.24\textwidth}
        \centering
        \includegraphics[width=\textwidth]{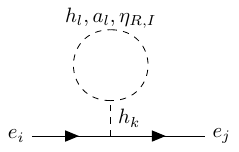}
    \end{subfigure}
    \begin{subfigure}{0.24\textwidth}
        \centering
        \includegraphics[width=\textwidth]{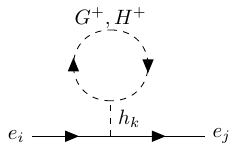}
    \end{subfigure}
    \begin{subfigure}{0.24\textwidth}
        \centering
        \includegraphics[width=\textwidth]{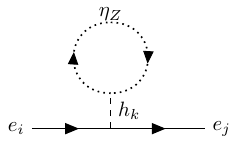}
    \end{subfigure}
    \begin{subfigure}{0.24\textwidth}
        \centering
        \includegraphics[width=\textwidth]{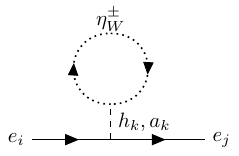}
    \end{subfigure}
    \begin{subfigure}{0.24\textwidth}
        \centering
        \includegraphics[width=\textwidth]{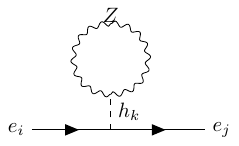}
    \end{subfigure}
    \begin{subfigure}{0.24\textwidth}
        \centering
        \includegraphics[width=\textwidth]{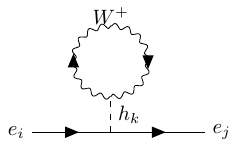}
    \end{subfigure}
    \caption{Feynman diagrams contributing to the charged lepton tadpole in the scotogenic model, where $i,j=1,2$. In the SM limit, $h_{i}$ and $a_{i}$ reduce to the SM Higgs boson $h$ and the would-be Goldstone boson $a$ of the $Z$ boson, respectively, while the contributions involving the $\chi, \eta_{R,I}$ and $H^{+}$ fields are absent.}
    \label{fig:llSE_tadpole}
\end{figure}
\endgroup
\begingroup
\begin{figure}[t]
    \centering
    \begin{subfigure}{0.24\textwidth}
        \centering
        \includegraphics[width=\textwidth]{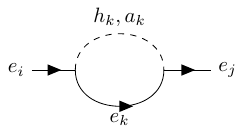}
    \end{subfigure}
    \begin{subfigure}{0.24\textwidth}
        \centering
        \includegraphics[width=\textwidth]{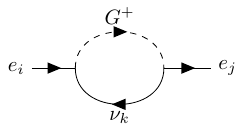}
    \end{subfigure}
    \begin{subfigure}{0.24\textwidth}
        \centering
        \includegraphics[width=\textwidth]{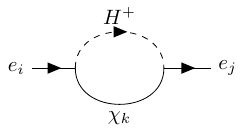}
    \end{subfigure}
    \begin{subfigure}{0.24\textwidth}
        \centering
        \includegraphics[width=\textwidth]{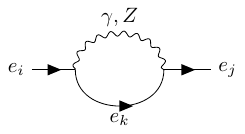}
    \end{subfigure}
    \begin{subfigure}{0.24\textwidth}
        \centering
        \includegraphics[width=\textwidth]{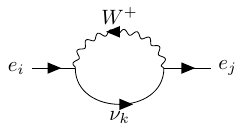}
    \end{subfigure}
    \caption{Feynman diagrams contributing to the charged lepton SE in the scotogenic model, where $i,j=1,2$. In the SM limit, $h_{i}$ and $a_{i}$ reduce to the SM Higgs boson $h$ and the would-be Goldstone boson $a$ of the $Z$ boson, respectively, while the contributions involving the $\chi, \eta_{R,I}$ and $H^{+}$ fields are absent.}
    \label{fig:llSE_SE}
\end{figure}
\endgroup
Following~\cite{Denner:1991kt}, the charged lepton propagator - with incoming and outgoing leptons indexed by $j$ and $i$, respectively - can be decomposed in terms of the projected self-energies with distinct Dirac structures as:
\begingroup
{\allowdisplaybreaks
\begin{align}
    \Gamma_{ij}^{f}\left( p \right) &= i \delta_{ij} \left( \slashed{p} - m_{i} \right) + i \left[ \slashed{p} P_{L} \Sigma_{ij}^{f,\gamma L}\left( p^{2} \right) + \slashed{p} P_{R} \Sigma_{ij}^{f,\gamma R}\left( p^{2} \right) + P_{L} \Sigma_{ij}^{f,L}\left( p^{2} \right) + P_{R} \Sigma_{ij}^{f,R}\left( p^{2} \right) \right]
\end{align}
}
\endgroup
where the mass terms appearing in front of $\Sigma_{ij}^{f,(L, R)}$ have been absorbed into the self-energy terms. The diagonal charged lepton field and mass CTs are defined as follows~\cite{Espriu:2002pz}:
\begingroup
{\allowdisplaybreaks
\begin{align}
    \delta Z_{ii}^{f, L} &= \Sigma_{ii}^{f, \gamma L} + X_{ii} + D_{ii}, \\
    \delta Z_{ii}^{f, R} &= \Sigma_{ii}^{f, \gamma R} - X_{ii} + D_{ii}, \\
    \overline{\delta Z}_{ii}^{f, L} &= \Sigma_{ii}^{f, \gamma L} - X_{ii} + D_{ii}, \\
    \overline{\delta Z}_{ii}^{f, R} &= \Sigma_{ii}^{f, \gamma R} + X_{ii} + D_{ii}, \\
    \delta m_{ii}^{f} &= -\frac{1}{2} \func{Re} \left( m_{i} \left( \Sigma_{ii}^{f, \gamma L}\left( m_{i}^{2} \right) + \Sigma_{ii}^{f, \gamma R}\left( m_{i}^{2} \right) \right) + \Sigma_{ii}^{f, L}\left( m_{i}^{2} \right) + \Sigma_{ii}^{f, R}\left( m_{i}^{2} \right) \right),
\end{align}
}
\endgroup
where
\begingroup
{\allowdisplaybreaks
\begin{align}
    X_{ii} &= \frac{1}{2} \frac{\Sigma_{ii}^{f, R}\left( m_{i}^{2} \right) - \Sigma_{ii}^{f, L}\left( m_{i}^{2} \right)}{m_{i}}, \\
    D_{ii} &= m_{i}^{2} \left( \Sigma_{ii}^{f, \gamma L, \prime}\left( m_{i}^{2} \right) + \Sigma_{ii}^{f, \gamma R, \prime}\left( m_{i}^{2} \right) \right) + m_{i} \left( \Sigma_{ii}^{f, L \prime}\left( m_{i}^{2} \right) + \Sigma_{ii}^{f, R \prime}\left( m_{i}^{2} \right) \right)
\end{align}
}
\endgroup
The off-diagonal ($i\neq j$) charged lepton field CTs are defined as follows:
\begingroup
{\allowdisplaybreaks
\begin{align}
    \delta Z_{ij}^{f, L} &= \frac{2}{m_{j}^{2}-m_{i}^{2}} \left[ \Sigma_{ij}^{f, \gamma R}\left( m_{j}^{2} \right) m_{i} m_{j} + \Sigma_{ij}^{f, \gamma L}\left( m_{j}^{2} \right) m_{j}^{2} + m_{i} \Sigma_{ij}^{f, L}\left( m_{j}^{2} \right) + m_{j} \Sigma_{ij}^{f, R}\left( m_{j}^{2} \right) \right], \\
    \delta Z_{ij}^{f, R} &= \frac{2}{m_{j}^{2}-m_{i}^{2}} \left[ \Sigma_{ij}^{f, \gamma L}\left( m_{j}^{2} \right) m_{i} m_{j} + \Sigma_{ij}^{f, \gamma R}\left( m_{j}^{2} \right) m_{j}^{2} + m_{i} \Sigma_{ij}^{f, R}\left( m_{j}^{2} \right) + m_{j} \Sigma_{ij}^{f, L}\left( m_{j}^{2} \right) \right], \\
    \overline{\delta Z}_{ij}^{f, L} &= \frac{2}{m_{i}^{2}-m_{j}^{2}} \left[ \Sigma_{ij}^{f, \gamma R}\left( m_{i}^{2} \right) m_{i} m_{j} + \Sigma_{ij}^{f, \gamma L}\left( m_{i}^{2} \right) m_{i}^{2} + m_{i} \Sigma_{ij}^{f, L}\left( m_{i}^{2} \right) + m_{j} \Sigma_{ij}^{f, R}\left( m_{i}^{2} \right) \right], \\
    \overline{\delta Z}_{ij}^{f, R} &= \frac{2}{m_{i}^{2}-m_{j}^{2}} \left[ \Sigma_{ij}^{f, \gamma L}\left( m_{i}^{2} \right) m_{i} m_{j} + \Sigma_{ij}^{f, \gamma R}\left( m_{i}^{2} \right) m_{i}^{2} + m_{i} \Sigma_{ij}^{f, R}\left( m_{i}^{2} \right) + m_{j} \Sigma_{ij}^{f, L}\left( m_{i}^{2} \right) \right].
\end{align}
}
\endgroup

\subsection{Neutrino self-energy}\label{subapp:vvSE}

In the neutrino sector, tadpole contributions are absent in both the SM and the BSM model, as there is no $h_{i}-v-v$ vertex. Consequently, only SE contributions arise. The diagrams contributing to the neutrino SE in the scotogenic model and the SM are shown in Figures~\ref{fig:vvSE_SE} and \ref{fig:SM_vvSE_SE}, respectively. 
\begingroup
\begin{figure}[t]
    \centering
    \begin{subfigure}{0.24\textwidth}
        \centering
        \includegraphics[width=\textwidth]{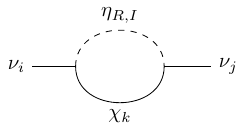}
    \end{subfigure}
    \begin{subfigure}{0.24\textwidth}
        \centering
        \includegraphics[width=\textwidth]{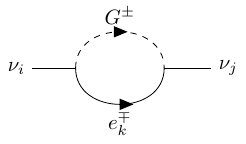}
    \end{subfigure}
    \begin{subfigure}{0.24\textwidth}
        \centering
        \includegraphics[width=\textwidth]{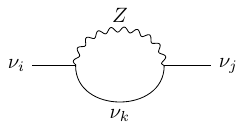}
    \end{subfigure}
    \begin{subfigure}{0.24\textwidth}
        \centering
        \includegraphics[width=\textwidth]{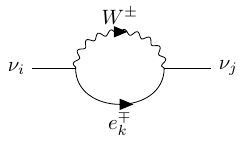}
    \end{subfigure}
    \caption{Feynman diagrams contributing to the neutrino SE in the scotogenic model, where $k=1,2,3$}
    \label{fig:vvSE_SE}
\end{figure}
\endgroup
\begingroup
\begin{figure}[t]
    \centering
    \begin{subfigure}{0.24\textwidth}
        \centering
        \includegraphics[width=\textwidth]{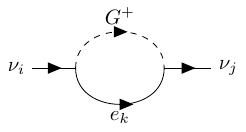}
    \end{subfigure}
    \begin{subfigure}{0.24\textwidth}
        \centering
        \includegraphics[width=\textwidth]{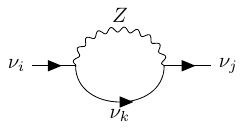}
    \end{subfigure}
    \begin{subfigure}{0.24\textwidth}
        \centering
        \includegraphics[width=\textwidth]{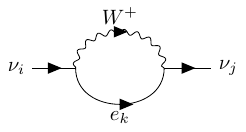}
    \end{subfigure}
    \caption{Feynman diagrams contributing to the neutrino SE in the SM, where $k=1,2,3$}
    \label{fig:SM_vvSE_SE}
\end{figure}
\endgroup
The diagonal and off-diagonal CTs for the neutrino mass and field renormalization are then derived following the same approach as in the charged lepton SE.

\section{Deriving bounded-from-below conditions}\label{app:bfb}

In this Section, we derive the BFB conditions discussed in the main body. We follow the approach of~\cite{Bhattacharyya:2015nca,CarcamoHernandez:2023wzf}. In order to derive the BFB conditions, it is enough to consider the quartic terms in the scalar potential:
\begingroup
\begin{equation}\label{eqn:sp_quartic}
\begin{split}
    V_{4} &= \frac{1}{4} \lambda_{1} ( H^{\dagger} H )^{2} + \frac{1}{4} \lambda_{2} ( \eta^{\dagger} \eta )^{2} + \lambda_{3} ( H^{\dagger} H ) ( \eta^{\dagger} \eta ) + \lambda_{4} ( H^{\dagger} \eta ) ( \eta^{\dagger} H ) \\
    &+ \frac{1}{4} \lambda_{5} \left[ \eta^{\dagger,2} H^2 + h.c. \right] + \frac{1}{4} \lambda_{6} ( S^{*} S )^{2} + \lambda_{7} ( H^{\dagger} H ) ( S^{*} S ) + \lambda_{8} ( \eta^{\dagger} \eta ) ( S^{*} S ),
\end{split}
\end{equation}
\endgroup
where the conditions $\lambda_{1}, \lambda_{2}, \lambda_{6} > 0$ follow from requiring the potential to be bounded from below along the individual field axes in the large-field limit. For the remaining conditions, it is convenient to parameterize each quartic term by a simple parameter as follows:
\begingroup
\begin{equation}\label{eqn:sp_quartic}
\begin{split}
    a &= H^{\dagger} H, \\
    b &= \eta^{\dagger} \eta, \\
    c &= S^{*} S, \\
    d &= \func{Re} ( H^{\dagger} \eta ), \\
    e &= \func{Im} ( H^{\dagger} \eta ), 
\end{split}
\end{equation}
\endgroup
where $a, b, c$ are by definition greater than $0$. Here, it is possible to define a relation between the parameters using Cauchy-Schwarz inequality equation~\ref{eqn:CS_inequality}.
\begingroup
\begin{equation} \label{eqn:CS_inequality}
    a b \geq d^{2} + e^{2}
\end{equation}
\endgroup
Under this circumstance, the scalar potential can be rewritten as follows:
\begingroup
\begin{equation}\label{eqn:sp_reduced}
\begin{split}
    V_{4} &= \frac{1}{8} \left( \sqrt{\lambda_{1}} a - \sqrt{\lambda_{2}} b \right)^{2} + \left( \frac{1}{4} \sqrt{\lambda_{1} \lambda_{2}} + \lambda_{3} \right) a b \\
    &+ \frac{1}{8} \left( \sqrt{\lambda_{1}} a - \sqrt{\lambda_{6}} c \right)^{2} + \left( \frac{1}{4} \sqrt{\lambda_{1} \lambda_{6}} + \lambda_{7} \right) a c \\
    &+ \frac{1}{8} \left( \sqrt{\lambda_{2}} b - \sqrt{\lambda_{6}} c \right)^{2} + \left( \frac{1}{4} \sqrt{\lambda_{2} \lambda_{6}} + \lambda_{8} \right) b c \\
    &+ \lambda_{4} \left( d^{2} + e^{2} \right) + \frac{1}{2} \lambda_{5} \left( d^{2} - e^{2} \right)
\end{split}
\end{equation}
\endgroup
We now analyze the scalar potential along the direction of each scalar field.

\subsection{\texorpdfstring{$\mathbf{a=0}$}{a is 0}}

When $a=0$, the parameters $d$ and $e$ vanish as a consequence of the inequality~\ref{eqn:CS_inequality}. The scalar potential then reduces to:
\begingroup
\begin{equation}
\begin{split}
    V_{4} \left( a = d = e = 0 \right) &= \frac{1}{4} \left( \sqrt{\lambda_{2}} b - \sqrt{\lambda_{6}} c \right)^{2} + \left( \frac{1}{2} \sqrt{\lambda_{2} \lambda_{6}} + \lambda_{8} \right) b c,
\end{split}
\end{equation}
\endgroup
which gives rise to the first condition:
\begingroup
\begin{equation}
    \lambda_{8} + \frac{1}{2} \sqrt{\lambda_{2} \lambda_{6}} > 0
\end{equation}
\endgroup

\subsection{\texorpdfstring{$\mathbf{b=0}$}{b is 0}}

When $b=0$, the derivation follows analogously to the previous case $a=0$.
\begingroup
\begin{equation}
\begin{split}
    V_{4} \left( b = d = e = 0 \right) &= \frac{1}{4} \left( \sqrt{\lambda_{1}} a - \sqrt{\lambda_{6}} c \right)^{2} + \left( \frac{1}{2} \sqrt{\lambda_{1} \lambda_{6}} + \lambda_{7} \right) a c,
\end{split}
\end{equation}
\endgroup
which yields the second condition:
\begingroup
\begin{equation}
    \lambda_{7} + \frac{1}{2} \sqrt{\lambda_{1} \lambda_{6}} > 0
\end{equation}
\endgroup

\subsection{\texorpdfstring{$\mathbf{c=0}$}{c is 0}}

When $c=0$, the reduced scalar potential is:
\begingroup
\begin{equation}
\begin{split}
    V_{4} \left( c = 0 \right) &= \frac{1}{4} \left( \sqrt{\lambda_{1}} a - \sqrt{\lambda_{2}} b \right)^{2} + \left( \frac{1}{2} \sqrt{\lambda_{1} \lambda_{2}} + \lambda_{3} \right) a b + \lambda_{4} \left( d^{2} + e^{2} \right) + \frac{1}{2} \lambda_{5} \left( d^{2} - e^{2} \right).
\end{split}
\end{equation}
\endgroup
To derive the BFB conditions, we need to examine additional field directions. First, we take the direction defined by $d = e = 0$. The scalar potential further reduces to:
\begingroup
\begin{equation}
\begin{split}
    V_{4} \left( c = d = e = 0 \right) &= \frac{1}{4} \left( \sqrt{\lambda_{1}} a - \sqrt{\lambda_{2}} b \right)^{2} + \left( \frac{1}{2} \sqrt{\lambda_{1} \lambda_{2}} + \lambda_{3} \right) a b,
\end{split}
\end{equation}
\endgroup
giving rise to the condition:
\begingroup
\begin{equation}
    \lambda_{3} + \frac{1}{2} \sqrt{\lambda_{1} \lambda_{2}} > 0
\end{equation}
\endgroup
Next, we consider the direction $a b = d^{2} + e^{2}$, which follows from the inequality. With this direction, the reduced scalar potential is:
\begingroup
\begin{equation}
\begin{split}
    V_{4} \left( c = 0, a b = d^{2} + e^{2} \right) &= \left( \lambda_{3} + \lambda_{4} + \frac{1}{2} \sqrt{\lambda_{1} \lambda_{2}} + \frac{1}{2} \lambda_{5} \right) d^{2} + \left( \lambda_{3} + \lambda_{4} + \frac{1}{2} \sqrt{\lambda_{1} \lambda_{2}} - \frac{1}{2} \lambda_{5} \right) e^{2}.
\end{split}
\end{equation}
\endgroup
Each coefficient must be non-negative. These two conditions can be combined into a single BFB condition:
\begingroup
\begin{equation}
\begin{split}
    \lambda_{3} + \lambda_{4} + \frac{1}{2} \sqrt{\lambda_{1} \lambda_{2}} > \frac{1}{2} \left| \lambda_{5} \right|.
\end{split}
\end{equation}
\endgroup

\subsection{\texorpdfstring{$\mathbf{a=\sqrt{\lambda_{6}/\lambda_{1}} c}$, $\mathbf{b=\sqrt{\lambda_{6}/\lambda_{2}} c}$}{some directions on a and b}}

With this direction, the scalar potential reduces to:
\begingroup
\begin{equation}
\begin{split}
    V_{4} \left( a = \sqrt{\frac{\lambda_{6}}{\lambda_{1}}} c, b = \sqrt{\frac{\lambda_{6}}{\lambda_{2}}} c \right) &= \left( \frac{3}{4} \lambda_{6} + \frac{\lambda_{3} \lambda_{6}}{\sqrt{\lambda_{1} \lambda_{2}}} + \lambda_{7} \sqrt{\frac{\lambda_{6}}{\lambda_{1}}} + \lambda_{8} \sqrt{\frac{\lambda_{6}}{\lambda_{2}}} \right) c^{2} \\
    &+ \lambda_{4} \left( d^{2} + e^{2} \right) + \frac{1}{2} \lambda_{5} \left( d^{2} - e^{2} \right) \\
    &\equiv \lambda_{a} c^{2} + \lambda_{4} \left( d^{2} + e^{2} \right) + \frac{1}{2} \lambda_{5} \left( d^{2} - e^{2} \right).
\end{split}
\end{equation}
\endgroup
In this direction, the inequality can be rewritten in terms of the parameter $c$ alone:
\begingroup
\begin{equation}
\begin{split}
    a b &\geq d^{2} + e^{2} \\
    \frac{\lambda_{6}}{\sqrt{\lambda_{1} \lambda_{2}}} c^{2} &\geq d^{2} + e^{2}.
\end{split}
\end{equation}
\endgroup
Substituting the inequality into the scalar potential, it then reduces to:
\begingroup
\begin{equation}
\begin{split}
    V_{4} \left( a = \sqrt{\frac{\lambda_{6}}{\lambda_{1}}} c, b = \sqrt{\frac{\lambda_{6}}{\lambda_{2}}} c \right) &= \lambda_{a} \frac{\sqrt{\lambda_{1} \lambda_{2}}}{\lambda_{6}} \left( d^{2} + e^{2} \right) + \lambda_{4} \left( d^{2} + e^{2} \right) + \frac{1}{2} \lambda_{5} \left( d^{2} - e^{2} \right) \\
    &= \left( \lambda_{a} \frac{\sqrt{\lambda_{1} \lambda_{2}}}{\lambda_{6}} + \lambda_{4} + \frac{1}{2} \lambda_{5} \right) d^{2} + \left( \lambda_{a} \frac{\sqrt{\lambda_{1} \lambda_{2}}}{\lambda_{6}} + \lambda_{4} - \frac{1}{2} \lambda_{5} \right) e^{2}.
\end{split}
\end{equation}
\endgroup
giving rise to the final BFB condition:
\begingroup
\begin{equation}
    \lambda_{a} \frac{\sqrt{\lambda_{1} \lambda_{2}}}{\lambda_{6}} + \lambda_{4} > \frac{1}{2} \left| \lambda_{5} \right|.
\end{equation}
\endgroup

\section{Renormalization Group Equations}\label{app:RGEs}

In this section, we collect all the one-loop RGEs generated by \texttt{SARAH}~\cite{Staub:2013tta,Staub:2015kfa}. The SM and Scotogenic RGEs are listed separately in the following subsections for ease of comparison. The convention for the one-loop RG equations adopted here is as follows:
\begingroup
\begin{equation}
    \beta\left( X \right) = \mu \frac{dX}{d\mu} \equiv \frac{1}{16\pi^{2}} \beta^{\left( 1 \right)} \left( X \right).
\end{equation}
\endgroup

\subsection{Standard Model}\label{subapp:SM_RGEs}
The SM RGEs are as follows:
\begingroup
{\allowdisplaybreaks
\begin{align}
    \beta^{\left( 1 \right)} \left( g_{1} \right) &= \frac{41}{10} g_{1}^{3} \\[6pt]
    \beta^{\left( 1 \right)} \left( g_{2} \right) &= -\frac{19}{6} g_{2}^{3} \\[6pt]
    \beta^{\left( 1 \right)} \left( g_{3} \right) &= -7 g_{3}^{3} \\[6pt]
    \beta^{\left( 1 \right)} ( y_{u}^{\left( i, j \right)} ) &= \left( -\frac{17}{20} g_{1}^{2} - \frac{9}{4} g_{2}^{2} - 8 g_{3}^{2} + 3 \func{Tr}\left[ y_{u} y_{u}^{\dagger} \right] + 3 \func{Tr}\left[ y_{d} y_{d}^{\dagger} \right] + \func{Tr}\left[ y_{e} y_{e}^{\dagger} \right] \right) y_{u}^{\left( i, j \right)} \notag \\
    &\quad - \frac{3}{2} \left( y_{u} y_{d}^{\dagger} y_{d} - y_{u} y_{u}^{\dagger} y_{u} \right)_{\left( i, j \right)} \\[6pt]
    \beta^{\left( 1 \right)} ( y_{d}^{\left( i, j \right)} ) &= \frac{1}{4} \left( -g_{1}^{2} - 9 g_{2}^{2} - 32 g_{3}^{2} + 12 \func{Tr}\left[ y_{d} y_{d}^{\dagger} \right] + 4 \func{Tr}\left[ y_{e} y_{e}^{\dagger} \right] + 12 \func{Tr}\left[ y_{u} y_{u}^{\dagger} \right] \right) y_{d}^{\left( i, j \right)} \notag \\
    &\quad + \frac{3}{2} \left( y_{d} y_{d}^{\dagger} y_{d} - y_{d} y_{u}^{\dagger} y_{u} \right)_{\left( i, j \right)} \\[6pt]
    \beta^{\left( 1 \right)} ( y_{e}^{\left( i, j \right)} ) &= \left( -\frac{9}{4} g_{1}^{2} - \frac{9}{4} g_{2}^{2} + 3 \func{Tr}\left[ y_{d} y_{d}^{\dagger} \right] + \func{Tr}\left[ y_{e} y_{e}^{\dagger} \right] + 3 \func{Tr}\left[ y_{u} y_{u}^{\dagger} \right] \right) y_{e}^{\left( i, j \right)} + \frac{3}{2} \left( y_{e} y_{e}^{\dagger} y_{e} \right)_{\left( i, j \right)} \\[6pt]
    \beta^{\left( 1 \right)} \left( \lambda \right) &= \frac{27}{50} g_{1}^{4} + \frac{9}{5} g_{1}^{2} g_{2}^{2} + \frac{9}{2} g_{2}^{4} - \frac{9}{5} g_{1}^{2} \lambda - 9 g_{2}^{2} \lambda + 6 \lambda^{2} \notag \\
    &\quad + 12 \lambda \func{Tr}\left[ y_{d} y_{d}^{\dagger} \right] + 4 \lambda \func{Tr}\left[ y_{e} y_{e}^{\dagger} \right] + 12 \lambda \func{Tr}\left[ y_{u} y_{u}^{\dagger} \right] \notag \\
    &\quad - 24 \func{Tr}\left[ y_{d} y_{d}^{\dagger} y_{d} y_{d}^{\dagger} \right] - 8 \func{Tr}\left[ y_{e} y_{e}^{\dagger} y_{e} y_{e}^{\dagger} \right] - 24 \func{Tr}\left[ y_{u} y_{u}^{\dagger} y_{u} y_{u}^{\dagger} \right]
\end{align}
}
\endgroup

\subsection{Scotogenic Model}\label{subapp:ScotoNHIH_RGEs}

The Scotogenic RGEs are as follows:

\begingroup
{\allowdisplaybreaks
\begin{align}
    \beta^{\left( 1 \right)} \left( g_{1} \right) &= \frac{21}{5} g_{1}^{3} \\[6pt]
    \beta^{\left( 1 \right)} \left( g_{2} \right) &= -3 g_{2}^{3} \\[6pt]
    \beta^{\left( 1 \right)} \left( g_{3} \right) &= -7 g_{3}^{3} \\[6pt]
    \beta^{\left( 1 \right)} ( y_{u}^{\left( i, j \right)} ) &= \left( -\frac{17}{20} g_{1}^{2} - \frac{9}{4} g_{2}^{2} - 8 g_{3}^{2} + 3 \func{Tr}\left[ y_{d} y_{d}^{\dagger} \right] + \func{Tr}\left[ y_{e} y_{e}^{\dagger} \right] + 3 \func{Tr}\left[ y_{u} y_{u}^{\dagger} \right] \right) y_{u}^{\left( i, j \right)} \notag \\
    &\quad - \frac{3}{2} \left( y_{u} y_{d}^{\dagger} y_{d} - y_{u} y_{u}^{\dagger} y_{u} \right)_{\left( i, j \right)} \\[6pt]
    \beta^{\left( 1 \right)} ( y_{d}^{\left( i, j \right)} ) &= \frac{1}{4} \left( -g_{1}^{2} - 9 g_{2}^{2} - 32 g_{3}^{2} + 12 \func{Tr}\left[ y_{d} y_{d}^{\dagger} \right] + 4 \func{Tr}\left[ y_{e} y_{e}^{\dagger} \right] + 12 \func{Tr}\left[ y_{u} y_{u}^{\dagger} \right] \right) y_{d}^{\left( i, j \right)} \notag \\
    &\quad + \frac{3}{2} \left( y_{d} y_{d}^{\dagger} y_{d} - y_{d} y_{u}^{\dagger} y_{u} \right)_{\left( i, j \right)} \\[6pt]
    \beta^{\left( 1 \right)} ( y_{e}^{\left( i, j \right)} ) &= \frac{1}{4} \left( -9 g_{1}^{2} - 9 g_{2}^{2} + 12 \func{Tr}\left[ y_{d} y_{d}^{\dagger} \right] + 4 \func{Tr}\left[ y_{e} y_{e}^{\dagger} \right] + 12 \func{Tr}\left[ y_{u} y_{u}^{\dagger} \right] \right) y_{e}^{\left( i, j \right)} \notag \\
    &\quad + \frac{1}{2} \left( g_{X} g_{X}^{\dagger} y_{e} \right)_{\left( i, j \right)} + \frac{3}{2} \left( y_{e} y_{e}^{\dagger} y_{e} \right)_{\left( i, j \right)} \\[6pt]
    \beta^{\left( 1 \right)} ( g_{X}^{\left( i, j \right)} ) &= \frac{1}{20} \left( -9 g_{1}^{2} - 45 g_{2}^{2} + 20 \func{Tr}\left[ g_{X} g_{X}^{\dagger} \right] \right) g_{X}^{\left( i, j \right)} \notag \\
    &\quad + \frac{1}{2} \left( g_{R} g_{R}^{*} g_{X} \right)_{\left( i, j \right)} + \frac{3}{2} \left( g_{X} g_{X}^{\dagger} g_{X} \right)_{\left( i, j \right)} + \frac{1}{20} \left( g_{X} y_{e}^{\dagger} y_{e} \right)_{\left( i, j \right)} \\[6pt]
    \beta^{\left( 1 \right)} ( g_{R}^{\left( i, j \right)} ) &= \frac{1}{2} \func{Tr}\left[ g_{R} g_{R}^{*} \right] g_{R}^{\left( i, j \right)} + \left( g_{R} g_{R}^{*} g_{R} \right)_{\left( i, j \right)} + \left( g_{R} g_{X}^{*} g_{X}^{T} \right)_{\left( i, j \right)} + \left( g_{X} g_{X}^{\dagger} g_{R} \right)_{\left( i, j \right)} \\[6pt]
    \beta^{\left( 1 \right)} \left( \lambda_{1} \right) &= \frac{27}{50} g_{1}^{4} + \frac{9}{5} g_{1}^{2} g_{2}^{2} + \frac{9}{2} g_{2}^{4} - \frac{9}{5} g_{1}^{2} \lambda_{1} - 9 g_{2}^{2} \lambda_{1} + 6 \lambda_{1}^{2} + 8 \lambda_{3}^{2} + 8 \lambda_{3} \lambda_{4} + 4 \lambda_{4}^{2} + \lambda_{5}^{2} + 4 \lambda_{7}^{2} \notag \\
    &\quad + 12 \lambda_{1} \func{Tr}\left[ y_{d} y_{d}^{\dagger} \right] + 4 \lambda_{1} \func{Tr}\left[ y_{e} y_{e}^{\dagger} \right] + 12 \lambda_{1} \func{Tr}\left[ y_{u} y_{u}^{\dagger} \right] \notag \\
    &\quad - 24 \func{Tr}\left[ y_{d} y_{d}^{\dagger} y_{d} y_{d}^{\dagger} \right] - 8 \func{Tr}\left[ y_{e} y_{e}^{\dagger} y_{e} y_{e}^{\dagger} \right] - 24 \func{Tr}\left[ y_{u} y_{u}^{\dagger} y_{u} y_{u}^{\dagger} \right] \\[6pt]
    \beta^{\left( 1 \right)} \left( \lambda_{2} \right) &= \frac{27}{50} g_{1}^{4} + \frac{9}{5} g_{1}^{2} g_{2}^{2} + \frac{9}{2} g_{2}^{4} - \frac{9}{5} g_{1}^{2} \lambda_{2} - 9 g_{2}^{2} \lambda_{2} + 6 \lambda_{2}^{2} + 8 \lambda_{3}^{2} + 8 \lambda_{3} \lambda_{4} + 4 \lambda_{4}^{2} + \lambda_{5}^{2} + 4 \lambda_{8}^{2} \notag \\
    &\quad + 4 \lambda_{2} \func{Tr}\left[ g_{X} g_{X}^{\dagger} \right] - 8 \func{Tr}\left[ g_{X} g_{X}^{\dagger} g_{X} g_{X}^{\dagger} \right] \\[6pt]
    \beta^{\left( 1 \right)} \left( \lambda_{3} \right) &= \frac{27}{100} g_{1}^{4} - \frac{9}{10} g_{1}^{2} g_{2}^{2} + \frac{9}{4} g_{2}^{4} - \frac{9}{5} g_{1}^{2} \lambda_{3} - 9 g_{2}^{2} \lambda_{3} + 3 \lambda_{1} \lambda_{3} + 3 \lambda_{2} \lambda_{3} + 4 \lambda_{3}^{2} \notag \\
    &\quad + \lambda_{1} \lambda_{4} + \lambda_{2} \lambda_{4} + 2 \lambda_{4}^{2} + \frac{1}{2} \lambda_{5}^{2} + 2 \lambda_{7} \lambda_{8} \notag \\
    &\quad + 2 \lambda_{3} \func{Tr}\left[ g_{X} g_{X}^{\dagger} \right] + 6 \lambda_{3} \func{Tr}\left[ y_{d} y_{d}^{\dagger} \right] + 2 \lambda_{3} \func{Tr}\left[ y_{e} y_{e}^{\dagger} \right] + 6 \lambda_{3} \func{Tr}\left[ y_{u} y_{u}^{\dagger} \right] \notag \\
    &\quad - 4 \func{Tr}\left[ g_{X} y_{e}^{\dagger} y_{e} g_{X}^{\dagger} \right] \\[6pt]
    \beta^{\left( 1 \right)} \left( \lambda_{4} \right) &= \frac{9}{5} g_{1}^{2} g_{2}^{2} - \frac{9}{5} g_{1}^{2} \lambda_{4} - 9 g_{2}^{2} \lambda_{4} + \lambda_{1} \lambda_{4} + \lambda_{2} \lambda_{4} + 8 \lambda_{3} \lambda_{4} + 4 \lambda_{4}^{2} + 2 \lambda_{5}^{2} \notag \\
    &\quad + 2 \lambda_{4} \func{Tr}\left[ g_{X} g_{X}^{\dagger} \right] + 6 \lambda_{4} \func{Tr}\left[ y_{d} y_{d}^{\dagger} \right] + 2 \lambda_{4} \func{Tr}\left[ y_{e} y_{e}^{\dagger} \right] + 6 \lambda_{4} \func{Tr}\left[ y_{u} y_{u}^{\dagger} \right] \notag \\
    &\quad + 4 \func{Tr}\left[ g_{X} y_{e}^{\dagger} y_{e} g_{X}^{\dagger} \right] \\[6pt]
    \beta^{\left( 1 \right)} \left( \lambda_{5} \right) &= - \frac{9}{5} g_{1}^{2} \lambda_{5} - 9 g_{2}^{2} \lambda_{5} + \lambda_{1} \lambda_{5} + \lambda_{2} \lambda_{5} + 8 \lambda_{3} \lambda_{5} + 12 \lambda_{4} \lambda_{5} \notag \\
    &\quad + 2 \lambda_{5} \func{Tr}\left[ g_{X} g_{X}^{\dagger} \right] + 6 \lambda_{5} \func{Tr}\left[ y_{d} y_{d}^{\dagger} \right] + 2 \lambda_{5} \func{Tr}\left[ y_{e} y_{e}^{\dagger} \right] + 6 \lambda_{5} \func{Tr}\left[ y_{u} y_{u}^{\dagger} \right] \\[6pt]
    \beta^{\left( 1 \right)} \left( \lambda_{6} \right) &= 5 \lambda_{6}^{2} + 8 \lambda_{7}^{2} + 8 \lambda_{8}^{2} + 2 \lambda_{6} \func{Tr}\left[ g_{R} g_{R}^{*} \right] - 4 \func{Tr}\left[ g_{R} g_{R}^{*} g_{R} g_{R}^{*} \right] \\[6pt]
    \beta^{\left( 1 \right)} \left( \lambda_{7} \right) &= - \frac{9}{10} g_{1}^{2} \lambda_{7} - \frac{9}{2} g_{2}^{2} \lambda_{7} + 3 \lambda_{1} \lambda_{7} + 2 \lambda_{6} \lambda_{7} + 4 \lambda_{7}^{2} + 4 \lambda_{3} \lambda_{8} + 2 \lambda_{4} \lambda_{8} \notag \\
    &\quad + \lambda_{7} \func{Tr}\left[ g_{R} g_{R}^{*} \right] + 6 \lambda_{7} \func{Tr}\left[ y_{d} y_{d}^{\dagger} \right] + 2 \lambda_{7} \func{Tr}\left[ y_{e} y_{e}^{\dagger} \right] + 6 \lambda_{7} \func{Tr}\left[ y_{u} y_{u}^{\dagger} \right] \\[6pt]
    \beta^{\left( 1 \right)} \left( \lambda_{8} \right) &= 4 \lambda_{3} \lambda_{7} + 2 \lambda_{4} \lambda_{7} - \frac{9}{10} g_{1}^{2} \lambda_{8} - \frac{9}{2} g_{2}^{2} \lambda_{8} + 3 \lambda_{2} \lambda_{8} + 2 \lambda_{6} \lambda_{8} + 4 \lambda_{8}^{2} \notag \\
    &\quad + \lambda_{8} \func{Tr}\left[ g_{R} g_{R}^{*} \right] + 2 \lambda_{8} \func{Tr}\left[ g_{X} g_{X}^{\dagger} \right] - 4 \func{Tr}\left[ g_{R} g_{R}^{*} g_{X} g_{X}^{\dagger} \right]
\end{align}
}
\endgroup

\bibliography{bibliography}

\end{document}